\documentclass[aps,pre,reprint,showpacs,superscriptaddress]{revtex4-2}

\usepackage{graphicx}
\usepackage{dcolumn}
\usepackage{bm}
\usepackage{xcolor}  
\expandafter\let\csname equation*\endcsname\relax
\expandafter\let\csname endequation*\endcsname\relax
\usepackage{amsmath}
\usepackage{graphicx}
\usepackage{soul}
\usepackage{float}
\usepackage{physics}
\emergencystretch=\maxdimen
\usepackage[export]{adjustbox}
\usepackage{hyperref}
\usepackage[mathlines]{lineno}

\begin{document}

\title{Stochastic thermodynamics of Fisher information}

\author{Pedro B. Melo}
\email{pedrobmelo@aluno.puc-rio.br}
 \affiliation{Departamento de F\'isica, Pontif\'icia Universidade Cat\'olica, 22452-970, Rio de Janeiro RJ, Brazil}
 \author{S\'ilvio M. \surname{Duarte~Queir\'os}}
 \affiliation{Centro Brasileiro de Pesquisas F\'isicas, 22290-180, Rio de Janeiro RJ, Brazil}
 \affiliation{INCT-Sistemas Complexos, Brazil}
\author{Welles A. M. Morgado}
\affiliation{Departamento de F\'isica, Pontif\'icia Universidade Cat\'olica, 22452-970, Rio de Janeiro RJ, Brazil}%
\affiliation{INCT-Sistemas Complexos, Brazil}

\date{\today}

\begin{abstract}
 In this manuscript, we investigate the stochastic thermodynamics of Fisher information (FI), meaning we characterize both the \textit{fluctuations} of FI, introducing a statistics of that quantity, and thermodynamic quantities. We introduce two initial conditions: an equilibrium initial condition and a minimum entropy initial condition, both under a protocol that drives the system to equilibrium. Its results indicate a dependence of the average FI on both the initial condition and path taken. Furthermore, the results point the chosen parameter directly affects the FI of thermodynamic quantities such as irreversible work and entropy, along with fluctuations of a stochastic FI. Last, we assess the further role of FI of the distribution of thermal quantities within the context of thermostatistical inequalities. \end{abstract}
\maketitle


\section{Introduction}

Although we can trace back to Maxwell and his demon~\cite{demon2002} the first hint over the thermodynamical framing of information via Shannon entropy, the interest in the subject has skyrocketed in recent years. Ultimately, besides findings such as the Landauer's principle~\cite{landauer1961} that one cannot delete information -- i.e., entropy increase -- without performing work, it prompted the analysis of information reservoirs~\cite{deffner2013,barato2014}, coexisting with the heat~\cite{callen1985} and work~\cite{niedenzu2016,morgado2016,medeiros2021} types of energy sources.

The thermodynamics-information tandem has also gained prominence due to the access and control of ever smaller systems which enables one to probe and bridge several thermodynamical and information theory theorems previously set forth as thought-experiments~\cite{Toyabe2010np}. The access to such small systems also has put the role of the fluctuations in the thermostatistical description of a system into the limelight~\cite{ciliberto2017}
and paved the way towards a probabilistic representation of thermodynamics~\cite{sekimoto2010,seifert2012,peliti2021,ParaguassuPRE2022}. This probabilistic description has extended to information matters; for instance, within stochastic thermodynamics of information, it has been determined the non-vanishing probability of deleting information whilst performing no work at all, which corresponds to an `informational free lunch' that would be impossible according to the canonical Landauer's principle~\cite{esposito2011,paraguassu2022}. 

Alongside $S$, which is interchangeably used in both thermodynamics and information, other important related forms enable the informational characterization of a system. Of particular interest is the Kullback-Leibler (KL) divergence~\cite{cover2006}
\begin{equation}
    \Delta I \equiv \mathcal{D}_{KL}\left[P(x)||P^{\prime}(x)\right] = \sum_{x} P(x) \ln \left[\frac{P(x)}{P^\prime (x)}\right].
\label{kl-distance}
\end{equation}
Although it cannot be formally considered a metric~\footnote{To begin with, it is not symmetric concerning the reference distribution.}, $\Delta I$ gives a divergence between probabilities in the probability space. An actual metric is obtained from the square root of a symmetric version of $\Delta I$ which admits different domains for $P (x)$ and $P^\prime (x)$ that is known as the Jensen-Shannon (JS) divergence. That geometrical interpretation has found a connection with thermodynamics, namely in establishing a way of measuring thermodynamic lengths~\cite{crooks2007} and information geometric inequalities in the thermodynamics of chemical reactions~\cite{yoshimura2021}. 

Another metric of sorts is the Fisher metric or Fisher information (FI)~\cite{fisher1922,Fisher1925,Nicholson2018,Ito2018}, 
\begin{eqnarray}
\mathcal{I}(\theta_i, \theta_j)\equiv \sum_{X} ~ P(X)~ \left( \pdv{\theta_i}\ln P(X)\right)\left( \pdv{\theta_j}\ln P(X)\right),~~~
\label{eq:Fisher information}
\end{eqnarray}
which is understood as Riemannian metric, giving a length measure $ds^2 = \sum_{ij}g_{ij}d\theta_{i}d\theta_{j}$, equivalent to the JS divergence ~\cite{fisher1922,Fisher1925}.

Informationally, $\mathcal{I}(\theta) \equiv \mathcal{I}(\theta_i, \theta_i)$ is also interpreted as the amount of information contained in a (continuous) stochastic variable $x$ on a parameter $\theta $ that defines the distribution $P(x;\theta)$, i.e., it allows for the best estimation of $\theta$. In practical terms, $\mathcal{I}(\theta)$ has been successfully employed in optimal experimental design in several areas of science and technology~\cite{atkinson1992,frieden2007} as well as a plethora of applications providing noteworthy insights; among those instances, we can mention biological~\cite{frank2009},  ecological~\cite{mayer2006}, quantum~\cite{min2022,hyllus2012,Toth2012,marvian2022}, and  financial~\cite{sahalia2008} systems to which we can add signal processing~\cite{pan2023},  computer and information thermodynamics~\cite{prokopenko2015,kirkpatrick2017,zhao2019} problems just to mention a few. For these cases, it is often asserted that for any $\theta$ parameter adopted, the Cram\'er-Rao bound \cite{Cramer_1946} imposes that the estimated quantity follows a distribution with variance $\sigma_{\theta}^{2}$,
\begin{eqnarray}
 \sigma_{\theta}^{2} \ge \frac{1}{\mathcal{I}(\theta)} \Leftrightarrow  \mathcal{I}(\theta)\ge \frac{1}{\sigma_{\theta}^{2}}.  
 \label{cramer-rao}
\end{eqnarray} 

When $\theta $ is the time, $t$, Eq.~(\ref{cramer-rao}) was recently used to establish a bound to the fluctuations of a time-dependent observable with the square root of $\mathcal{I}$ being a proxy for the intrinsic speed of evolution allowing the determination of hidden states~\cite{ito2020}. Also, it was shown that the thermodynamic uncertainty relations are particular cases of the Cram\'er-Rao bound \cite{TanVanVu_Uncertainty}.

In casting a non-equilibrium framework, stochastic thermodynamics is systematically studied considering changes in the state of the system by applying a given protocol that shifts a given set of its parameters into new values and hence a new state. Therefrom, stochastic thermodynamical relations have been derived often in the form of generating functions~\cite{jarzynski1997,hatano2001,seifert2005,speck2005} shown to be independent of the traits of the protocol driving the system~\cite{jarzynski2007}.

Heeding our knowledge about the probabilistic nature of the variation of entropy -- \textit{i.e.}, the metric gauging the distance between states when we particularly refer to $I$ -- of a non-equilibrium system, it is logical to question ourselves what occurs to the FI measure. Does it hinge on the path taken?  Does it depend on the condition defining the initial state, $\rho_0 (x_0)$? 
How do the fluctuations of the instantaneous FI behave? Answering these questions represents the core of the present manuscript that is organized as follows: in Sec.~\ref{sec-dynprob} we introduce our overdamped model and establish its probabilistics regarding position, and irreversible work and entropy production; in Sec.~\ref{sec-StochFI} define stochastic fisher information and provide it with a physical interpretation; in Sec.~\ref{sec-thermoFI} we evaluate further thermodynamical quantities for the stochastic FI. Last, in Sec.~\ref{sec-remarks} we present an overview of our results, explore their implications and hint future avenues of research on this subject.



\section{Dynamics and Probabilistics}
\label{sec-dynprob}
We start from an overdamped Langevin equation,
\begin{equation}
    \gamma \dot{x}(t) = -kx(t) + F(t) + \eta(t),
    \label{Langevin_eq}
\end{equation}
where $\langle \eta(t)\eta(t') \rangle = 2\gamma\beta^{-1}\delta(t - t')$, with  $\beta ^{-1}  \equiv k_{B}T$, 
ruling the dynamics of a particle in contact with a heat reservoir at temperature, $T$,
subjected to a harmonic potential, and that is under the influence of the well-liked driving force protocol, $F(t) = F_{0}(1 - e^{-t/\tau})$, see e.g., Ref.~\cite{morgado2010exact}. This protocol is chosen because it is easily tractable, with $\tau \rightarrow \infty$ representing the reversible work process. Herein we assume $k_{B} = 1$.

Using path integral methods~\cite{Wio_Path}, as described in Appendix \ref{AppA}, we obtain the transition probability $P[x_{t},t|x_{0},0]$ for the system, as detailed in the supplementary material (SM), yielding
\begin{eqnarray}
    P[x_{t},t|x_{0},0] = \sqrt{\frac{k \beta \Theta_{+}(t)}{4\pi}} \exp \left[-\frac{\beta e^{-\frac{2 t}{\tau }} \Theta _{-}(t)}{4 k \Gamma_{\tau}^2} \left(F_{0} \, \tau \, k \,e^{\frac{k t}{\gamma }}\right.\right.\nonumber\\\left.\left.-e^{t/\tau } (\gamma  F_{0}-\Gamma_{\tau} k x_{0})+(F_{0}-k x_{t}) \Gamma_{\tau} e^{t \left(\frac{k}{\gamma }+\frac{1}{\tau }\right)}\right)^2\right],~~~~~
\label{propagator}
\end{eqnarray}
with $\Theta _{\pm}(t) = \left(\coth \left(\frac{k t}{\gamma }\right)\pm1\right)$, and $\Gamma_{\tau} = (\gamma - k\tau)$. For $t \rightarrow \infty$, the system evolves into equilibrium; ie,
\begin{equation}
\lim_{t\rightarrow\infty} P[x_{t},t|x_{0},0] = \sqrt{\frac{\beta k }{2 \pi }} \exp \left[-\frac{(F_{0}-k x_{t})^2}{2 k T} \right].   
\end{equation}
%
%
Another interesting limit is the Heaviside step function limit at $t = 0$ protocol, $\tau \rightarrow 0^{+}$. The conditional probability tends to
\begin{eqnarray}
\lim_{\tau \rightarrow 0^{+}} P[x_{t},t|x_{0},0] & = & \sqrt{\frac{k \beta \Theta_{+}(t)}{4\pi}} \times \nonumber \\
&& \\
& \times &\exp \left[-\frac{\beta \gamma^2 e^{\frac{2k t}{ \gamma}}\Theta _{-}(t)}{4 k \gamma} (\gamma e^{-\frac{k t}{ \gamma}}\Bar{F}_{0} +  \Bar{F}_{t})^2 \right], \nonumber
\end{eqnarray}
with $\Bar{F}_{t} \equiv (F_0 - kx_t)$.

%
%
The joint probability of having positions $x_t$ at time $t$ and $x_0$ at $t=0$ is naturally  $P(x_{t},t;x_{0},0) = P[x_{t},t|x_{0},0]\rho_{0}(x_{0})$.
%

As stated, we assume two scenarios for the initial state of the system: either it is at equilibrium (maximal entropy state),
\begin{equation}
    \rho_{0}(x_{0}) = \sqrt{\frac{\beta k}{2\pi}}e^{-\frac{1}{2}\beta kx_{0}^2};
    \label{Initial_equilibrium_dist}
\end{equation}
or at an exact initial position (minimal entropy state),
\begin{equation}
    \rho_{0}(x_{0}) = \lim_{\varepsilon\rightarrow 0} \frac{\sqrt{\beta}}{\sqrt{4\pi\gamma \varepsilon}}\exp\left(-\frac{\beta x_{0}^{2}}{4\gamma \varepsilon}\right) = \delta(x_{0}),
    \label{diracdeltastate}
\end{equation}
matching a Dirac delta distribution. From each initial condition, we  obtain the position distribution function (PDF) $P(x_t, t) = \int \dd x_0 P[x_t,t|x_0,0] \rho_0(x_0)$. The resulting $P(x_t, t)$ are defined in the SM for both cases.

In a stochastic thermodynamics account of a system, energetic and informational quantities are described by probability distributions obtained using the propagator in Eq.~(\ref{propagator}), namely the stochastic irreversible work, $w_{irr}$, -- \textit{i.e.} the work that is not converted into equilibrium free energy $\Delta \mathcal{F}^{eq}$ \cite{esposito2011} -- and the stochastic variation of irreversible entropy $\Delta_{i} s$ performed in applying $F(t)$.

We start by defining the former as 
\begin{equation}
    w_{irr} \equiv w[x (t)] - \Delta \mathcal{F}^{eq} = w[x (t)] + \frac{F_{0}^2  \left(1-e^{-t/\tau }\right)^2}{2 k},
\end{equation}
where,
\begin{equation}
    w[x(t)] = -\int_{0}^{\tau} \dot{F}(t)~ x ~\dd t = -\frac{F_{0}}{\tau}\int_{0}^{\tau}e^{-\frac{t}{\tau}}x~\dd t.
\end{equation}
%
The average dissipated work made on the system, $W_{irr} \equiv \langle w[x(t)]\rangle_{eq} - \Delta \mathcal{F}^{eq}$, is bounded by the 2nd law of thermodynamics $W_{irr} \ge 0$~\footnote{
A more general case is obtained via the Landauer principle \cite{landauer1961}, $ W_{irr} \ge \frac{\Delta I}{\beta}$;
is the variation of the distance between a non-equilibrium distribution, $P(x)$, and its equilibrium counterpart, $\rho_{eq}(x)$.}. 
Herein, $\langle \dots \rangle_{eq}$ means that the average is computed over the equilibrium distribution that matches Eq.~(\ref{Initial_equilibrium_dist}).

The application of the external protocol $F(t)$ to a system at equilibrium Eq.~(\ref{Initial_equilibrium_dist}) at $t=0$ for a span equal to $t$ corresponds to the performance of an amount of irreversible work following a distribution that reads [details in the Supplementary Material (SM)], 
\begin{equation}
    P_{j}(w_{irr}) = \frac{1}{\sqrt{2\pi\sigma_{w_{irr},j}^2}}e^{\left(-\frac{(w_{irr} - \mu_{w_{irr},j})^2}{2\sigma_{w_{irr},j}^2}\right)},
\end{equation}
with $j = \{\mathcal{G}\equiv\mbox{ equilibrium},\delta\equiv\mbox{ delta function}\}$ representing the index for the initial condition of the distribution, which changes accordingly as described in the SM.
\begin{figure*}[t!]
    \centering
    \includegraphics[width=0.8\textwidth]{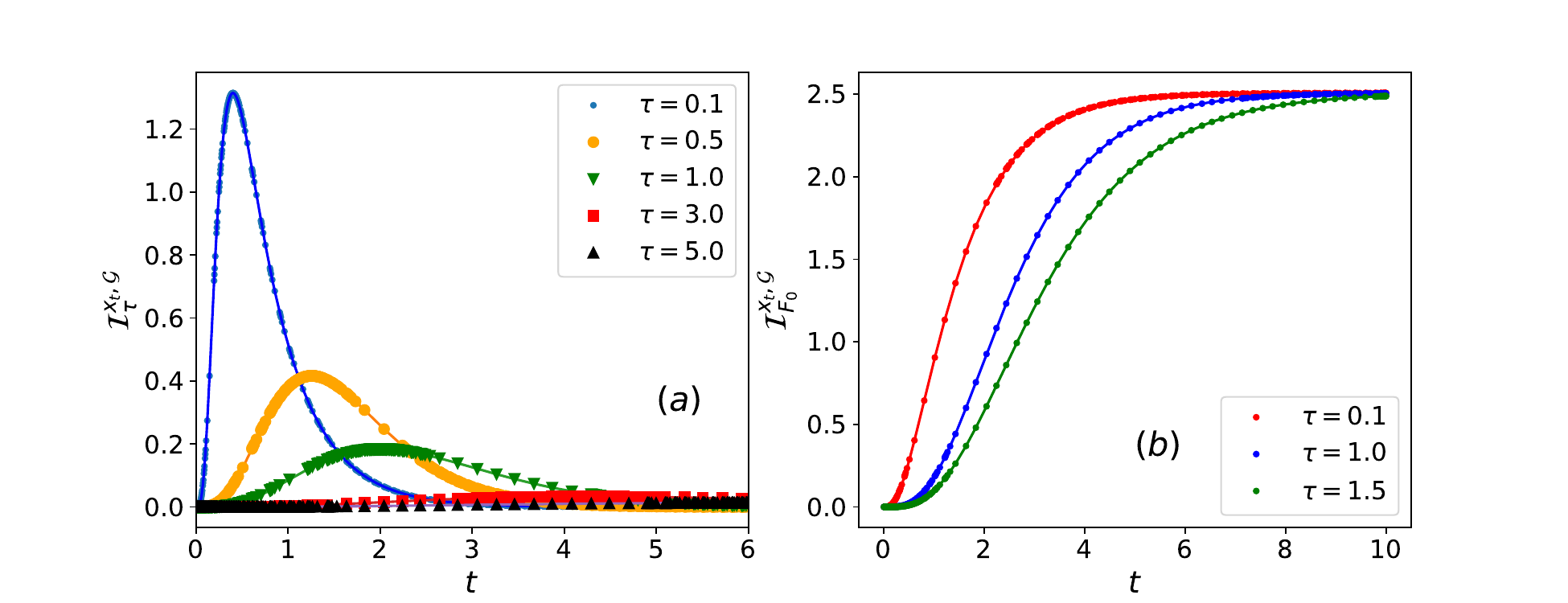}
    \caption{Main diagonal FI for $P(x_t, t)$. (a) Average FI $\mathcal{I}_{\tau}^{x_{t},\mathcal{G}}$ varying with $t$, and (b) $\mathcal{I}_{F_0}^{x_{t},\mathcal{G}}$ varying with $t$ for different characteristic times $\tau$, both with all other parameters set to $1$ (We avoid repetition of the sub-indexes $\tau$ or $F_0$ for the main diagonal terms of FI). FI for $\tau$ initially has a maximum at $t \propto \tau$, which varies as the time increases, and flattens, tending to a constant null FI in the infinite $\tau$ limit. Such behavior happens because, metrologically, the system becomes independent of the driving force $F(t)$ in that limit, meaning there cannot be any extractable information for its parameters. For $F_0$, FI does not variate with the parameter $F_0$, meaning there is always a finite variance for the estimation of $F_0$. Such behavior happens because there is no maxima FI for such a parameter, however, the maximal FI with time is directly related to $\tau$. However, in the infinite $\tau$ limit, such behavior does not occur as the system will become independent of $F(t)$.}
    \label{fig:FI_Avg}
\end{figure*}

%

\medskip
Out of equilibrium, one can define the counterpart to the equilibrium free energy, the non-equilibrium free energy,
\begin{equation}
\Delta f = \Delta U - \Delta s/\beta,    
\end{equation}
where $\Delta U$ is the internal energy variation, and $\Delta s$ is the stochastic entropy variation. The minimized version of $\Delta f$ is $\Delta \mathcal{F}^{eq}$. 
Accordingly, the irreversible entropy production,
\begin{equation}
\Delta_{i}s[x] \equiv \beta(w[x] - \Delta f),    
\end{equation}
for Eq.~(\ref{Langevin_eq}) equals (detailed algebra in SM),
\begin{equation}
    \frac{\Delta_{i} s[x_{t},t]}{\beta} =  -\Delta f[x_{t},t] - \frac{F_{0}}{k}\int_{0}^{t}\dd t'e^{-t'/\tau}x(t').
\end{equation}
Considering the initial condition of the maximum entropy of the system, given by a Gaussian initial distribution, the probability distribution of $\Delta_{i}s/\beta$ is given by
\begin{equation}
    P_{j}\left(\frac{\Delta_{i}s}{\beta}\right) = \frac{1}{\sqrt{2\pi\sigma_{s,j}^2}}\exp\left(-\frac{(\frac{\Delta_{i}s}{\beta} - \mu_{s,j})^2}{2\sigma_{s,j}^2}\right),
    \label{eq:Entropy distribution gaussian}
\end{equation}
where again $j = \{\mathcal{G},\delta\}$.

\section{Stochastic Fisher Information}
\label{sec-StochFI}
We proceed looking at FI given by Eq.~(\ref{eq:Fisher information}) for the dynamical observable, $x$. As our driving process is defined by two parameters, we define $\theta \equiv \{F_{0}, \tau\}$. Explicitly, $F_{0}$ is the scale parameter of $F(t)$, and $\tau$ the characteristic time scale of the protocol. The time-dependent FI is a symmetric matrix given by,
\begin{equation}
    \mathcal{I}(t) = \mqty(\mathcal{I}_{F_{0},F_0}&\mathcal{I}_{F_{0},\tau}\\\mathcal{I}_{F_{0},\tau}&\mathcal{I}_{\tau,\tau}),
    \label{fisher_matrix}
\end{equation}
where for the PDF $P(x_{t},t;\theta_{i})$, each term reads
\begin{equation}
    \mathcal{I}_{\theta_{i},\theta_{j}}^{x_{t},h} = \left\langle\left(\partial_{\theta_i}\ln P(x_{t},t;\theta_{i})\right)\left(\partial_{\theta_{j}} \ln P(x_{t},t;\theta_{j})\right)\right\rangle, \label{Fisher information for position}
\end{equation}
with $h = \mathcal{G}$ for the equilibrium initial condition or $h = \delta$ for a Dirac-delta initial condition.

\begin{figure*}[t!]
    \centering
    \includegraphics[width=1.0\textwidth]{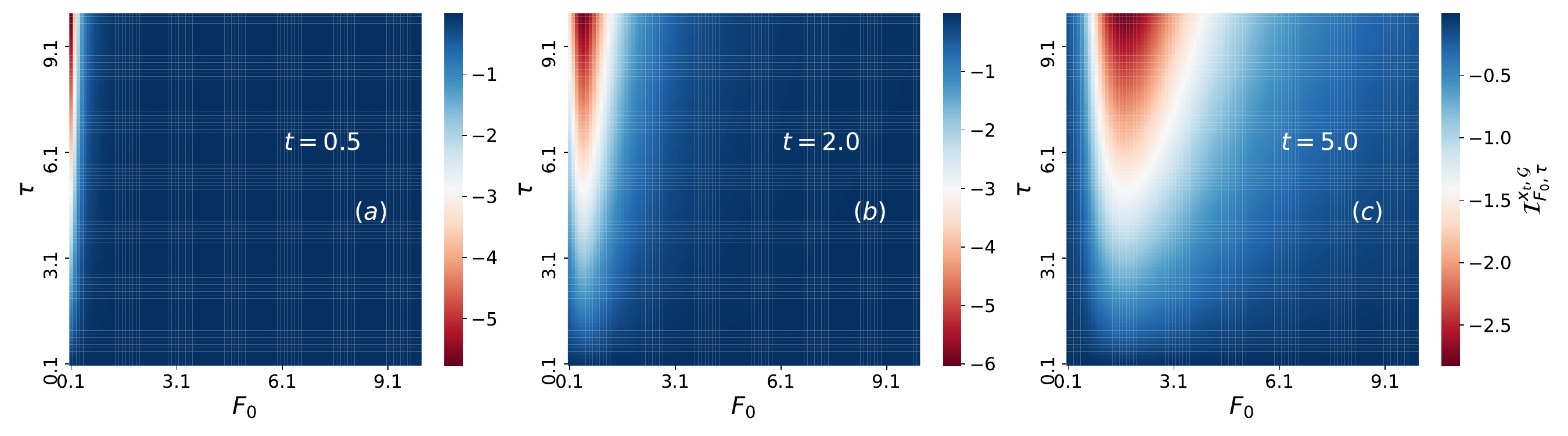}
    \caption{ Off-diagonal FI for $P(x_t, t)$ and equilibrium initial condition. The color indicates the value of $\mathcal{I}_{F_0,\tau}^{x_t,\mathcal{G}}$ for every figure. (a) $\mathcal{I}_{F_0,\tau}^{x_t,\mathcal{G}}$ for $t = 0.5$, (b) $\mathcal{I}_{F_0,\tau}^{x_t,\mathcal{G}}$ for $t = 2.0$, and (c) $\mathcal{I}_{F_0,\tau}^{x_t,\mathcal{G}}$ for $t = 5.0$, all heat maps varying with $\tau$ and $F_0$, with all other parameters set to $1$. Even though FI exhibits no divergence, its minima vary with time from (a) to (c), along with the variance growth of the extreme region. The presence of an extrema region suggests the presence of correlations between the parameters for the position PDF.}
    \label{fig:FI_Avg_off_diagonal}
\end{figure*}

For an initial equilibrium distribution, the variation of FI is independent of $F_{0}$ and equal to, 
\begin{equation}
    \mathcal{I}^{x_{t}\mathcal{G}}_{F_{0},F_{0}} = \mathcal{I}^{x_{t}\mathcal{G}}_{F_{0}} = \frac{ \beta \left(\Gamma _{\tau }  + k \tau e^{-\frac{t}{\tau}}-\gamma e^{-\frac{kt}{\gamma}}\right)^2}{k \Gamma _{\tau }^2},\label{eq16}
\end{equation}
reflecting the fact that at long times the protocol $F(t)$ displaces the equilibrium position whilst keeping the variance of the estimation finite. The FI, for $\tau$, yields
\begin{equation}
    \mathcal{I}^{x_{t}\mathcal{G}}_{\tau,\tau} = \mathcal{I}^{x_{t}\mathcal{G}}_{\tau} = \frac{ \beta  k F_{0}^2\left(\gamma   \tau  e^{-\frac{kt}{\gamma} }- e^{-\frac{t}{\tau }} (\gamma  \tau + t\Gamma_{\tau})\right)^2}{\tau ^2\Gamma_{\tau}^4}.
\end{equation}

Figure \ref{fig:FI_Avg} (a) displays FI diagonal terms, $\mathcal{I}^{x_{t}\mathcal{G}}_{\tau}$, and Fig.~\ref{fig:FI_Avg} (b) displays $\mathcal{I}^{x_{t}\mathcal{G}}_{F_0}$ varying with $\tau$ and $t$, respectively.~For $\mathcal{I}^{x_t,\mathcal{G}}_{\tau}$, a maximum forms at the value of the system time, which diverges for $\tau = 0$, and gets broad as $\tau$ grows. This behavior suggests the noisy environment via fluctuations from heat bath destroys the precision of the measurements. For $\mathcal{I}^{x_t,\mathcal{G}}_{F_0}$, FI grows with time reaching a stable maximum value regardless of $F_0$, for which the time of growth depends on $\tau$. In the long-term we verify $\lim_{t \rightarrow \infty}\mathcal{I}^{x_t,\mathcal{G}}_{F_0}(t) = \frac{\beta}{k}$.

On the other hand, the off-diagonal terms are equal to,
\begin{eqnarray}
 \mathcal{I}_{F_{0},\tau}^{x_{t}\mathcal{G}} = \mathcal{I}_{\tau,F_{0}}^{x_{t}\mathcal{G}} = -\frac{\beta  F_{0}}{\tau \Gamma_{\tau}^3}  \left(k \tau  e^{-\frac{t}{\tau}}+\Gamma_{\tau} -\gamma e^{-\frac{kt}{\gamma} }\right)\nonumber\\\left(\gamma \tau  e^{-\frac{kt}{\gamma}}-e^{-\frac{t}{\tau}} (\gamma\tau +t \Gamma_{\tau})\right),
\end{eqnarray}
displayed in Fig.~\ref{fig:FI_Avg_off_diagonal} for instants $t = 0.5, 2.0,$ and $5.0$. The plots exhibit a variation of FI with time, showing there is a minima region, where $F_0$ and $\tau$ can be more precisely measured, that disperses with time much likely the behavior of $\mathcal{I}_{\tau}^{x_t,\mathcal{G}}$. This result suggests that there are non-trivial correlations between $F_0$ and $\tau$ in the minima region.

Regarding matrix invariants, we determine the determinant and trace of FI matrix. While the determinant of the FI matrix for position PDF is trivial $\det \mathcal{I}^{x_{t},\mathcal{G}}(t) = 0$ for a Gaussian initial condition, the trace of $\mathcal{I}^{x_{t},\mathcal{G}}(t)$ equals
\begin{eqnarray}
    \Tr \mathcal{I}^{x_{t},\mathcal{G}}(t) = \frac{\beta}{k\Gamma _{\tau }^4} \left(\frac{k^2}{\tau ^2} \left(\gamma   \tau  e^{-\frac{kt}{\gamma} }- e^{-\frac{
   t}{\tau }} (\gamma  \tau + t\Gamma_{\tau})\right)^2\right.\nonumber\\\left.+\Gamma _{\tau }^2 \left(\Gamma _{\tau }  + k \tau e^{-\frac{t}{\tau}}-\gamma e^{-\frac{kt}{\gamma}}\right)^2\right),~~~
\end{eqnarray}
whereas Eq.~(\ref{Fisher information for position}) is the average of a Stochastic Fisher information (SFI),
\begin{equation}
\iota _{\theta_{i},\theta_j}^{\alpha\beta} \equiv \left(\frac{\partial_{\theta_i}P_{\beta}(\alpha)}{P_{\beta}(\alpha)} \right)\left(\frac{\partial_{\theta_j}P_{\beta}(\alpha)}{P_{\beta}(\alpha)}\right),    \label{stochasticfisherinfo}
\end{equation}
obtained for applications of the controlled protocol or measurements in some non-equilibrium system for which we are willing to find out the respective $\theta $ ($\alpha$ indicates the physical quantity used to extract the SFI and $\beta$ indicates the initial condition taken). It should be noticed that the definition of the SFI follows a parallel definition in stochastic thermodynamics where by taking the average of the stochastic entropy we obtain the Shannon entropy of the system. That is to say, Eq.~\ref{stochasticfisherinfo} is the product of derivatives of the stochastic entropy of the system).

\subsection{Physical interpretation of the SFI} 
The stochastic entropy $\Delta s$ indicates the total entropy variation \textit{i.e.} the entropy of the system and the bath \cite{peliti2021}. This entropy can be decomposed into two parts, the system entropy and the bath entropy, given by
\begin{eqnarray}
    \Delta s[x(t)] = \Delta s_{sys}(x) + \Delta s_{bath}[x(t)] \nonumber\\= -k_{B}\ln\frac{P(x_t,t)}{P(x_{0},0)} + \beta ~q[x(t)],
\end{eqnarray}
where the system entropy variation $\Delta s_{sys}(x)$ depends only on the initial and final states, and the bath entropy variation depends on the stochastic heat $q[x(t)]$ and on the bath temperature $T = k_{B}/\beta^{-1}$.

Geometrically, the SFI represents the metric of $\Delta s_{sys}(x)$, concerning some $\theta$ parameters, meaning one can define a metric of \textit{fluctuating} paths in the probability space, analogous to previously proposed fluctuating spacetime metrics \cite{Stochastic_metric}. On a thermodynamic-informational framework, it states that for a set of experiments that are made on a noisy environment, the amount of Fisher information obtained for a given realization of the environment follows a distribution $P(\iota)$ such that the average information in that distribution is the ``conventional'' Fisher information, obeying the Cram\'er-Rao bound on the average.

\subsection{Distribution of SFI} The Stochastic Fisher information~(\ref{stochasticfisherinfo}) follows an associated PDF, $P(\iota _{\theta_{i},\theta_j}^{\alpha\beta} )$ and non-vanishing statistical cumulants such as the variance $\sigma^{2~x_{t} \mathcal{G}}_{\theta_{i},\theta_{j}}$ of a SFI distribution of the position PDF are given by 
\begin{equation}
\sigma^{2~ x_{t}\mathcal{G}}_{\theta_{i},\theta_{j}} = (\frac{3}{\sqrt{2 \pi }}-1)\mathcal{I}_{\theta_i,\theta_j}^{x_{t}\mathcal{G}~2}. 
\end{equation}
For all $P(\iota _{\theta_{i},\theta_j}^{x_{t}\mathcal{G}})$, the skewness is 
\begin{equation}
\Tilde{\mu}_{3,\theta_{i}\theta_{j}}^{x_{t}\mathcal{G}} = \frac{\sqrt{2} \left(15+4 \pi -9 \sqrt{2 \pi }\right)}{\left(3 \sqrt{2}-2 \sqrt{\pi }\right)^{3/2} \sqrt[4]{\pi }},
\end{equation}
and the excess kurtosis is
\begin{equation}
\kappa^{x_{t}\mathcal{G}}_{\theta_{i},\theta_{j}} = \frac{3 \left(40-35 \sqrt{\frac{2}{\pi }}+4 \pi -12 \sqrt{2 \pi }\right)}{2 \left(-9-2 \pi +6 \sqrt{2 \pi }\right)}-3.
\end{equation}

For an initial delta distribution, the FI matrix of $x_{t}$ PDF is similar to the FI matrix for the equilibrium initial condition of $x_{t}$ PDF, up to a $\frac{\Theta_{+}(t)}{\sqrt{2\pi}}$ factor, and for the off-diagonal term as
\begin{equation}
    \mathcal{I}_{F_0,\tau}^{x_{t},\delta} = \frac{\Theta_{-}(t)}{2\sqrt{2\pi}} \mathcal{I}_{F_0,\tau}^{x_{t},\mathcal{G}}.
    \label{FI-DeltaDirac}
\end{equation}
Still for the $\delta$ initial condition, the variance depends on the $\Theta$ term; however, and similarly to the case of an initial equilibrium condition, it is always proportional to $\mathcal{I}_{\theta_i,\theta_j}^{x_{t} \delta ~2}$. For instance, $\sigma_{\theta_{i},\theta_{i}}^{2~ x_{t}\delta} = 2 \mathcal{I}_{\theta_{i},\theta_{i}}^{x_{t}\delta 
~2}$, for $\theta_{i} = \{F_{0},\tau\}$. The off-diagonal SFI has variance given by,
\begin{equation}
\sigma_{F_0,\tau}^{2~x_{t}\delta} = \left[6 \coth \left(\frac{k t}{\gamma }\right)+3 \csch^2\left(\frac{k t}{\gamma }\right)+5 \right]\mathcal{I}_{F_0,\tau}^{x_{t}\delta~2}.    
\end{equation}
The skewness of all the $P(\iota _{\theta_{i},\theta_j}^{x_{t}\delta})$ is given by $\Tilde{\mu}_{3}^{x_{t}\delta} = 5\sqrt{2}$, the excess kurtosis is $\kappa^{x_{t}\delta} = 12$ for all distributions. From the numerical simulation of the model and applying the Meerschaert-Scheffler method~\cite{m-s-method} we verified that for large values of the stochastic FI, $P(\iota _{\theta_{i},\theta_j}^{x_{t}\delta}) \sim \iota ^{-\nu}$, with $\nu = 2.91$.

The trace of $\mathcal{I}^{x_{t}\delta}(t)$ equals,
\begin{eqnarray}
    \Tr \mathcal{I}^{x_{t},\delta}(t) =\frac{\beta}{2
   k \Gamma _{\tau }^4}\nonumber\\ \times \left(\frac{F_0^2 k^2 \Theta _+(t)}{\tau
   ^2} \left(\gamma  \tau  e^{-\frac{k t}{\gamma }}-e^{-\frac{2 t}{\tau }} (\gamma  \tau -k t \tau +\gamma  t)\right)^2\right.\nonumber\\\left.+\Gamma _{\tau }^2 \Theta _+(t) \left(\Gamma _{\tau }-\gamma  e^{-\frac{k t}{\gamma }}+k \tau  e^{-\frac{t}{\tau }}\right)^2\right).~~
\end{eqnarray}
The determinant of $\mathcal{I}^{x_{t}\delta}(t)$ is non-zero given by
\begin{eqnarray}
    \det \mathcal{I}^{x_{t}\delta}(t) = \frac{\beta ^2 F_0^2 \left(\Theta_+^{2}(t)-1\right) e^{-4 t \left(\frac{k}{\gamma }+\frac{1}{\tau }\right)}}{4 \tau ^2 \Gamma _{\tau }^6}\nonumber\\\times\left(\Gamma _{\tau } e^{t \left(\frac{k}{\gamma }+\frac{1}{\tau }\right)}+k \tau  e^{\frac{k t}{\gamma
   }}-\gamma e^{t/\tau }\right)^2 \nonumber\\\times\left(\gamma  \tau  e^{t/\tau }-e^{\frac{k t}{\gamma }} \left(\gamma  \tau +t \Gamma _{\tau }\right)\right)^2.
\end{eqnarray}
In the long-term limit, $t/ \tau \gg 1$, we obtain $\det \mathcal{I}^{\delta}(t) \rightarrow 0$.

\begin{figure*}[t!]
    \centering
    \includegraphics[width=0.6\textwidth]{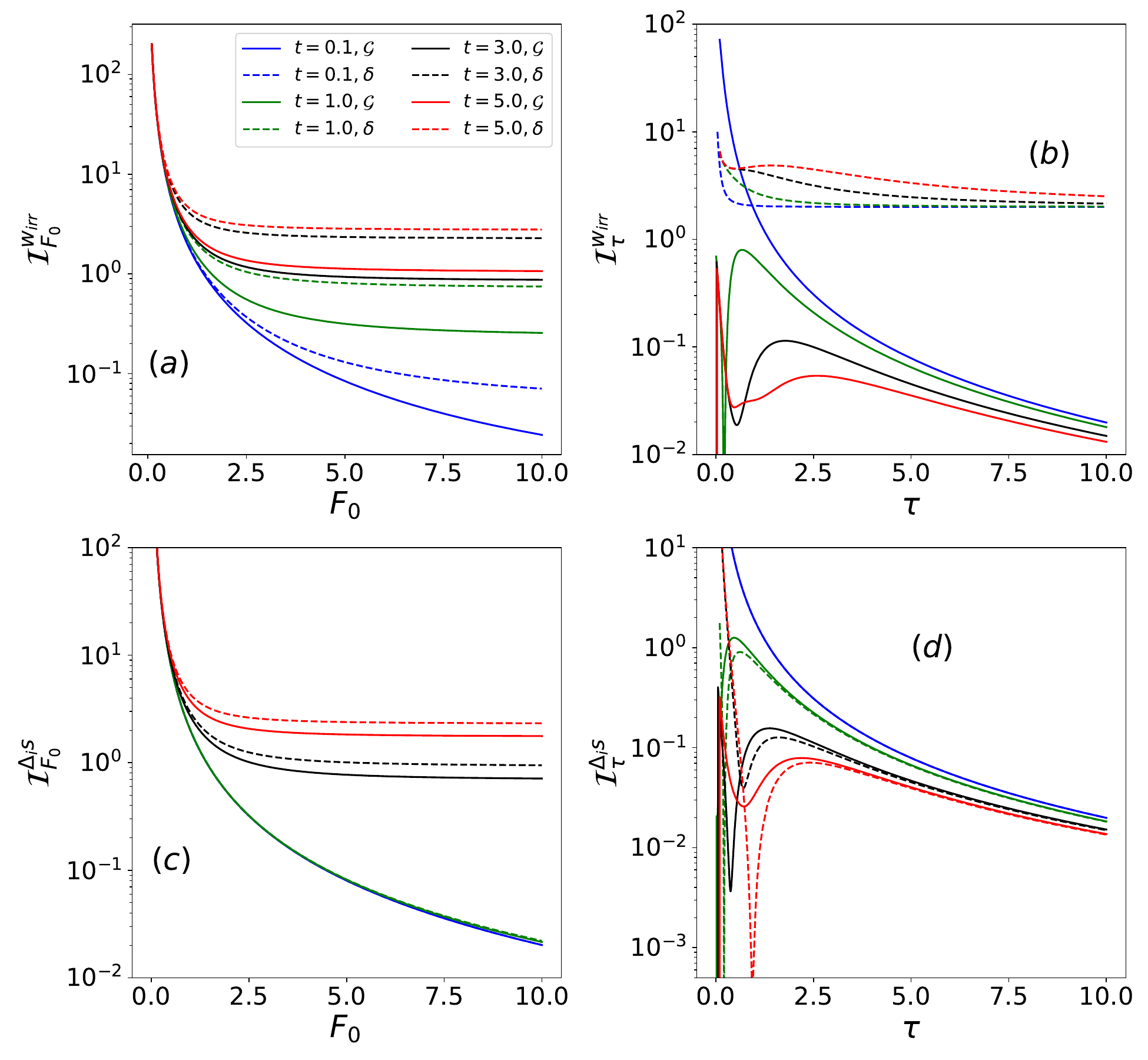}
    \caption{FI for $w_{irr}$ and $\Delta_{i} s$ at different times for both initial conditions, $\mathcal{G}$ for an equilibrium (Gaussian) initial condition and $\delta$ for an initial Dirac delta condition. (a) $\mathcal{I}_{F_0,F_0}^{w_{irr}}$ varying with $F_0$ for both initial conditions, indicated by solid lines (equilibrium initial condition), and dashed lines (Dirac delta initial condition). (b) $\mathcal{I}_{\tau,\tau}^{w_{irr}}$ varying with $\tau$ for both initial conditions, indicated in the same manner as in (a). (c) $\mathcal{I}_{F_0, F_0}^{\Delta_i s}$ varying with $F_0$, considering the same initial conditions as (a) and (b). (d) $\mathcal{I}_{\tau,\tau}^{\Delta_i s}$ varying with $\tau$, again with the same initial conditions as previous subfigures. $\mathcal{I}_{F_0,F_0}$ shows a universal behavior, given by Eq.~(\ref{eq:fisher_information_thermo}). Regarding $\mathcal{I}_{\tau, \tau}$, for $w_{irr}$ it exhibits a peak variation for an initial equilibrium condition, whereas in the Dirac delta initial condition, the divergence at $\tau = 0$ persists, along with a second finite peak that appears as time passes. For $\Delta_i s$, $\mathcal{I}_{\tau,\tau}^{\Delta_i s}$ is qualitatively similar to the equilibrium initial condition for $\mathcal{I}_{\tau,\tau}^{w_{irr}}$, with no difference between both initial conditions.}
    \label{fig:Fisher wirr}
\end{figure*}
\medskip

\section{Thermodynamics of Fisher Information}
\label{sec-thermoFI}

Figure~\ref{fig:Fisher wirr} synthesizes the behavior of     Eq.~(\ref{fisher_matrix}), where $\theta =\{F_0,\tau \}$, for both $w_{irr}$ and $\Delta_i s$, with either of the initial conditions. It is interesting to ask about how are FI averages and central moments of the distributions $P(\iota_{\theta_i,\theta_j})$ related to the production of $w_{irr}$ and $\Delta_i s$.

Our results indicate that the analytical expression for both FI of $F_{0}$ for both initial conditions is given by
\begin{equation}
    \mathcal{I}^{i,j}_{F_{0},F_{0}} = \frac{2}{F_{0}^2} + \alpha_{ij}(t)~(i = \{w_{irr}, \Delta_{i} s\},~j=\{\mathcal{G},\delta\}). \label{eq:fisher_information_thermo}
\end{equation}
Note that the $\mathcal{I}^{i,j}_{F_{0},F_{0}}$ above describes the FI of a SFI distribution, and it is not of the form of Eq.~(\ref{eq16}).
In Eq.~(\ref{eq:fisher_information_thermo}), $\alpha_{ij}(t)$ is a time-dependent quantity dependent on $i$ the thermodynamical quantity and the $j$ initial conditions, with all $\alpha_{ij}(t)$ detailed in the SM. The term $\alpha_{ij}(t)$ has the asymptotic limit,
\begin{equation}
\lim_{t \rightarrow \infty}\alpha_{ij}(t) = \frac{\beta (\gamma +2 k \tau )^2}{\pi  k^2 \tau },
\end{equation}
that is independent of both the thermodynamic quantity and the initial condition. Figures~\ref{fig:Fisher wirr} (a) and (c) indicate FI for $\mathcal{I}_{F_{0}}$, for different times, with equilibrium and minimal entropy initial conditions (solid lines and dashed lines, respectively) at different times. Both figures indicate a similar, universal behavior of FI, which diverges for $F_{0} = 0$, and its plateau asymptotic raises as $t$ grows, achieving the limit mentioned above.

For $\tau$, FI exhibits different behavior for both $w_{irr}$ and $\Delta_i s$. For these two quantities, FI exhibits a divergence for small times at $\tau \rightarrow 0$, as shown in Figs.~\ref{fig:Fisher wirr}(b) and ~\ref{fig:Fisher wirr}(d). For $\mathcal{I}_{\tau,\tau}^{w_{irr}}$ with equilibrium initial conditions, the time evolution ends the divergence at $\tau \rightarrow 0$, and imposes a two-peak regime, one close to $\tau = 0$, and the other that varies (and broadens) with time. The same is not true for the case with initial delta condition, as the divergence holds for all time values. However, a second peak appears at time passes and the asymptotic behavior is at a different value of $\mathcal{I}_{\tau,\tau}^{w_{irr}}$. $\mathcal{I}_{\tau,\tau}^{\Delta_{i}s}$ shows the same qualitative behavior as $\mathcal{I}_{\tau,\tau}^{w_{irr}}$ for the peak structure, however the asymptotic behavior for both initial conditions is the same.

\begin{figure}[t!]
    \centering
    \includegraphics[width=0.5\textwidth]{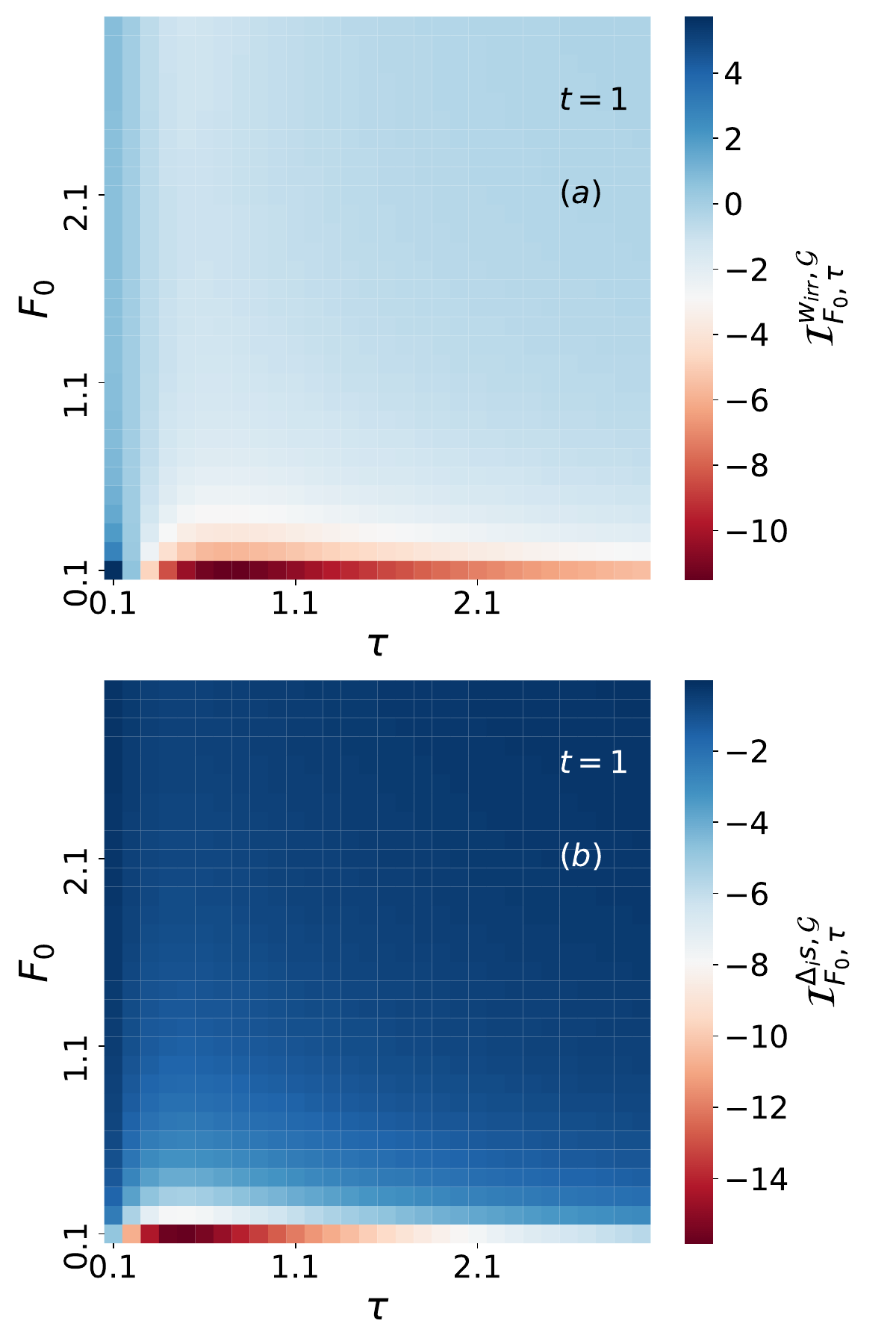}
    \caption{Off-diagonal FI for an initial equilibrium condition. (a) $\mathcal{I}_{F_0, \tau}^{w_{irr}, \mathcal{G}}$, for each set of parameters $(F_0, \tau)$. (b) $\mathcal{I}_{F_0,\tau}^{\Delta_i s~\mathcal{G}}$, for each set of parameters $(F_0, \tau)$. In (a), the off-diagonal FI is maximum for $F_0 \rightarrow 0$, with $\tau \approx t = 1$. A similar behavior happens for $\Delta_{i} s$ in (b), with the difference that for $F_{0} > 0$, $w_{irr}$ has a off-diagonal FI with non-trivial values. Our results indicate that the off-diagonal FI represents a correlation between the parameters, which is stronger as $F_0$ tends to $0$ in the region of $\tau \approx t$, but still exhibits weaker signatures for bigger values of $F_0$.}
    \label{fig:Fisher_wirr_off}
\end{figure}

The off-diagonal terms $\mathcal{I}_{F_0, \tau}^{\{w_{irr},\Delta_i s\}, \mathcal{G}}$ are presented in Figs.~\ref{fig:Fisher_wirr_off}(a) and \ref{fig:Fisher_wirr_off}(b), with expressions detailed in sections III and IV of the SM. For (a) and (b), our results display similar qualitative behavior, where a possible interpretation to such behavior is that $\mathcal{I}_{F_0, \tau}$ represents the correlation between the parameters $F_0$ and $\tau$. For both quantities, a maxima of FI around $\tau \approx t$ indicates a behavior a similar to the off-diagonal FI of $P(x_t)$. Besides that, for both $w_{irr}$ and $\Delta_i s$ there is a region for $F_0 > 0$ with weaker signatures of correlations.

\section{Concluding remarks}
\label{sec-remarks}

In this manuscript, we have studied the Fisher information, $\mathcal{I}_{\theta_i, \theta_j}^{\alpha\beta} \equiv \langle \iota_{\theta_i,\theta_j}^{\alpha\beta}\rangle$, of the distribution for the position, $x_{t}$, as well as stochastic energetic quantities like irreversible work, $w_{irr}$, and produced entropy, $\Delta_{i}s$, given the driving protocol applied to an overdamped system. We have done so by taking into consideration the stochastic nature of $\iota_{\theta_i, \theta_j}^{\alpha\beta}$, which opens a further avenue of study over how fluctuations in parameter estimation are bounded for systems out-of-equilibrium. 

For $P(x_t, t)$, the FI is parameter-free when $\theta_{i} = \theta_{j} = F_{0}$. This implies the variance of the estimator for the force $F_0$ reaches a constant maximal value; thus, the parameter $F_0$ is measured with a dwindling uncertainty (see figure~\ref{fig:FI_Avg}) that converges to a minimum non-zero value as time elapses. This means that the maximum attainability of information about $F_{0}$ happens for any value of such parameter, but needs to surpass the characteristic time $\tau$.

For $\theta_i = \theta_j = \tau$, the ideal characteristic time evolves with time. However, the minimal uncertainty diminishes indicating equilibrium tends to make estimations of $\tau$ equally probable. The peak in FI in time is achieved after the system achieves the characteristic time $\tau$, and decreases to zero. However, the height of the peak is inversely proportional to $\tau$, as expected, due to noise.

Off-diagonal terms, $\mathcal{I}_{F_0,\tau}$, depend on both parameters and give a geometry that presents maximal values changing in time; it can also present negative values as it is not the average of a quadratic quantity. This suggests the off-diagonal FI represents correlations induced between the parameters, being maxima when the correlations are stronger, and null when there are not any correlations between parameters.

FI is also sensible to the initial condition as Eq.~(\ref{FI-DeltaDirac}) highlights. Another interesting point is that all the FI are directly proportional to the inverse temperature $\beta$, meaning temperature destroys FI as it increases. Although this result is not surprising since the higher the temperature, the higher the fluctuations, it is worth stressing that such intuition is analytically verified from our approach.

We also calculated the other central moments of Fisher distributions $P(\iota_{\theta_i,\theta_j})$. The results show that the variance is proportional to the square of $\mathcal{I}$, and that both skewness and excess kurtosis exhibit constant values, which are also dependent on the initial condition. The interpretation of the SFI is discussed from two viewpoints: One geometric, where we speculate that it represents a stochastic metric, and another informational, where SFI represents the distribution ensemble of possible results of estimated FI on noisy environments. It is worth noticing that in terms of Shannon theory, the SFI represents the -- squared -- rate of surprisal \cite{Shannon1948}, for instance for $\theta_{i} = \theta_{j}$ corresponding to any time-representative variable.

The results inndicate that even though there is a dynamics of FI dependent on initial conditions, non-zero entropy states will always carry an intrinsic uncertainty to the estimation of parameters, as expected. Furthermore, the initial condition heavily influences the average FI. Also, there is a relation between the chosen parameter, determinism, and the appearance of extreme behavior on FI. For instance, let us examine the case of the universal $\frac{2}{F_0^2} + \alpha(t)$ behavior of $\mathcal{I}_{F_0, F_0}^{\{w_{irr}, \Delta_i s\},\{\mathcal{G},\delta\}
}$ when $F_0 \rightarrow 0$, $\mathcal{I}_{F_0,F_0} \rightarrow \infty$. When the driving force that drives the system initially out-of-equilibrium tends to $0$, our system becomes approximately deterministic in what concerns $F_0$, and the variance of estimation of $F_0$ tends to $0$. That means the fluctuations from the heat bath will be reflected only on the fluctuations of $\iota_{\theta_i, \theta_j}$ distributions. This issue is to be explored assuming  driving protocols different to that employed herein.

Last, the relevance of FI of other distributions other than $P(x,t)$ is further understood as follows: in systems akin to Eq.~(\ref{Langevin_eq}), the total heat, $Q$, is composed of injected, $Q_j $ and dissipated, $Q_d,$ contributions $\int q_{j,d}\, P_{q_{j,d}} \, dq_{j,d} $~\cite{Morgado2014}. Comparing to previous results where $\theta = t$~\cite{nicholson2020}, we describe the respective powers, $dQ_{j,d}/dt$, as a function of the associated FI, $\mathcal{I} ^{q_{j,d}} _{t,t}$; namely, after shifting the injected heat, $q_j(t) \equiv \int _{t_0}^t v(t) \, \eta_t  \, dt $, by its average, $Q_j(t)$, we get   
$\left|dQ_j/dt \right|= \left\langle q_j \, \sqrt{\iota_{t,t}  ^{q_j}} \right\rangle \le \sigma _{q_j} \, \sqrt{\mathcal{I}^{q_j}_{t,t}}$. 
Since, $\left[ \sigma _{q_j} / \left|(dQ_j/dt) \right| \right] = \left[ \mathrm{time} \right]$, there is a scale, $\tau _j $, corresponding to the time the reservoir needs to inject into the system a quantity of heat of the order of the respective fluctuation. That scale cannot be less than $1/\sqrt{\mathcal{I}^{q_j}_{t,t}}$, which is the (inverse) of the fluctuation scale of $\sqrt{\iota_{t,t} ^{q_j}}$, i.e., the average square entropy rate, $\partial (- \ln P_{q_j})/\partial t $. Still, with this inequality in mind we go back to Eq.~(\ref{cramer-rao}) and retrieve its parameter statistical inference original goal: blending both inequalities we can gauge a proxy to the time precision of an experiment described by Eq.~(\ref{Langevin_eq}) from the injected heat fluctuations as 
$ \sigma _t \sim \sigma _{q_j} \, \left|(dQ_j/dt) \right|^{-1} $.
\\ On the other hand, $Q$ can be also written as the superposition of excess, $Q_{ex}$, and housekeeping, $Q_{hk}$, heats geometrically studied in Ref.~\cite{dechant2022}. The present work can be similarly extended to these quantities, especially by surveying FI of each distribution and the relation between their scales of variation with other energetic quantities.

For subsequent work, we plan to survey the precise estimation of possible simultaneously hidden or eigen- parameters of a thermostatistical system as well.
\medskip

\section*{Acknowledgments}
P.B.M. acknowledges Pedro V. Paraguass\'{u} and Diogo O. S.-Pinto for insightful comments. The authors acknowledge Ruynet M. Lima for useful discussions. This work is supported by the Brazilian agencies Conselho Nacional de Desenvolvimento Científico e Tecnológico (CNPq), Coordena\c{c}\~{a}o de Aperfei\c{c}oamento de Pessoal de Ensino Superior (CAPES), and Funda\c{c}\~{a}o Carlos Chagas de Apoio \`a Pesquisa do Estado do Rio de Janeiro (FAPERJ). PBM acknowledges CAPES' scholarship, finance code 001. SMDQ acknowledges CNPq grant No. 302348/2022-0.  WAMM acknowledges CNPq grant No. 308560/2022-1.
\appendix
\section{Transitional probability distribution}\label{AppA}

Our starting point is the transitional probability $P[x_{t},t|x_{0},0]$, computed by
\begin{equation}
    P[x_{t},t|x_{0},0] = \int\mathcal{D}x~e^{S[x]},
\end{equation}
with a stochastic action given by
\begin{equation}
    S[x] = -\frac{\beta}{4\gamma}\int_{0}^{\tau} \dd t~(\gamma \dot{x} + kx - F(t))^2
    \label{path_integral}.
\end{equation}

The exact solution of Eq.(\ref{path_integral}) is carried as follows. One can write the variable $x(t) = x_{c}(t) + y(t)$, where $x_{c}(t)$ is the solution of the optimization, and $y(t)$ the contribution from fluctuations. Both contributions carry boundary conditions $x_{c}(t') = x_{t'}$, $x_{c}(0) = x_{0}$ and $y(t') = y(0) = 0$. The path integral can be written as
\begin{equation}
    P[x_{t'},t'|x_{0},0] = e^{S[x_{c}]}\int\mathcal{D}y e^{S[y]}.
\end{equation}

The solution to the Euler-Lagrange equation, that gives us $x_{c}$ via optimization, is obtained via input of the formula for $F(t) = F_{0}(1 - e^{-\frac{t}{\tau}})$, given by
\begin{eqnarray}
    \delta S = 0,\nonumber\\  \gamma ^2 \ddot{x}(t) +k (F_{0}-k x(t)) -\frac{F_{0} e^{-\frac{t}{\tau }} (\gamma +k \tau )}{\tau } =0,
\end{eqnarray}
From normalization of eq.~(\ref{path_integral}), one notices that 
\begin{equation}
    \int\mathcal{D}y~e^{S[y]} = \frac{1}{\int_{-\infty}^{\infty}e^{S[x_{c}]}dx_{t}}.
\end{equation}

The expressions of $x_{c}(t)$, $S[x_{c}]$, and $P[x_t, t| x_{0}, 0]$ are all written in the SM, section I.

\providecommand{\newblock}{}
\bibliography{refs}




\end{document}


\title{Supplemental Material to "Stochastic thermodynamics of Fisher information"}

\author{Pedro B. Melo}
 \affiliation{Departamento de F\'isica, Pontif\'icia Universidade Cat\'olica, 22452-970, Rio de Janeiro RJ, Brazil}
\email{pedrobmelo@aluno.puc-rio.br}
 \author{S\'ilvio M. \surname{Duarte~Queir\'os}}
 \affiliation{Centro Brasileiro de Pesquisas F\'isicas, 22290-180, Rio de Janeiro RJ, Brazil}
 \affiliation{INCT-Sistemas Complexos, Brazil}
\author{Welles A. M. Morgado}
\affiliation{Departamento de F\'isica, Pontif\'icia Universidade Cat\'olica, 22452-970, Rio de Janeiro RJ, Brazil}%
\affiliation{INCT-Sistemas Complexos, Brazil}

\date{\today}

\maketitle

\onecolumngrid
\section{Transitional probability distribution}

Our starting point is the transitional probability $P[x_{t},t|x_{0},0]$, computed by
\begin{equation}
    P[x_{t},t|x_{0},0] = \int\mathcal{D}x~e^{S[x]},
\end{equation}
with a stochastic action given by
\begin{equation}
    S[x] = -\frac{\beta}{4\gamma}\int_{0}^{\tau} \dd t~(\gamma \dot{x} + kx - F(t))^2
    \label{path_integral}.
\end{equation}

The exact solution of Eq.(\ref{path_integral}) is carried as follows. One can write the variable $x(t) = x_{c}(t) + y(t)$, where $x_{c}(t)$ is the solution of the optimization, and $y(t)$ the contribution from fluctuations. Both contributions carry boundary conditions $x_{c}(t') = x_{t'}$, $x_{c}(0) = x_{0}$ and $y(t') = y(0) = 0$. The path integral can be written as
\begin{equation}
    P[x_{t'},t'|x_{0},0] = e^{S[x_{c}]}\int\mathcal{D}y e^{S[y]}.
\end{equation}

The solution to the Euler-Lagrange equation, that gives us $x_{c}$ via optimization, is obtained via input of the formula for $F(t) = F_{0}(1 - e^{-\frac{t}{\tau}})$, given by
\begin{equation}
    \delta S = 0 \rightarrow -\frac{F_{0} e^{-\frac{t}{\tau }} (\gamma +k \tau )}{\tau }+k (F_{0}-k x(t))+\gamma ^2 \ddot{x}(t)=0.
\end{equation}
The solution is
\begin{eqnarray}
    x_{c}(t) = \frac{e^{-\frac{k t}{\gamma }-\frac{t'+t}{\tau }} \left(\coth \left(\frac{k t'}{\gamma }\right)-1\right)}{2 k (k\tau -\gamma )} \left(F_{0} k \tau  \left(e^{\frac{2 k t}{\gamma }}-1\right) e^{\frac{k t'}{\gamma }+\frac{t}{\tau }}+e^{\frac{t'}{\tau }} \left(-F_{0} k \tau  e^{\frac{k \left(2 t'+t\right)}{\gamma }}+F_{0} (k \tau -\gamma ) e^{\frac{k \left(2 t'+t\right)}{\gamma}+\frac{t}{\tau }}\right.\right.\nonumber\\\left.\left.+(\gamma  F_{0}+k x_{0} (k \tau -\gamma )) e^{\frac{2 k t'}{\gamma }+\frac{t}{\tau }}+(F_{0}-k x_{t'}) (\gamma -k \tau ) e^{\frac{k \left(t'+2t\right)}{\gamma }+\frac{t}{\tau }}+(k x_{t'}-F_{0}) (\gamma -k \tau ) e^{\frac{k t'}{\gamma }+\frac{t}{\tau }}+F_{0} k \tau  e^{\frac{k t}{\gamma }}\right.\right.\nonumber\\\left.\left.+F_{0} (\gamma -k \tau ) e^{t\left(\frac{k}{\gamma }+\frac{1}{\tau }\right)}-e^{t \left(\frac{2 k}{\gamma }+\frac{1}{\tau }\right)} (\gamma  F_{0}+k x_{0} (k \tau -\gamma ))\right)\right)~~
\end{eqnarray}

Upon substitution on $S[x]$ and integration over $t$, we obtain (making $t' \rightarrow t$)
\begin{equation}
    S[x_{c}] = \frac{\gamma  e^{-\frac{2 t}{\tau }} \Theta_{-}(t) \left(F_0 k \tau  e^{\frac{k t}{\gamma }}-e^{t/\tau } \left(\gamma  F_0-k x_0\Gamma_{\tau}\right)+\Gamma_{\tau} \left(F_0-k x_t\right) e^{t \left(\frac{k}{\gamma }+\frac{1}{\tau }\right)}\right){}^2}{k\Gamma_{\tau}^2}.
\end{equation}

From normalization of eq.~(\ref{path_integral}), one notices that 
\begin{equation}
    \int\mathcal{D}y~e^{S[y]} = \frac{1}{\int_{-\infty}^{\infty}e^{S[x_{c}]}dx_{t}}.
\end{equation}
We obtain
\begin{eqnarray}
    P[x_{t},t|x_{0},0] = \sqrt{\frac{k\beta \Theta_{+}(t)}{4\pi}}\exp \left(-\frac{\beta  e^{-\frac{2 t}{\tau }} \Theta_{-}(t) \left(F_0 k \tau  e^{\frac{k t}{\gamma }}-e^{t/\tau } \left(\gamma  F_0 - k x_0\Gamma_{\tau}\right)+\Gamma_{\tau} \left(F_0-k
   x_t\right) e^{t \left(\frac{k}{\gamma }+\frac{1}{\tau }\right)}\right)^2}{4 k \Gamma_{\tau}^2}\right)
\end{eqnarray}

\section{Fisher information for the position distribution}

To start, we calculate the positional distribution $P(x_{t},t)$, given by
\begin{equation}
    P(x_{t},t) = \int P[x_{t},t|x_{0},0] \rho_{0}(x_{0})\dd x_{0}.
\end{equation}
Assuming equilibrium initial distribution, one gets
\begin{equation}
    P(x_{t},t) = \sqrt{\beta  k} \exp \left(-\frac{\beta  e^{-\frac{2 t}{\tau }} \left(F_0 k \tau  e^{\frac{k t}{\gamma }}+(\gamma -k \tau ) \left(F_0-k x_t\right) e^{t
   \left(\frac{k}{\gamma }+\frac{1}{\tau }\right)}+\gamma  F_0 \left(-e^{t/\tau }\right)\right)^2}{2 (\gamma -k \tau )^2 \left(k \left(e^{\frac{2 k t}{\gamma
   }}-1\right)+k\right)}\right).
\end{equation}

Expanding the Logarithm of $P(x_{t},t)$ it reads
\begin{equation}
    \ln{P(x_{t},t)} = \frac{1}{2} (\log (\beta )+\log (k))-\frac{\beta  e^{-\frac{2 t}{\tau }} \left(F_0 k \tau  e^{\frac{k t}{\gamma }}+(\gamma -k \tau ) \left(F_0-k x_t\right) e^{t
   \left(\frac{k}{\gamma }+\frac{1}{\tau }\right)}+\gamma  F_0 \left(-e^{t/\tau }\right)\right){}^2}{2 (\gamma -k \tau )^2 \left(k \left(e^{\frac{2 k t}{\gamma
   }}-1\right)+k\right)}
\end{equation}

For the case of an initial Dirac distribution (null entropy initial condition), the positional distribution reads
\begin{equation}
    P(x_{t},t) = \sqrt{\frac{k\beta \left(\coth \left(\frac{k t}{\gamma }\right)+1\right)}{4\pi}} \exp \left(-\frac{\beta 
   e^{-\frac{2 t}{\tau }} \Theta_{-}(t) \left(F_0 k \tau  e^{\frac{k t}{\gamma
   }}+\Gamma_{\tau} \left(F_0-k x_t\right) e^{t \left(\frac{k}{\gamma }+\frac{1}{\tau }\right)}+\gamma  F_0
   \left(-e^{t/\tau }\right)\right){}^2}{4 k \Gamma_{\tau}^2}\right)
\end{equation}

\subsection*{Calculation for Fisher information with an equilibrium initial distribution}

In a general way, the Fisher information is written as 
\begin{equation}
     \mathcal{I}_{\theta_{i},\theta_{j}}^{x_{t},j} = \left\langle\left(\frac{\partial_{\theta_{i}}P(x_{t},t;\theta_{i})}{P(x_{t},t;\theta_{i})}\right)\left(\frac{\partial_{\theta_{j}}P(x_{t},t;\theta_{j})}{P(x_{t},t;\theta_{j})}\right)\right\rangle,
\end{equation}
with $j = \{\mathcal{G},\delta\}$, where for $F_{0}$ it reads
\begin{equation}
     \mathcal{I}_{F_{0}} = \left\langle\left(\frac{\partial_{F_{0}}P(x_{t},t;F_{0})}{P(x_{t},t;F_{0})}\right)^{2}\right\rangle = \int P(x_{t},t;F_{0})\left(\pdv{\ln(P(x_{t},t;F_{0}))}{F_{0}}\right)^2 \dd x_{t} . 
\end{equation}

For an equilibrium initial distribution is
\begin{equation}
    \mathcal{I}^{x_{t},\mathcal{G}}_{F_{0}} = \frac{ \beta \left(\Gamma _{\tau }  + k \tau e^{-\frac{t}{\tau}}-\gamma e^{-\frac{kt}{\gamma}}\right)^2}{k \Gamma _{\tau }^2}
\end{equation}

The Fisher information for $\tau$ reads
\begin{equation}
   \mathcal{I}^{x_{t}~\mathcal{G}}_{\tau} = \frac{ \beta  k F_{0}^2\left(\gamma   \tau  e^{-\frac{kt}{\gamma} }- e^{-\frac{
   t}{\tau }} (\gamma  \tau + t\Gamma(\tau))\right)^2}{\tau ^2\Gamma_{\tau}^4}.
\end{equation}

The off-diagonal term $\mathcal{I}^{x_{t}~\mathcal{G}}_{F_{0},\tau} = \mathcal{I}^{x_t~\mathcal{G}}_{\tau,F_{0}}$ is given as
\begin{equation}
    \mathcal{I}_{F_{0}, \tau}^{x_{t},\{\mathcal{G},\delta\}} = \left\langle\left(\frac{\partial_{F_{0}}P(x_{t},t;F_{0})}{P(x_{t},t;F_{0})}\right)\left(\frac{\partial_{\tau}P(x_{t},t;\tau)}{P(x_{t},t;\tau)}\right)\right\rangle = \int P(x_{t},t;F_{0},\tau)\left(\pdv{\ln(P(x_{t},t;F_{0}))}{F_{0}}\right)\left(\pdv{\ln(P(x_{t},t;\tau))}{\tau}\right) \dd x_{t} .
\end{equation}

It results in
\begin{equation}
    \mathcal{I}_{\tau,F_{0}}^{x_t, \mathcal{G}} = -\frac{\beta  F_{0}}{\tau \Gamma_{\tau}^3}  \left(k \tau  e^{-\frac{t}{\tau}}+\Gamma_{\tau} -\gamma e^{-\frac{kt}{\gamma} }\right) \left(\gamma \tau  e^{-\frac{kt}{\gamma}}-e^{-\frac{t}{\tau}} (\gamma\tau +t \Gamma_{\tau})\right).
\end{equation}

For an initial delta function (minimal entropy), the Fisher information is identical to the initial equilibrium distribution, up to a $\frac{\Theta_{+}(t)}{\sqrt{2\pi}}$ factor, for $\mathcal{I}_{F_0}^{x_t~\delta}$ and $\mathcal{I}_{\tau}^{x_t~\delta}$, and to a $\frac{\Theta_{-}(t)}{2\sqrt{2\pi}}$ factor for the off-diagonal $\mathcal{I}_{F_{0},\tau}^{x_t~\delta}$ term.

We could also obtain a numerical simulation form of the distributions of the stochastic FI $\iota_{\theta_i}^{x_t~\mathcal{G}}$, with $\mathcal{I}_{F_0}^{x_t~\mathcal{G}} = \langle \iota_{F_0}^{x_t~\mathcal{G}}\rangle$ and $\mathcal{I}{\tau}^{x_t~\mathcal{G}} = \langle\iota_{\tau}^{x_t~\mathcal{G}} \rangle$.
\begin{figure}[t!]
    \centering
    \includegraphics[width=.7\textwidth]{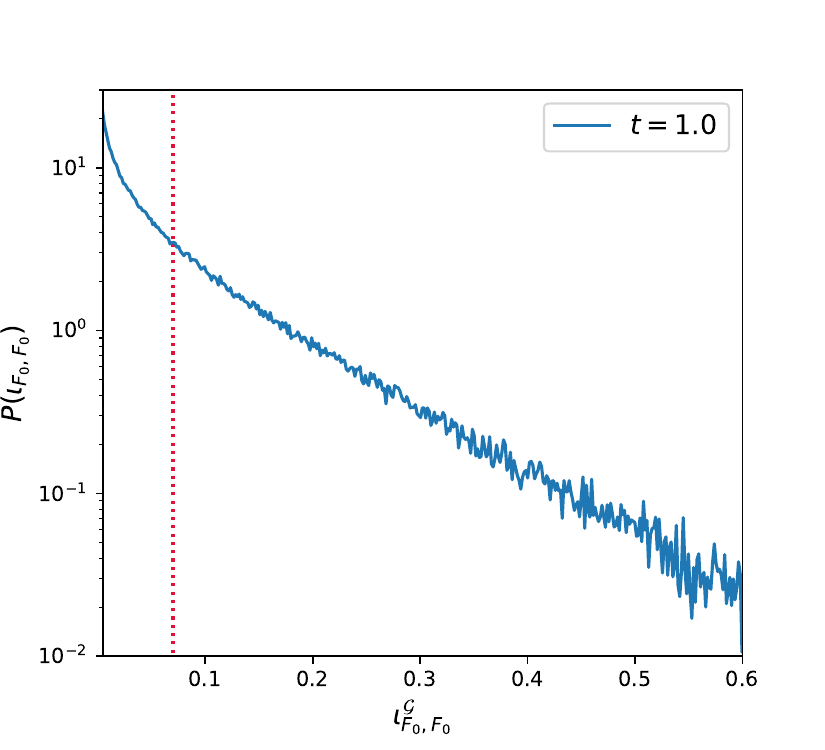}
    \caption{$P(\iota_{F_0, F_0})$ for positive values of $\iota_{F_0, F_0}$ for $t = 1$ and all parameters set to $1$. The distribution presents no negative values, as $\iota_{F_0, F_0}$ is a quadratic quantity. The pink dashed line is the numerical average $\mathcal{I}_{F_0,F_0}^{\mathcal{G}}$, which has a relative error $\abs{1 - \frac{\mathcal{I}^{N}}{\mathcal{I}^{A}}}\lesssim 0.5\%$, where $N$ stands to numeric and $A$ to analytic.}
    \label{fig:dist_f0}
\end{figure}

\begin{figure}[t!]
    \centering
    \includegraphics[width=.7\textwidth]{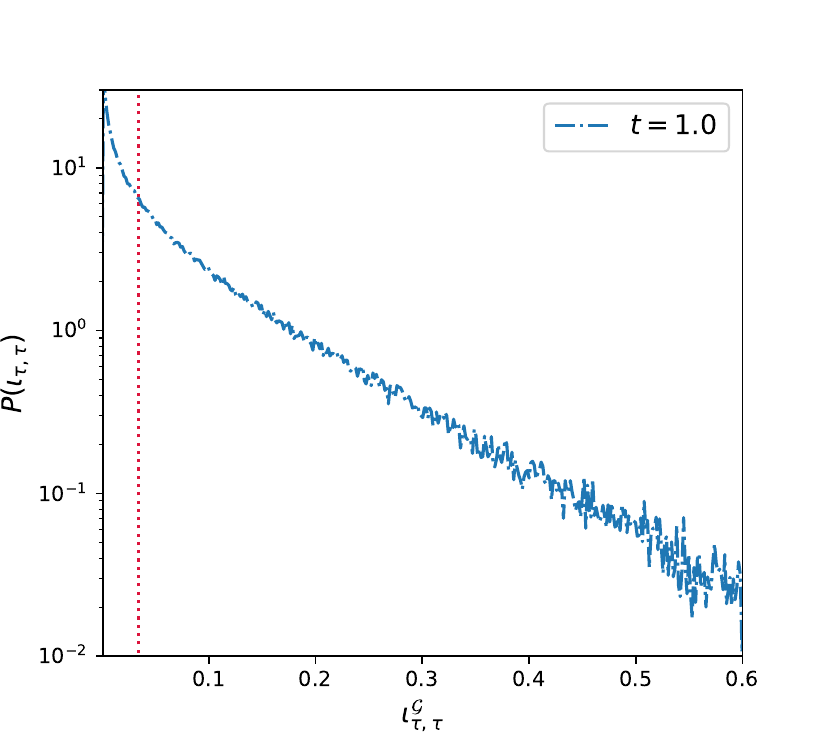}
    \caption{$P(\iota_{\tau, \tau})$ for positive values of $\iota_{\tau, \tau}$ for $t = 1$ and all parameters set to $1$. The distribution presents no negative values, as $\iota_{\tau, \tau}$ is a quadratic quantity. The pink dashed line is the numerical average $\mathcal{I}_{\tau, \tau}^{\mathcal{G}}$, which has a relative error $\abs{1 - \frac{\mathcal{I}^{N}}{\mathcal{I}^{A}}}\lesssim 0.5\%$, where $N$ stands to numeric and $A$ to analytic.}
    \label{fig:dist_tau}
\end{figure}

Figure \ref{fig:dist_f0} and Figure \ref{fig:dist_tau} display the distribution of stochastic Fisher information $\iota_{F_0}^{x_t~\mathcal{G}}$ and $\iota_{\tau}^{x_t~\mathcal{G}}$ respectively. The distributions have a gamma-like behavior as depicted in the figures, with the average stochastic FI coinciding to the FI calculated analytically up to $99.5\%$ precision.

\section{Irreversible work distribution}

We now turn to irreversible work production. Applying the Gaussian path integral method, we get
\begin{equation}
    P(w_{irr}) = \int \dd x_{0}~\dd x_{t} \rho(x_{0})\int\mathcal{D}x e^{S[x]}\delta(w_{irr} - w_{irr}[x]),
\end{equation}
with
\begin{equation}
     w_{irr} = w[x] - \Delta \mathcal{F}^{eq} = -\int_{0}^{t} \dot{F}(t')~ x ~\dd t' - \Delta \mathcal{F}^{eq} = -\frac{F_{0}}{\tau}\int_{0}^{t}e^{-\frac{t'}{\tau}}x~\dd t' + \frac{F_{0}^2 e^{-\frac{2 t}{\tau }} \left(e^{t/\tau }-1\right)^2}{2 k}.
\end{equation}

As $\delta(w_{irr} - w_{irr}[x])$ has an integral form in Fourier space, this allows us defining the characteristic function $Z(\lambda)$
\begin{equation}
    Z(\lambda) = e^{-\lambda \frac{F_{0}^2}{2k}}\int \dd x_{0}~\dd x_{t} \rho_{0}(x_{0})\int\mathcal{D}x~e^{S[x] + i\lambda\int_{0}^{t}\dd t'~( e^{-\frac{2 t'}{\tau }} \left(e^{t'/\tau }-1\right)^2)},
\end{equation}
where the probability is the inverse Fourier transform
\begin{equation}
    P(w_{irr}) = \int_{-\infty}^{\infty}e^{i\lambda w_{irr}}Z(\lambda).
\end{equation}
We solve the path integral in $Z(\lambda)$ following the same procedure for the following Gaussian path integral
\begin{equation}
    I_{w} = \int\mathcal{D}x~e^{S[x] + i\lambda\frac{F_0^2}{k^2}\int_{0}^{t}\dd t'~\left(  \left(e^{-t'/\tau }-1\right)^2\right)},
    \label{iw}
\end{equation}
where we now have a shifted action, given by
\begin{equation}
    S'[x] = -\frac{\beta}{4\pi}\int_{0}^{t'}\dd t  \left(F_0 \left(1-e^{-\frac{t}{\tau }}\right)+k x(t)+\gamma  \dot{x}(t)\right){}^2+\frac{i\lambda F_0}{\tau }\int_{0}^{t'}\dd t   e^{-\frac{t}{\tau }} x(t).
\end{equation}

The optimized solution is obtained via Euler-Lagrange equations
\begin{equation}
    \beta  e^{t/\tau } \left(k \left(F_0+k x(t)\right)-\gamma ^2 \ddot{x}(t)\right)=\frac{F_0 (\beta  (\gamma +k \tau )+2 i \pi  \lambda )}{\tau },
\end{equation}
that gives the extremised solution
\begin{eqnarray}
    x_{c}(t) = -\frac{e^{-\frac{k t}{\gamma }-\frac{t'+t}{\tau }}\Theta_{-}(t')}{2 \beta
    k \Gamma_{\tau}\Gamma_{\tau}^{*}}  \left(F_0 k \tau  (\beta\Gamma_{\tau}^{*}+2 i \pi  \lambda ) \left(\left(e^{\frac{2 k t'}{\gamma
   }}-1\right) e^{\frac{k t}{\gamma }+\frac{t'}{\tau }}-\left(e^{\frac{2 k t}{\gamma }}-1\right) e^{\frac{k t'}{\gamma }+\frac{t}{\tau }}\right)\right.\nonumber\\\left.+2 e^{\frac{\left(t'+t\right) (\gamma +k \tau )}{\gamma  \tau }}
   \left(\beta  F_0 \Gamma_{\tau}\Gamma_{\tau}^{*} \sinh \left(\frac{k t'}{\gamma }\right)+\sinh \left(\frac{k \left(t-t'\right)}{\gamma }\right) \left(\beta  \gamma  F_0 (\gamma +k \tau )+2 i \pi  F_0 k
   \lambda  \tau +\beta  k x_{0}\Gamma_{\tau}\Gamma_{\tau}^{*}\right)\right.\right.\nonumber\\\left.\left.-\beta  \left(F_0+k x_{t}\right) \Gamma_{\tau}\Gamma_{\tau}^{*}  \sinh \left(\frac{k t}{\gamma }\right)\right)\right)
\end{eqnarray}
with $\Gamma_{\tau}^{*} = (\gamma +k \tau )$.
\subsection{$P(w_{irr})$ and $\mathcal{I}^{w_{irr},\{\mathcal{G},\delta\}}_{\{F_0,\tau\}}$ for an equilibrium initial condition}
By repeating the process made to obtain $P[x_{t},t|x_{0},0]$ we obtain $S[x_{c}]$, and obtain $I_{w}$ and $Z(\lambda)$ for a Gaussian $\rho_{0}(x_{0})$. The characteristic function is given by
\begin{eqnarray}
   Z(\lambda) = \exp \left(-\frac{F_0^2 \lambda  e^{-2 t \left(\frac{k}{\gamma }+\frac{1}{\tau }\right)}}{2 \beta  k\Gamma_{\tau}\Gamma_{\tau}^{*2}} \left(-(\pi -\gamma ) \gamma  \lambda  \Gamma_{\tau}+\Gamma_{\tau} e^{2 t \left(\frac{k}{\gamma }+\frac{1}{\tau}\right)} \left(\gamma ^2 \lambda +i \beta  \gamma  k \tau +k \tau  (\pi  \lambda +i \beta  k \tau )\right)\right.\right.\nonumber\\\left.\left.+2 \gamma  e^{t \left(\frac{k}{\gamma }+\frac{1}{\tau }\right)} \left(-\lambda  \left(\gamma ^2-\gamma k \tau +2 \pi  k \tau \right)+i \beta  \gamma  (\gamma +k \tau )\right)+(\gamma +k \tau )^2 e^{\frac{2 k t}{\gamma }} (\pi  \lambda -i \beta  k \tau )-2 i \beta  (\gamma -k \tau ) (\gamma +k \tau )^2 e^{t\left(\frac{2 k}{\gamma }+\frac{1}{\tau }\right)}\right)\right.\nonumber\\\left.-\frac{i F_0^2 \lambda  e^{-\frac{2 t}{\tau }} \left(e^{t/\tau }-1\right)^2}{2 k}+i \lambda 
   w_{irr}\right).~~~~
\end{eqnarray}
By inverse-Fourier transforming $Z(\lambda)$, we obtain 
\begin{equation}
    P(w_{irr}) =  \frac{e^{\left(-\frac{(w_{irr} - \mu_{w}(t;F_{0}))^2}{2\sigma_{w}(t;F_{0})^2}\right)}}{\sqrt{2\pi\sigma_{w}(t;F_{0})^2}},
\end{equation}
where 
\begin{equation}
    \mu_{w}(t) = \frac{\gamma  F_0^2 \left(\gamma -2 \gamma  e^{-t \left(\frac{k}{\gamma }+\frac{1}{\tau }\right)}+e^{-\frac{2 t}{\tau }} (\gamma +k \tau )-k \tau \right)}{2 \gamma ^2 k-2 k^3 \tau ^2},
\end{equation}
and
\begin{eqnarray}
    \sigma_{w}^2(t) = -\frac{F_0^2 e^{-2 t \left(\frac{k}{\gamma }+\frac{1}{\tau }\right)}}{\beta  k \Gamma_{\tau}\Gamma_{\tau}^{*2}} \left(\pi  \left(-k^2 \tau ^2 \left(e^{\frac{2 t}{\tau }}-1\right) e^{\frac{2 k t}{\gamma }}+\gamma ^2 \left(e^{\frac{2 k t}{\gamma
   }}-1\right)+\gamma  k \tau  \left(-4 e^{t \left(\frac{k}{\gamma }+\frac{1}{\tau }\right)}+e^{2 t \left(\frac{k}{\gamma }+\frac{1}{\tau }\right)}+2 e^{\frac{2 k t}{\gamma }}+1\right)\right)\right.\nonumber\\\left.+\gamma ^2 \Gamma_{\tau} \left(e^{t \left(\frac{k}{\gamma }+\frac{1}{\tau }\right)}-1\right)^2\right).~~~
\end{eqnarray}

The Fisher information for $F_{0}$ takes a general form
\begin{equation}
    \mathcal{I}_{F_{0}}^{w_{irr},\mathcal{G}} = \frac{2 \sigma_w (F_{0}) \dot{\mu}^2_w(F_{0})+\dot{\sigma}^2_w(F_{0})}{2 \sigma_w^2(F_{0})} = \frac{2}{F_{0}^2} + \alpha_{w_{irr}~\mathcal{G}}(t),\label{fisher_F0}
\end{equation}
where
\begin{eqnarray}
    \alpha_{w_{irr}~\mathcal{G}}^{-1}(t) =  \frac{2 k \Gamma _{\tau } \left(\pi\left(-k^2 \tau ^2 \left(e^{\frac{2 t}{\tau }}-1\right) e^{\frac{2 k t}{\gamma }}+\gamma ^2 \left(e^{\frac{2 k t}{\gamma}}-1\right)+\gamma  k \tau  \left(-4 e^{t \left(\frac{k}{\gamma }+\frac{1}{\tau }\right)}+e^{2 t \left(\frac{k}{\gamma }+\frac{1}{\tau }\right)}+2 e^{\frac{2 k t}{\gamma }}+1\right)\right)\right)}{\beta  e^{-\frac{2 t}{\tau }} \left(\Gamma _{\tau } (\gamma +2 k \tau ) e^{\frac{k t}{\gamma }+\frac{2 t}{\tau }}+(\gamma +k \tau ) e^{\frac{k t}{\gamma }} \left(\gamma -2 k \tau -4 \Gamma _{\tau } e^{t/\tau}\right)+2 \gamma ^2 e^{t/\tau }\right)^2}~~\nonumber\\+\frac{\left(2 k \gamma ^2 \Gamma^2_{\tau} \left(e^{t \left(\frac{k}{\gamma }+\frac{1}{\tau }\right)}-1\right)^2\right)}{\beta  e^{-\frac{2 t}{\tau }} \left(\Gamma _{\tau } (\gamma +2 k \tau ) e^{\frac{k t}{\gamma }+\frac{2 t}{\tau }}+(\gamma +k \tau ) e^{\frac{k t}{\gamma }} \left(\gamma -2 k \tau -4 \Gamma _{\tau } e^{t/\tau}\right)+2 \gamma ^2 e^{t/\tau }\right)^2}.~
\end{eqnarray}

The Fisher information of $\tau$ has the same form of $\mathcal{I}_{F_{0}}$ in terms of $\sigma_{w}$ and $\mu_{w}$. However, it does not have the same universal behavior $\frac{2}{\theta^2} + \alpha(t)$. We have
\begin{equation}
    \mathcal{I}_{\tau}^{w_{irr},\mathcal{G}} = \alpha_{1}(\tau)\left(\left(\beta_{1}(\tau)-\gamma_{1}(\tau)\right)^2-\delta_{1}(\tau)\right)
\end{equation}
With
\begin{eqnarray}
    \alpha_{1}^{(-1)}(\tau) = \frac{8}{b}\left(c + d\right)^2 F_0^4,\\
   \beta_{1}(\tau) = ef
   \gamma_{1}(\tau) = g \left(h + i\right) F_0^2\\
   \delta_{1}(\tau) = jlmF_0^6
\end{eqnarray}

\begin{eqnarray}
    b = e^{4 t \left(\frac{k}{\gamma }+\frac{1}{\tau }\right)} k^2 \beta ^2 (k \tau -\gamma )^2 (\gamma +k \tau )^4\\
    c = \left(-1+e^{t \left(\frac{k}{\gamma }+\frac{1}{\tau }\right)}\right)^2 (\gamma -k \tau ) \gamma ^2,\\
    d = \pi 
   \left(\left(-1+e^{\frac{2 k t}{\gamma }}\right) \gamma ^2+\left(1+2 e^{\frac{2 k t}{\gamma }}-4 e^{t \left(\frac{k}{\gamma }+\frac{1}{\tau }\right)}+e^{2 t \left(\frac{k}{\gamma }+\frac{1}{\tau }\right)}\right) k
   \tau  \gamma -e^{\frac{2 k t}{\gamma }} \left(-1+e^{\frac{2 t}{\tau }}\right) k^2 \tau ^2\right),\\
   e = \frac{2 e^{-2 t \left(\frac{k}{\gamma }+\frac{1}{\tau }\right)} \left(3 k^2 \tau ^3-k (2
   k t+\gamma ) \tau ^2+2 t \gamma ^2\right)}{k \beta  \tau ^2 (\gamma -k \tau )^2 (\gamma +k \tau )^3},\\
   f = \left(\left(-1+e^{t \left(\frac{k}{\gamma }+\frac{1}{\tau }\right)}\right)^2 (\gamma -k \tau ) \gamma ^2\right.\nonumber\\\left.+\pi  \left(\left(-1+e^{\frac{2 k t}{\gamma }}\right) \gamma
   ^2+\left(1+2 e^{\frac{2 k t}{\gamma }}-4 e^{t \left(\frac{k}{\gamma }+\frac{1}{\tau }\right)}+e^{2 t \left(\frac{k}{\gamma }+\frac{1}{\tau }\right)}\right) k \tau  \gamma -e^{\frac{2 k t}{\gamma }}
   \left(-1+e^{\frac{2 t}{\tau }}\right) k^2 \tau ^2\right)\right) F_0^2,\\
   g = \frac{2 e^{-2 t \left(\frac{k}{\gamma }+\frac{1}{\tau }\right)}}{k \beta  (k \tau -\gamma ) (\gamma +k \tau )^2},\\
   h = k (\pi
   -\gamma ) \gamma +\frac{2 e^{t \left(\frac{k}{\gamma }+\frac{1}{\tau }\right)} \left(k (\gamma -2 \pi ) \tau ^2+t \left(\gamma ^2-k \tau  \gamma +2 k \pi  \tau \right)\right) \gamma }{\tau ^2}+2 e^{\frac{2 k
   t}{\gamma }} k \pi  (\gamma +k \tau ),\\
   i = -\frac{e^{2 t \left(\frac{k}{\gamma }+\frac{1}{\tau }\right)} \left(k \left(\gamma ^2-\pi  \gamma +2 k \pi  \tau \right) \tau ^2+2 t (\gamma -k \tau ) \left(\gamma ^2+k \pi 
   \tau \right)\right)}{\tau ^2},\\
   j = \frac{e^{-2 \left(\frac{k}{\gamma }+\frac{1}{\tau }\right) t-\frac{2 k t}{\gamma }-\frac{4 t}{\tau }}}{k^3 \beta  \tau ^4 (\gamma -k \tau )^4 (k \tau -\gamma ) (\gamma +k \tau )^6},\\
   l = \left(\left(-1+e^{t \left(\frac{k}{\gamma }+\frac{1}{\tau }\right)}\right)^2 (\gamma -k \tau ) \gamma ^2\right.\nonumber\\\left.+\pi  \left(\left(-1+e^{\frac{2 k t}{\gamma }}\right) \gamma ^2+\left(1+2 e^{\frac{2 k t}{\gamma }}-4 e^{t
   \left(\frac{k}{\gamma }+\frac{1}{\tau }\right)}+e^{2 t \left(\frac{k}{\gamma }+\frac{1}{\tau }\right)}\right) k \tau  \gamma -e^{\frac{2 k t}{\gamma }} \left(-1+e^{\frac{2 t}{\tau }}\right) k^2 \tau
   ^2\right)\right),\\
   m = \left(2 e^{t/\tau } \left(2 k^2 \tau ^3+t (\gamma -k \tau ) (\gamma +k \tau )\right) \gamma ^2+e^{\frac{k t}{\gamma }+\frac{2 t}{\tau }} k \tau ^2 (\gamma -k \tau )^2 \gamma \right.\nonumber\\\left.-4 e^{t
   \left(\frac{k}{\gamma }+\frac{1}{\tau }\right)} t \left(\gamma ^2-k^2 \tau ^2\right)^2+e^{\frac{k t}{\gamma }} (\gamma +k \tau )^2 \left(2 t (\gamma -2 k \tau ) (\gamma -k \tau )-k \gamma  \tau ^2\right)\right)^2.
\end{eqnarray}
The off-diagonal term Fisher information of $w_{irr}$ is given by
\begin{equation}
\mathcal{I}_{F_{0},\tau}^{w_{irr},\mathcal{G}} = \frac{2\sigma_{w}^{2}\partial_{F_{0}}\mu_{w}\partial_{\tau}\mu_{w} + \partial_{F_{0}}\sigma_{w}^{2}\partial_{\tau}\sigma_{w}^{2}}{2\sigma_{w}^{2}},\label{Fisher offdiag}
\end{equation}
whereby substitution one gets
\begin{eqnarray}
    \mathcal{I}_{F_{0},\tau}^{w_{irr},\mathcal{G}} = \frac{e^{2 t \left(\frac{k}{\gamma }+\frac{1}{\tau }\right)}}{a} \left(bc + d(\pi e+f)\right).
\end{eqnarray}
\begin{eqnarray}
    a = 4 F_0^3 k \tau ^2 \Gamma _{\tau }^2 (\gamma +k \tau ) \left(\pi  \left(-k^2 \tau ^2 \left(e^{\frac{2 t}{\tau
   }}-1\right) e^{\frac{2 k t}{\gamma }}+\gamma ^2 \left(e^{\frac{2 k t}{\gamma }}-1\right)\right.\right.\nonumber\\\left.\left.+\gamma  k \tau  \left(-4 e^{t \left(\frac{k}{\gamma }+\frac{1}{\tau
   }\right)}+e^{2 t \left(\frac{k}{\gamma }+\frac{1}{\tau }\right)}+2 e^{\frac{2 k t}{\gamma }}+1\right)\right)+\gamma ^2 \Gamma _{\tau } \left(e^{t \left(\frac{k}{\gamma
   }+\frac{1}{\tau }\right)}-1\right)^2\right),\\
   b = \beta  F_0^4 e^{-\frac{2 k t}{\gamma }-\frac{4 t}{\tau }} \left(\Gamma _{\tau } (\gamma +2 k \tau )
   e^{\frac{k t}{\gamma }+\frac{2 t}{\tau }}+(\gamma +k \tau ) e^{\frac{k t}{\gamma }} \left(\gamma -2 k \tau -4 \Gamma _{\tau } e^{t/\tau }\right)+2 \gamma ^2 e^{t/\tau
   }\right),\\
   c = \left(2 \gamma ^2 e^{t/\tau } \left(2 k^2 \tau ^3+t \Gamma _{\tau } (\gamma +k \tau )\right)+\gamma  k \tau ^2 \Gamma _{\tau }^2 e^{\frac{k t}{\gamma }+\frac{2
   t}{\tau }}\right.\nonumber\\\left.+(\gamma +k \tau )^2 e^{\frac{k t}{\gamma }} \left(\gamma  (-k) \tau ^2-2 t \Gamma _{\tau } \left(-\gamma +2 k \tau +2 \Gamma _{\tau } e^{t/\tau
   }\right)\right)\right),\\
   d = -4 F_0^2 k (k \tau -\gamma ) e^{-2 t \left(\frac{k}{\gamma }+\frac{1}{\tau }\right)},\\
   e = 2 t \Gamma _{\tau } (\gamma +k \tau )
   \left(k^2 \tau ^2 e^{\frac{2 k t}{\gamma }}+\gamma ^2 \left(e^{\frac{2 k t}{\gamma }}-1\right)+\gamma  k \tau  \left(-2 e^{t \left(\frac{k}{\gamma }+\frac{1}{\tau
   }\right)}+2 e^{\frac{2 k t}{\gamma }}+1\right)\right)\nonumber\\+k \tau ^2 \left(2 \gamma ^3+k^3 \tau ^3 e^{\frac{2 k t}{\gamma }}+2 \gamma  k^2 \tau ^2-4 \gamma  \left(\gamma ^2+2
   k^2 \tau ^2-\gamma  k \tau \right) e^{t \left(\frac{k}{\gamma }+\frac{1}{\tau }\right)}+3 \gamma  k^2 \tau ^2 e^{\frac{2 k t}{\gamma }}-4 \gamma ^2 k \tau +\gamma ^3
   e^{\frac{2 k t}{\gamma }}\right.\nonumber\\\left.+3 \gamma ^2 k \tau  e^{\frac{2 k t}{\gamma }}+\Gamma _{\tau }^3 e^{2 t \left(\frac{k}{\gamma }+\frac{1}{\tau }\right)}\right),\\
   f = -2 \gamma ^2
   \Gamma _{\tau }^2 \left(e^{t \left(\frac{k}{\gamma }+\frac{1}{\tau }\right)}-1\right) \left(k \tau ^2 \left(e^{t \left(\frac{k}{\gamma }+\frac{1}{\tau
   }\right)}-1\right)+t (\gamma +k \tau )\right).
\end{eqnarray}

\subsection{$P(w_{irr})$ and $\mathcal{I}^{w_{irr}~\delta}_{\{F_0,\tau\}}$ for an minimum entropy initial condition}

For an initial delta condition centered at the origin, the characteristic function is given by
\begin{eqnarray}
    Z(\lambda) = \exp \left(-\frac{F_0^2 \lambda  e^{-2 t \left(\frac{k}{\gamma }+\frac{1}{\tau }\right)}}{2 \beta  k \Gamma_{\tau}\Gamma_{\tau}^{*2}} \left(\pi  \gamma  \lambda  (k \tau -\gamma )+(\gamma +k \tau )^2 e^{\frac{2 k
   t}{\gamma }} (\pi  \lambda +i \beta  (\gamma -2 k \tau )) +2 i \gamma  e^{t \left(\frac{k}{\gamma }+\frac{1}{\tau }\right)} (\beta  \gamma  (\gamma +k \tau )+2 i \pi  k
   \lambda  \tau )\right.\right.\nonumber\\\left.\left. +(\gamma -k \tau ) e^{2 t \left(\frac{k}{\gamma }+\frac{1}{\tau }\right)} (\pi  k \lambda  \tau +i \beta  (\gamma +k \tau ) (\gamma +2 k \tau ))-4 i \beta
    (\gamma -k \tau ) (\gamma +k \tau )^2 e^{t \left(\frac{2 k}{\gamma }+\frac{1}{\tau }\right)}\right)+i \lambda 
   w_{irr}\right).~~~~~~
\end{eqnarray}

We also obtain a Gaussian $P(w_{irr})$, but with
\begin{equation}
    \mu_{w}^{\delta} = -\frac{F_0^2 e^{-\frac{k t}{\gamma }-\frac{2 t}{\tau }} \left((\gamma -k \tau ) (\gamma +2 k \tau ) e^{\frac{k t}{\gamma }+\frac{2 t}{\tau }}+(\gamma +k \tau ) e^{\frac{k
   t}{\gamma }} \left(\gamma -4 e^{t/\tau } (\gamma -k \tau )-2 k \tau \right)+2 \gamma ^2 e^{t/\tau }\right)}{2 k (k \tau -\gamma ) (\gamma +k \tau )},
\end{equation}
and 
\begin{equation}
    \sigma_{w}^{\delta 2} = -\frac{\pi  F_0^2 e^{-2 t \left(\frac{k}{\gamma }+\frac{1}{\tau }\right)} \left(-k^2 \tau ^2 \left(e^{\frac{2 t}{\tau }}-1\right) e^{\frac{2 k t}{\gamma }}+\gamma ^2
   \left(e^{\frac{2 k t}{\gamma }}-1\right)+\gamma  k \tau  \left(-4 e^{t \left(\frac{k}{\gamma }+\frac{1}{\tau }\right)}+e^{2 t \left(\frac{k}{\gamma }+\frac{1}{\tau
   }\right)}+2 e^{\frac{2 k t}{\gamma }}+1\right)\right)}{ \beta  k (k \tau -\gamma ) (\gamma +k \tau )^2}.
\end{equation}

Fisher information for $F_{0}$ takes the same form as presented in eq.(\ref{fisher_F0}), with
\begin{equation}
    \alpha_{w_{irr}~\delta}(t) = \frac{\beta  e^{-\frac{2 t}{\tau }} \left(\Gamma _{\tau } (\gamma +2 k \tau ) e^{\frac{k t}{\gamma }+\frac{2 t}{\tau }}+(\gamma +k \tau )
   e^{\frac{k t}{\gamma }} \left(\gamma -2 k \tau -4 \Gamma _{\tau } e^{t/\tau }\right)+2 \gamma ^2 e^{t/\tau }\right)^2}{\pi  k (k \tau
   -\gamma ) \left(\gamma ^2+k^2 \tau ^2 e^{2 t \left(\frac{k}{\gamma }+\frac{1}{\tau }\right)}-\gamma  k \tau +4 \gamma  k \tau  e^{t
   \left(\frac{k}{\gamma }+\frac{1}{\tau }\right)}-\gamma  k \tau  e^{2 t \left(\frac{k}{\gamma }+\frac{1}{\tau }\right)}-(\gamma +k \tau )^2
   e^{\frac{2 k t}{\gamma }}\right)}.
\end{equation}

Fisher information for $\tau$ is given by
\begin{eqnarray}
    \mathcal{I}_{\tau}^{w_{irr},\delta}(t)= \frac{\beta ^2 k^2 \Gamma_{\tau }^{* 4}\Gamma_{\tau}^2 e^{4 t \left(\frac{k}{\gamma }+\frac{1}{\tau
   }\right)}}{2 \pi ^2 F_0^4 \left(-k^2 \tau ^2
   \left(e^{\frac{2 t}{\tau }}-1\right) e^{\frac{2 k t}{\gamma }}+\gamma ^2 \left(e^{\frac{2 k t}{\gamma }}-1\right)+\gamma  k \tau  \left(-4
   e^{t \left(\frac{k}{\gamma }+\frac{1}{\tau }\right)}+e^{2 t \left(\frac{k}{\gamma }+\frac{1}{\tau }\right)}+2 e^{\frac{2 k t}{\gamma
   }}+1\right)\right)^2}\nonumber\\ \times\left(\frac{4 \pi ^2 F_0^2 e^{-4 t \left(\frac{k}{\gamma }+\frac{1}{\tau }\right)}}{\beta ^2 k^2 \Gamma_{\tau}^2\Gamma _{\tau }^{* 4}} \left(-k^2 \tau ^2 \left(e^{\frac{2 t}{\tau
   }}-1\right) e^{\frac{2 k t}{\gamma }}+\gamma ^2 \left(e^{\frac{2 k t}{\gamma }}-1\right)+\gamma  k \tau  \left(-4 e^{t
   \left(\frac{k}{\gamma }+\frac{1}{\tau }\right)}+e^{2 t \left(\frac{k}{\gamma }+\frac{1}{\tau }\right)}+2 e^{\frac{2 k t}{\gamma}}+1\right)\right)^2\right.\nonumber\\\left.+\frac{2 \pi  F_0^4 e^{-2 t
   \left(\frac{k}{\gamma }+\frac{1}{\tau }\right)}}{\beta  k \Gamma _{\tau
   }^{* 2}\Gamma_{\tau} \left(\gamma ^2 k-k^3 \tau ^2\right)^2} \left(-2 k^2 \tau ^2 e^{-\frac{2 t}{\tau }} \left(e^{t/\tau }-1\right)^2+\gamma ^2 \left(2
   e^{-t \left(\frac{k}{\gamma }+\frac{1}{\tau }\right)}+e^{-\frac{2 t}{\tau }}-4 e^{-\frac{t}{\tau }}+1\right)+\gamma  k \tau 
   \left(1-e^{-\frac{2 t}{\tau }}\right)\right)^2\right.\nonumber\\\left. \times\left(-k^2 \tau ^2 \left(e^{\frac{2 t}{\tau }}-1\right) e^{\frac{2 k t}{\gamma }}+\gamma ^2
   \left(e^{\frac{2 k t}{\gamma }}-1\right)+\gamma  k \tau  \left(-4 e^{t \left(\frac{k}{\gamma }+\frac{1}{\tau }\right)}+e^{2 t
   \left(\frac{k}{\gamma }+\frac{1}{\tau }\right)}+2 e^{\frac{2 k t}{\gamma }}+1\right)\right)\right).~~
\end{eqnarray}

Following the lines of Eq.~(\ref{Fisher offdiag}), one gets the off-diagonal term for the Dirac delta initial condition
\begin{eqnarray}
    \mathcal{I}_{F_0,\tau}^{w_{irr},\delta} = \frac{a}{b}(cdF_0^5 + fg(h+i) + j)
\end{eqnarray}
where
\begin{eqnarray}
    a = e^{4 t \left(\frac{k}{\gamma }+\frac{1}{\tau }\right)} k^2 \beta ^2 (k \tau -\gamma )^2 \left(\Gamma_{\tau}^{*}\right)^4,\\
    b = 2 \pi ^2 \left(\left(-1+e^{\frac{2 k t}{\gamma }}\right) \gamma ^2+\left(1+2 e^{\frac{2 k t}{\gamma
   }}-4 e^{t \left(\frac{k}{\gamma }+\frac{1}{\tau }\right)}+e^{2 t \left(\frac{k}{\gamma }+\frac{1}{\tau }\right)}\right) k \tau  \gamma -e^{\frac{2 k t}{\gamma }}
   \left(-1+e^{\frac{2 t}{\tau }}\right) k^2 \tau ^2\right)^2 F_0^4,~~~\\
   c = \frac{e^{-2
   \left(\frac{k}{\gamma }+\frac{1}{\tau }\right) t-\frac{2 k t}{\gamma }-\frac{4 t}{\tau }} \pi}{k^3 \beta  \tau ^2 (k \tau -\gamma )^2 \Gamma _{\tau }^2 \Gamma_{\tau
   }^{* 5}}  \left(\left(-1+e^{\frac{2 k t}{\gamma }}\right) \gamma ^2+\left(1+2
   e^{\frac{2 k t}{\gamma }}-4 e^{t \left(\frac{k}{\gamma }+\frac{1}{\tau }\right)}+e^{2 t \left(\frac{k}{\gamma }+\frac{1}{\tau }\right)}\right) k \tau  \gamma \right.\nonumber\\\left.-e^{\frac{2
   k t}{\gamma }} \left(-1+e^{\frac{2 t}{\tau }}\right) k^2 \tau ^2\right),~~\\
   d = \left(2 e^{t/\tau } \gamma ^2+e^{\frac{k t}{\gamma }+\frac{2 t}{\tau }} (\gamma +2 k \tau ) \Gamma
   _{\tau }+e^{\frac{k t}{\gamma }} \left(\gamma -2 k \tau -4 e^{t/\tau } \Gamma _{\tau }\right) \Gamma_{\tau}^{*}\right),~~\\
   e = \left(2 e^{t/\tau } \left(2 k^2 \tau
   ^3+t \Gamma _{\tau } \Gamma_{\tau}^{*}\right) \gamma ^2+e^{\frac{k t}{\gamma }+\frac{2 t}{\tau }} k \tau ^2 \Gamma _{\tau }^2 \gamma -4 e^{t
   \left(\frac{k}{\gamma }+\frac{1}{\tau }\right)} t \left(\gamma ^2-k^2 \tau ^2\right)^2+e^{\frac{k t}{\gamma }} \left(2 t (\gamma -2 k \tau ) \Gamma _{\tau }-k \gamma 
   \tau ^2\right) \Gamma_{\tau}^{* 2}\right),~~\\
    f = \frac{2 e^{-4 t \left(\frac{k}{\gamma }+\frac{1}{\tau }\right)} \pi ^2 }{k^2 \beta ^2 \tau ^2 (k \tau -\gamma ) \Gamma_{\tau }^2\Gamma_{\tau}^{* 5}},~~\\
   g = \left(-1+e^{\frac{2 k t}{\gamma }}\right) \gamma ^2+\left(1+2
   e^{\frac{2 k t}{\gamma }}-4 e^{t \left(\frac{k}{\gamma }+\frac{1}{\tau }\right)}+e^{2 t \left(\frac{k}{\gamma }+\frac{1}{\tau }\right)}\right) k \tau  \gamma -e^{\frac{2
   k t}{\gamma }} \left(-1+e^{\frac{2 t}{\tau }}\right) k^2 \tau ^2,~~\\
   h = k \tau ^2 (-e^{\frac{2 k t}{\gamma }} \gamma ^3-2 \gamma ^3-3 e^{\frac{2 k t}{\gamma }}
   k \tau  \gamma ^2+4 k \tau  \gamma ^2-3 e^{\frac{2 k t}{\gamma }} k^2 \tau ^2 \gamma -2 k^2 \tau ^2 \gamma +4 e^{t \left(\frac{k}{\gamma }+\frac{1}{\tau }\right)}
   \left(\gamma ^2-k \tau  \gamma +2 k^2 \tau ^2\right)) \gamma,~~\\
   i = -e^{\frac{2 k t}{\gamma }} k^3 \tau ^3-e^{2 t \left(\frac{k}{\gamma }+\frac{1}{\tau }\right)} \Gamma^{3}_{\tau},~~\\
   j = -2 t \left(\left(-1+e^{\frac{2 k t}{\gamma }}\right) \gamma ^2+\left(1+2 e^{\frac{2 k t}{\gamma }}-2 e^{t \left(\frac{k}{\gamma }+\frac{1}{\tau }\right)}\right)k \tau  \gamma +e^{\frac{2 k t}{\gamma }} k^2 \tau ^2\right) \Gamma_{\tau} \Gamma_{\tau}^{*} F_0^3.~~
\end{eqnarray}

\section{Irreversible entropy distribution}

Again, in the lines of path integral method, the irreversible entropy distribution is given by
\begin{equation}
    P(\Delta_{i}s) = \int \dd x_{0}~\dd x_{t} \rho(x_{0})\int\mathcal{D}x e^{S[x]}\delta(\Delta_i s - \Delta_i s[x]),
\end{equation}
where 
\begin{equation}
    \frac{\Delta_{i} s[x_{t},t]}{\beta} =  -\Delta f[x_{t},t] - \frac{F_{0}}{k}\int_{0}^{t}\dd t'e^{-t'/\tau}x(t'),
\end{equation}
\begin{equation}
    \Delta f(t) = \frac{F_{0} \left(2 \gamma  x_{t} \Gamma \left(e^{-\frac{t}{\tau }}-e^{-\frac{k t}{\gamma }}\right)\right)}{2 \Gamma^2}- \left(\frac{ \left(F_{0} k \tau  e^{\frac{-t}{\tau }}+(F_{0}-k
   x_{t})\Gamma - \gamma F_{0}e^{-kt/\gamma }\right)^2}{2k \Gamma^2}- \frac{  k x_{0}^2}{2}\right).
\end{equation}

Fourier-transforming $P(\Delta_i s)$, one obtains the characteristic function 
\begin{equation}
    Z(\lambda) = e^{i\lambda\Delta f[x_{t},t] }\int \dd x_{0}~\dd x_{t} \rho_{0}(x_{0})\int\mathcal{D}x~e^{S[x] + i\lambda \frac{F^2_{0}}{k^2}\int_{0}^{t}\dd t'~ \left(e^{-t'/\tau }-1\right)^2},
\end{equation}
with $Z(\lambda)$ equivalent to
\begin{equation}
    Z(\lambda) = e^{i\lambda\Delta f[x_{t},t] }\int \dd x_{0}~\dd x_{t} \rho_{0}(x_{0}) I_{w},
\end{equation}
with $I_w$ calculated as Eq.~(\ref{iw}). As in the case for $P(w_{irr})$, we have a shifted action equivalent to that case. 

\subsection{$P(\Delta_i s)$ and $\mathcal{I}^{\Delta_i s~\mathcal{G}}_{\{F_0,\tau\}}$ for an equilibrium initial condition}

For an equilibrium initial condition, the characteristic function is given by
\begin{eqnarray}
   \log (Z(\lambda)) = \left(-\frac{F_0^2 \lambda  e^{-2 t \left(\frac{2 k}{\gamma }+\frac{1}{\tau }\right)}}{2 \beta  k \Gamma _{\tau }^2 \Gamma_{\tau
   }^{* 2}} \left(-4 (\pi -\gamma ) \gamma  k^2 \lambda  \tau ^2 e^{\frac{2 k t}{\gamma }}+4
   \gamma  k \tau  e^{t \left(\frac{3 k}{\gamma }+\frac{1}{\tau }\right)} \left(\lambda  \left(\gamma ^2-\gamma  k \tau +2 \pi  k \tau \right)\right.\right.\right.\nonumber\\\left.\left.\left.-i \beta  \Gamma_{\tau
   }^* (\gamma +2 k \tau )\right)+\Gamma _{\tau }^2 e^{2 t \left(\frac{2 k}{\gamma }+\frac{1}{\tau }\right)} \left(\lambda  \left(\gamma ^2+\pi  k \tau \right)+i
   \beta  \gamma  \Gamma_{\tau}^{*}\right)+i k \tau  \Gamma_{\tau}^{* 2} e^{\frac{4 k t}{\gamma }} (\beta  (\gamma +2 k \tau
   )+i \pi  \lambda )\right.\right.\nonumber\\\left.\left.+\gamma  \Gamma_{\tau}^{*} e^{2 t \left(\frac{k}{\gamma }+\frac{1}{\tau }\right)} \left(-2 \gamma  \lambda  \Gamma _{\tau }+i \beta 
   \Gamma_{\tau}^{*} (\gamma +2 k \tau )+\pi  \lambda  (\gamma -3 k \tau )\right)-2 i \beta  k \tau  \Gamma_{\tau}^{* 2} (k
   \tau -\gamma ) e^{t \left(\frac{4 k}{\gamma }+\frac{1}{\tau }\right)}\right.\right.\nonumber\\\left.\left.-2 i \beta  \gamma  \Gamma _{\tau } \Gamma_{\tau}^{* 2} e^{\frac{3 k
   t}{\gamma }+\frac{2 t}{\tau }}+4 (\pi -\gamma ) \gamma  k \lambda  \tau  \Gamma_{\tau}^{*} e^{t \left(\frac{k}{\gamma }+\frac{1}{\tau }\right)}+\gamma 
   (\gamma -\pi ) \lambda  \Gamma_{\tau}^{* 2} e^{\frac{2 t}{\tau }}\right)\right).
\end{eqnarray}

Again, Fourier transforming $Z(\lambda)$ we obtain $P(\frac{\Delta_i s}{\beta})$, where 
\begin{eqnarray}
    \mu_{s}(t) = \frac{F_0^2 e^{-2 t \left(\frac{2 k}{\gamma }+\frac{1}{\tau }\right)} }{2
   k \Gamma _{\tau }^2 \Gamma_{\tau}^{*}}\left(\left(2 \gamma ^2 k \tau -2 k^3 \tau ^3\right) e^{t \left(\frac{4 k}{\gamma }+\frac{1}{\tau
   }\right)}+\gamma  \Gamma _{\tau }^2 e^{2 t \left(\frac{2 k}{\gamma }+\frac{1}{\tau }\right)}-2 \gamma  \Gamma_{\tau}^{*} \Gamma _{\tau } e^{\frac{3 k
   t}{\gamma }+\frac{2 t}{\tau }} +\gamma  \Gamma_{\tau}^{*} (\gamma +2 k \tau ) e^{2 t \left(\frac{k}{\gamma }+\frac{1}{\tau }\right)}\right.\nonumber\\\left.+k \tau  \Gamma
   _{\tau }^* (\gamma +2 k \tau ) e^{\frac{4 k t}{\gamma }}-4 \gamma  k \tau  (\gamma +2 k \tau ) e^{t \left(\frac{3 k}{\gamma }+\frac{1}{\tau }\right)}\right).~~~~
\end{eqnarray}

And
\begin{eqnarray}
    \sigma_{s}^{2}(t) = -\frac{2 F_0^2 e^{-2 t \left(\frac{2 k}{\gamma }+\frac{1}{\tau }\right)}}{\beta  k \Gamma _{\tau }^2 \Gamma _{\tau }^{* 2}} \left(-4 (\pi -\gamma ) \gamma  k^2 \tau ^2 e^{\frac{2 k t}{\gamma }}+\Gamma _{\tau }^2
   \left(\gamma ^2+\pi  k \tau \right) e^{2 t \left(\frac{2 k}{\gamma }+\frac{1}{\tau }\right)}-4 \gamma  k \tau  \left(-\gamma ^2+\gamma  k \tau -2 \pi  k \tau \right)
   e^{t \left(\frac{3 k}{\gamma }+\frac{1}{\tau }\right)}\right.\nonumber\\\left.-\pi  k \tau  \Gamma_{\tau}^{* 2} e^{\frac{4 k t}{\gamma }}+4 (\pi -\gamma ) \gamma 
   k \tau  \Gamma_{\tau}^{*} e^{t \left(\frac{k}{\gamma }+\frac{1}{\tau }\right)}+\gamma  \Gamma_{\tau}^{*} \left(\pi  (\gamma -3 k \tau )-2
   \gamma  \Gamma _{\tau }\right) e^{2 t \left(\frac{k}{\gamma }+\frac{1}{\tau }\right)}\right.\nonumber\\\left.+\gamma  (\gamma -\pi ) \Gamma _{\tau }^{* 2} e^{\frac{2
   t}{\tau }}\right).~~~~
\end{eqnarray}

Fisher information for $F_0$ has the general form of Eq.(\ref{fisher_F0}). In this case,
\begin{eqnarray}
    \alpha_{\Delta_i s~ \mathcal{G}}(t) =  \frac{\beta  e^{-\frac{2 t}{\tau }} a}{k \Gamma _{\tau }^2 b}
\end{eqnarray}
with
\begin{eqnarray}
    a = \left(\left(4 \gamma ^2 k \tau -4 k^3 \tau ^3\right) e^{t \left(\frac{2 k}{\gamma }+\frac{1}{\tau }\right)}+3 \gamma ^2 e^{\frac{2
   t}{\tau }} (\gamma +k \tau )-4 \gamma  \Gamma _{\tau } (\gamma +k \tau ) e^{\frac{k t}{\gamma }+\frac{2 t}{\tau }}+\Gamma_{\tau }^2 (\gamma +2 k \tau ) e^{2 t
   \left(\frac{k}{\gamma }+\frac{1}{\tau }\right)}\right.\nonumber\\\left.-4 \gamma  k \tau  (2 \gamma +k \tau ) e^{t \left(\frac{k}{\gamma }+\frac{1}{\tau }\right)}+k \tau  (\gamma +k \tau )
   (\gamma +2 k \tau ) e^{\frac{2 k t}{\gamma }}\right)^2,\\
   b = \left(-4 (\pi -\gamma ) \gamma  k^2 \tau ^2 e^{\frac{2 k t}{\gamma }}+\Gamma _{\tau }^2
   \left(\gamma ^2+\pi  k \tau \right) e^{2 t \left(\frac{2 k}{\gamma }+\frac{1}{\tau }\right)}-4 \gamma  k \tau  \left(-\gamma ^2+\gamma  k \tau -2 \pi  k \tau \right) e^{t
   \left(\frac{3 k}{\gamma }+\frac{1}{\tau }\right)}\right.\nonumber\\\left.+\gamma  (\gamma +k \tau ) \left(-2 \gamma ^2+\pi  \gamma +2 \gamma  k \tau -3 \pi  k \tau \right) e^{2 t
   \left(\frac{k}{\gamma }+\frac{1}{\tau }\right)}-\pi  k \tau  (\gamma +k \tau )^2 e^{\frac{4 k t}{\gamma }}\right.\nonumber\\\left.+4 (\pi -\gamma ) \gamma  k \tau  (\gamma +k \tau ) e^{t
   \left(\frac{k}{\gamma }+\frac{1}{\tau }\right)}+\gamma  (\gamma -\pi ) e^{\frac{2 t}{\tau }} (\gamma +k \tau )^2\right).
\end{eqnarray}

The $\tau$ dependent term is given by
\begin{equation}
    \mathcal{I}_{\tau}^{\Delta_i s~\mathcal{G}} = \frac{a(j(c + d + e + f)^2 g + (h + i)^2)}{b}
\end{equation}
with
\begin{eqnarray}
    a = \beta ^2 k^2 \Gamma _{\tau }^4 \Gamma_{\tau }^{*~4} \left(e^{\frac{2 k t}{\gamma }}-1\right)^4 e^{\frac{8 k t}{\gamma }+\frac{12 t}{\tau }},
\end{eqnarray}
\begin{eqnarray}
    b = 2 F_0^4 \left(2 k \tau  \left(\pi  \left(\gamma ^2+k^2 \tau ^2\right)+2 \gamma ^2 k \tau \right) e^{\frac{6 k t}{\gamma }+\frac{4 t}{\tau }}-k \tau  \left(\pi  \left(\gamma ^2+k^2 \tau ^2-6 \gamma  k \tau \right)+8\gamma ^2 k \tau \right) e^{4 t \left(\frac{k}{\gamma }+\frac{1}{\tau }\right)}\right.\nonumber\\\left.-4 (\pi -\gamma ) \gamma  k^2 \tau ^2 e^{\frac{2 k t}{\gamma }+\frac{4 t}{\tau }}+\left((\pi -4 \gamma ) \gamma ^3-2 \pi  k^3 \tau^3+\pi  \gamma  k^2 \tau ^2+4 (\gamma -\pi ) \gamma ^2 k \tau \right) e^{6 t \left(\frac{k}{\gamma }+\frac{1}{\tau }\right)}\right.\nonumber\\\left.+\left(-3 (\pi -2 \gamma ) \gamma ^3+\pi  k^3 \tau ^3+(3 \pi -2 \gamma ) \gamma  k^2 \tau ^2+3 \pi  \gamma ^2 k \tau \right) e^{\frac{4 k t}{\gamma }+\frac{6 t}{\tau }}+4 \gamma ^2 k \tau  (3 \gamma +k \tau -2 \pi ) e^{\frac{3 k t}{\gamma }+\frac{5 t}{\tau }}\right.\nonumber\\\left.-\gamma  (\gamma +k \tau ) \left(4\gamma ^2-3 \pi  \gamma +\pi  k \tau \right) e^{\frac{2 k t}{\gamma }+\frac{6 t}{\tau }}-4 \gamma  k \tau  \left(-\gamma ^2+\gamma  k \tau -2 \pi  k \tau \right) e^{\frac{7 k t}{\gamma }+\frac{5 t}{\tau}}+(\gamma -k \tau )^2 \left(\gamma ^2+\pi  k \tau \right) e^{\frac{8 k t}{\gamma }+\frac{6 t}{\tau }}\right.\nonumber\\\left.+(\gamma -\pi ) \gamma  e^{\frac{6 t}{\tau }} (\gamma +k \tau )^2+4 (\pi -\gamma ) \gamma  k \tau  (\gamma +k\tau ) e^{\frac{k t}{\gamma }+\frac{5 t}{\tau }}+4 \gamma  k \tau  (\pi  (\gamma -3 k \tau )+\gamma  (k \tau -3 \gamma )) e^{5 t \left(\frac{k}{\gamma }+\frac{1}{\tau }\right)}\right.\nonumber\\\left.-\pi  k \tau  (\gamma +k \tau )^2 e^{4 t \left(\frac{2 k}{\gamma }+\frac{1}{\tau }\right)}\right)^2
\end{eqnarray}

\begin{eqnarray}
    j = \frac{2 F_0^2 e^{-\frac{4 k t}{\gamma }-\frac{6 t}{\tau }}}{\beta  k \Gamma _{\tau }^2 \left(\left(\Gamma _{\tau }\right){}^*\right){}^2 \left(e^{\frac{2 k t}{\gamma }}-1\right)^2},
\end{eqnarray}
\begin{eqnarray}
    c = -\frac{F_0^2 e^{-2 t \left(\frac{k}{\gamma }+\frac{1}{\tau }\right)}}{2 k \Gamma _{\tau }^2 \left(\Gamma _{\tau }\right)^*} \left(-2 k^2 \tau  \left(\Gamma _{\tau }\right){}^* e^{\frac{2 k t}{\gamma }}+4 k^2 \tau  \left(\Gamma _{\tau }\right){}^* e^{t \left(\frac{2k}{\gamma }+\frac{1}{\tau }\right)}+4 \gamma  k^2 \tau  e^{t \left(\frac{k}{\gamma }+\frac{1}{\tau }\right)}+4 k^2 \tau  (k \tau -\gamma ) e^{t \left(\frac{2 k}{\gamma }+\frac{1}{\tau }\right)}\right.\nonumber\\\left.-k^2 \tau  (\gamma+2 k \tau ) e^{\frac{2 k t}{\gamma }}-3 \gamma ^2 k e^{\frac{2 t}{\tau }}+\frac{2 t \Gamma _{\tau }^2 (\gamma +2 k \tau ) e^{2 t \left(\frac{k}{\gamma }+\frac{1}{\tau }\right)}}{\tau ^2}-\frac{8 \gamma  t \Gamma_{\tau } \left(\Gamma _{\tau }\right){}^* e^{\frac{k t}{\gamma }+\frac{2 t}{\tau }}}{\tau ^2}-2 k \Gamma _{\tau }^2 e^{2 t \left(\frac{k}{\gamma }+\frac{1}{\tau }\right)}\right.\nonumber\\\left.+4 \gamma  k \Gamma _{\tau } e^{\frac{k t}{\gamma }+\frac{2 t}{\tau }}+2 k \Gamma _{\tau } (\gamma +2 k \tau ) e^{2 t \left(\frac{k}{\gamma }+\frac{1}{\tau }\right)}-4 \gamma  k \left(\Gamma _{\tau }\right){}^* e^{\frac{k t}{\gamma }+\frac{2 t}{\tau}}+4 k \left(\Gamma _{\tau }\right){}^* (k \tau -\gamma ) e^{t \left(\frac{2 k}{\gamma }+\frac{1}{\tau }\right)}\right.\nonumber\\\left.-\frac{4 k t \left(\Gamma _{\tau }\right){}^* (k \tau -\gamma ) e^{t \left(\frac{2 k}{\gamma}+\frac{1}{\tau }\right)}}{\tau }-k \left(\Gamma _{\tau }\right){}^* (\gamma +2 k \tau ) e^{\frac{2 k t}{\gamma }}+4 \gamma  k (2 \gamma +k \tau ) e^{t \left(\frac{k}{\gamma }+\frac{1}{\tau }\right)}\right.\nonumber\\\left.-\frac{4 \gamma  k t (2 \gamma +k \tau ) e^{t \left(\frac{k}{\gamma }+\frac{1}{\tau }\right)}}{\tau }+\frac{6 \gamma ^2 t \left(\Gamma _{\tau }\right){}^* e^{\frac{2 t}{\tau }}}{\tau ^2}\right),
\end{eqnarray}
\begin{eqnarray}
    d = \frac{F_0^2 e^{-2 t \left(\frac{k}{\gamma }+\frac{1}{\tau }\right)}}{2 \Gamma _{\tau }^2 \left(\left(\Gamma _{\tau }\right){}^*\right){}^2} \left(4 \gamma  \Gamma _{\tau } \left(\Gamma _{\tau }\right){}^* e^{\frac{k t}{\gamma }+\frac{2 t}{\tau }}-\Gamma _{\tau }^2 (\gamma +2 k \tau )e^{2 t \left(\frac{k}{\gamma }+\frac{1}{\tau }\right)}+4 k \tau  \left(\Gamma _{\tau }\right){}^* (k \tau -\gamma ) e^{t \left(\frac{2 k}{\gamma }+\frac{1}{\tau }\right)}\right.\nonumber\\\left.-k \tau  \left(\Gamma _{\tau }\right){}^*(\gamma +2 k \tau ) e^{\frac{2 k t}{\gamma }}+4 \gamma  k \tau  (2 \gamma +k \tau ) e^{t \left(\frac{k}{\gamma }+\frac{1}{\tau }\right)}-3 \gamma ^2 \left(\Gamma _{\tau }\right){}^* e^{\frac{2 t}{\tau}}\right),
\end{eqnarray}
\begin{eqnarray}
    e = -\frac{F_0^2 e^{-2 t \left(\frac{k}{\gamma }+\frac{1}{\tau }\right)}}{\Gamma _{\tau }^3 \left(\Gamma _{\tau }\right){}^*} \left(4 \gamma  \Gamma _{\tau } \left(\Gamma _{\tau }\right){}^* e^{\frac{k t}{\gamma }+\frac{2 t}{\tau }}-\Gamma _{\tau }^2 (\gamma +2 k \tau )e^{2 t \left(\frac{k}{\gamma }+\frac{1}{\tau }\right)}+4 k \tau  \left(\Gamma _{\tau }\right){}^* (k \tau -\gamma ) e^{t \left(\frac{2 k}{\gamma }+\frac{1}{\tau }\right)}\right.\nonumber\\\left.-k \tau  \left(\Gamma _{\tau }\right){}^*(\gamma +2 k \tau ) e^{\frac{2 k t}{\gamma }}+4 \gamma  k \tau  (2 \gamma +k \tau ) e^{t \left(\frac{k}{\gamma }+\frac{1}{\tau }\right)}-3 \gamma ^2 \left(\Gamma _{\tau }\right){}^* e^{\frac{2 t}{\tau}}\right).
\end{eqnarray}
\begin{eqnarray}
    f = -\frac{F_0^2 t e^{-2 t \left(\frac{k}{\gamma }+\frac{1}{\tau }\right)}}{k \tau ^2 \Gamma _{\tau }^2 \left(\Gamma _{\tau }\right){}^*} \left(4 \gamma  \Gamma _{\tau } \left(\Gamma _{\tau }\right){}^* e^{\frac{k t}{\gamma }+\frac{2 t}{\tau }}-\Gamma _{\tau }^2 (\gamma +2 k \tau ) e^{2 t \left(\frac{k}{\gamma }+\frac{1}{\tau }\right)}+4 k \tau  \left(\Gamma _{\tau }\right){}^* (k \tau -\gamma ) e^{t \left(\frac{2 k}{\gamma }+\frac{1}{\tau }\right)}\right.\nonumber\\\left.-k \tau  \left(\Gamma _{\tau }\right){}^*(\gamma +2 k \tau ) e^{\frac{2 k t}{\gamma }}+4 \gamma  k \tau  (2 \gamma +k \tau ) e^{t \left(\frac{k}{\gamma }+\frac{1}{\tau }\right)}-3 \gamma ^2 \left(\Gamma _{\tau }\right){}^* e^{\frac{2 t}{\tau}}\right)
\end{eqnarray}
\begin{eqnarray}
g = 2 k \tau  \left(\pi  \left(\gamma ^2+k^2 \tau ^2\right)+2 \gamma ^2 k \tau \right) e^{\frac{6 k t}{\gamma }+\frac{4 t}{\tau }}-k \tau  \left(\pi  \left(\gamma ^2+k^2 \tau ^2-6 \gamma  k \tau \right)+8 \gamma ^2 k
   \tau \right) e^{4 t \left(\frac{k}{\gamma }+\frac{1}{\tau }\right)}\nonumber\\-4 (\pi -\gamma ) \gamma  k^2 \tau ^2 e^{\frac{2 k t}{\gamma }+\frac{4 t}{\tau }}+\left((\pi -4 \gamma ) \gamma ^3-2 \pi  k^3 \tau ^3+\pi  \gamma
    k^2 \tau ^2+4 (\gamma -\pi ) \gamma ^2 k \tau \right) e^{6 t \left(\frac{k}{\gamma }+\frac{1}{\tau }\right)}\nonumber\\+\left(-3 (\pi -2 \gamma ) \gamma ^3+\pi  k^3 \tau ^3+(3 \pi -2 \gamma ) \gamma  k^2 \tau ^2+3 \pi 
   \gamma ^2 k \tau \right) e^{\frac{4 k t}{\gamma }+\frac{6 t}{\tau }}-\gamma  \left(\Gamma _{\tau }\right){}^* \left(4 \gamma ^2-3 \pi  \gamma +\pi  k \tau \right) e^{\frac{2 k t}{\gamma }+\frac{6 t}{\tau
   }}\nonumber\\+\Gamma _{\tau }^2 \left(\gamma ^2+\pi  k \tau \right) e^{\frac{8 k t}{\gamma }+\frac{6 t}{\tau }}+4 \gamma ^2 k \tau  (3 \gamma +k \tau -2 \pi ) e^{\frac{3 k t}{\gamma }+\frac{5 t}{\tau }}-4 \gamma  k \tau 
   \left(-\gamma ^2+\gamma  k \tau -2 \pi  k \tau \right) e^{\frac{7 k t}{\gamma }+\frac{5 t}{\tau }}\nonumber\\+4 (\pi -\gamma ) \gamma  k \tau  \left(\Gamma _{\tau }\right){}^* e^{\frac{k t}{\gamma }+\frac{5 t}{\tau }}-\pi 
   k \tau  \left(\left(\Gamma _{\tau }\right){}^*\right){}^2 e^{4 t \left(\frac{2 k}{\gamma }+\frac{1}{\tau }\right)}+4 \gamma  k \tau  (\pi  (\gamma -3 k \tau )+\gamma  (k \tau -3 \gamma )) e^{5 t
   \left(\frac{k}{\gamma }+\frac{1}{\tau }\right)}\nonumber\\+(\gamma -\pi ) \gamma  \left(\left(\Gamma _{\tau }\right){}^*\right){}^2 e^{\frac{6 t}{\tau }}
   \end{eqnarray}

\begin{eqnarray}
    h = -\frac{e^{-\frac{4 k t}{\gamma }-\frac{6 t}{\tau }} F_0^2}{\left(-1+e^{\frac{2 k t}{\gamma }}\right)^2 k \beta  \tau ^2 \Gamma _{\tau }^2 \left(\left(\Gamma _{\tau }\right){}^*\right){}^2} \left(-4 e^{\frac{3 k t}{\gamma }+\frac{5 t}{\tau }} k \tau  (2 \pi  (5 t-\tau )-5 t (3 \gamma +k \tau )+\tau  (3 \gamma +2 k \tau )) \gamma ^2\right.\nonumber\\\left.-8 e^{\frac{2
   k t}{\gamma }+\frac{4 t}{\tau }} k^2 (\pi -\gamma ) (2 t-\tau ) \tau ^2 \gamma +4 e^{\frac{7 k t}{\gamma }+\frac{5 t}{\tau }} k \tau  \left(5 t \left(\gamma ^2-k \tau  \gamma +2 k \pi  \tau \right)+\tau 
   \left(-\gamma ^2+2 k \tau  \gamma -4 k \pi  \tau \right)\right) \gamma \right.\nonumber\\\left.-4 e^{5 t \left(\frac{k}{\gamma }+\frac{1}{\tau }\right)} k \tau  \left(\pi  \tau  (\gamma -6 k \tau )-5 \pi  t (\gamma -3 k \tau )+\gamma 
   \left(2 k \tau ^2-5 k t \tau -3 \gamma  \tau +15 t \gamma \right)\right) \gamma \right.\nonumber\\\left.-4 e^{\frac{k t}{\gamma }+\frac{5 t}{\tau }} k (\pi -\gamma ) \tau  \left(\tau  (\gamma +2 k \tau )-5 t \left(\Gamma _{\tau
   }\right){}^*\right) \gamma -2 e^{\frac{6 t}{\tau }} (\pi -\gamma ) \left(\Gamma _{\tau }\right){}^* \left(3 t \left(\Gamma _{\tau }\right){}^*-k \tau ^2\right) \gamma \right.\nonumber\\\left.+2 e^{\frac{2 k t}{\gamma }+\frac{6 t}{\tau
   }} \left(2 \left(k \tau ^2-6 t \left(\Gamma _{\tau }\right){}^*\right) \gamma ^2+\pi  \left(k (k \tau -\gamma ) \tau ^2+3 t (3 \gamma -k \tau ) \left(\Gamma _{\tau }\right){}^*\right)\right) \gamma \right.\nonumber\\\left.+2 e^{6 t
   \left(\frac{k}{\gamma }+\frac{1}{\tau }\right)} \left(3 k^3 \pi  \tau ^4-k^2 \pi  (6 k t+\gamma ) \tau ^3+k \gamma  (3 k \pi  t+2 (\pi -\gamma ) \gamma ) \tau ^2+12 k t \gamma ^2 (\gamma -\pi ) \tau +3 t (\pi -4
   \gamma ) \gamma ^3\right)\right.\nonumber\\\left.-2 e^{\frac{6 k t}{\gamma }+\frac{4 t}{\tau }} k \tau  \left(4 k \tau  (\tau -2 t) \gamma ^2-4 \pi  t \left(\gamma ^2+k^2 \tau ^2\right)+\pi  \tau  \left(\gamma ^2+3 k^2 \tau
   ^2\right)\right)\right.\nonumber\\\left.+e^{4 t \left(\frac{k}{\gamma }+\frac{1}{\tau }\right)} k \tau  \left(16 k \tau  (\tau -2 t) \gamma ^2+\pi  \left(3 k^2 \tau ^3-4 k (k t+3 \gamma ) \tau ^2+\gamma  (24 k t+\gamma ) \tau -4 t
   \gamma ^2\right)\right)\right.\nonumber\\\left.+e^{\frac{8 k t}{\gamma }+\frac{6 t}{\tau }} \Gamma _{\tau } \left(k \left(2 \gamma ^2-\pi  \gamma +3 k \pi  \tau \right) \tau ^2+6 t \left(\gamma ^2+k \pi  \tau \right) \Gamma _{\tau
   }\right)+e^{4 t \left(\frac{2 k}{\gamma }+\frac{1}{\tau }\right)} k \pi  \tau  \left(\Gamma _{\tau }\right){}^* \left(\tau  (\gamma +3 k \tau )-4 t \left(\Gamma _{\tau }\right){}^*\right)\right.\nonumber\\\left.+e^{\frac{4 k t}{\gamma
   }+\frac{6 t}{\tau }} \left(4 \left(k^2 \tau ^3-3 k^2 t \tau ^2+9 t \gamma ^2\right) \gamma ^2-3 k \pi  \tau ^2 \left(\left(\Gamma _{\tau }\right){}^*\right){}^2+6 \pi  t \left(-3 \gamma ^3+3 k \tau  \gamma ^2+3
   k^2 \tau ^2 \gamma +k^3 \tau ^3\right)\right)\right),
\end{eqnarray}
and
\begin{eqnarray}
    i = -\frac{2 F_0^2 e^{-\frac{4 k t}{\gamma }-\frac{6 t}{\tau }}}{\beta  k \tau ^2 \left(\left(\Gamma _{\tau }\right){}^*\right){}^3 (k \tau -\gamma )^3 \left(e^{\frac{2 k t}{\gamma }}-1\right)^2} \left(2 k^2 \tau ^3+3 t \Gamma _{\tau } \left(\Gamma _{\tau }\right){}^*\right) \left(2 k \tau  \left(\pi  \left(\gamma ^2+k^2 \tau ^2\right)+2 \gamma ^2 k
   \tau \right) e^{\frac{6 k t}{\gamma }+\frac{4 t}{\tau }}\right.\nonumber\\\left.-k \tau  \left(\pi  \left(\gamma ^2+k^2 \tau ^2-6 \gamma  k \tau \right)+8 \gamma ^2 k \tau \right) e^{4 t \left(\frac{k}{\gamma }+\frac{1}{\tau }\right)}-4
   (\pi -\gamma ) \gamma  k^2 \tau ^2 e^{\frac{2 k t}{\gamma }+\frac{4 t}{\tau }}\right.\nonumber\\\left.+\left((\pi -4 \gamma ) \gamma ^3-2 \pi  k^3 \tau ^3+\pi  \gamma  k^2 \tau ^2+4 (\gamma -\pi ) \gamma ^2 k \tau \right) e^{6 t
   \left(\frac{k}{\gamma }+\frac{1}{\tau }\right)}\right.\nonumber\\\left.+\left(-3 (\pi -2 \gamma ) \gamma ^3+\pi  k^3 \tau ^3+(3 \pi -2 \gamma ) \gamma  k^2 \tau ^2+3 \pi  \gamma ^2 k \tau \right) e^{\frac{4 k t}{\gamma }+\frac{6 t}{\tau
   }}-\gamma  \left(\Gamma _{\tau }\right){}^* \left(4 \gamma ^2-3 \pi  \gamma +\pi  k \tau \right) e^{\frac{2 k t}{\gamma }+\frac{6 t}{\tau }}\right.\nonumber\\\left.+\Gamma _{\tau }^2 \left(\gamma ^2+\pi  k \tau \right) e^{\frac{8 k
   t}{\gamma }+\frac{6 t}{\tau }}+4 \gamma ^2 k \tau  (3 \gamma +k \tau -2 \pi ) e^{\frac{3 k t}{\gamma }+\frac{5 t}{\tau }}-4 \gamma  k \tau  \left(-\gamma ^2+\gamma  k \tau -2 \pi  k \tau \right) e^{\frac{7 k
   t}{\gamma }+\frac{5 t}{\tau }}\right.\nonumber\\\left.+4 (\pi -\gamma ) \gamma  k \tau  \left(\Gamma _{\tau }\right){}^* e^{\frac{k t}{\gamma }+\frac{5 t}{\tau }}-\pi  k \tau  \left(\left(\Gamma _{\tau }\right){}^*\right){}^2 e^{4 t
   \left(\frac{2 k}{\gamma }+\frac{1}{\tau }\right)}+4 \gamma  k \tau  (\pi  (\gamma -3 k \tau )+\gamma  (k \tau -3 \gamma )) e^{5 t \left(\frac{k}{\gamma }+\frac{1}{\tau }\right)}\right.\nonumber\\\left.+(\gamma -\pi ) \gamma 
   \left(\left(\Gamma _{\tau }\right){}^*\right){}^2 e^{\frac{6 t}{\tau }}\right).
\end{eqnarray}

The off-diagonal term is given by

\begin{equation}
    \mathcal{I}_{F_0, \tau}^{\Delta_i s~\mathcal{G}} = \frac{b \left(c h (d+e+f+g)^2+ i(j+k+l + m)\right)}{a}
\end{equation}

with
\begin{eqnarray}
    a = 2 F_0^4 \left(2 k \tau  \left(\pi  \left(\gamma ^2+k^2 \tau ^2\right)+2 \gamma ^2 k \tau \right) e^{\frac{6 k t}{\gamma }+\frac{4 t}{\tau }}-k \tau  \left(\pi  \left(\gamma ^2+k^2 \tau ^2-6 \gamma  k \tau \right)+8
   \gamma ^2 k \tau \right) e^{4 t \left(\frac{k}{\gamma }+\frac{1}{\tau }\right)}\right.\nonumber\\\left.-4 (\pi -\gamma ) \gamma  k^2 \tau ^2 e^{\frac{2 k t}{\gamma }+\frac{4 t}{\tau }}+\left((\pi -4 \gamma ) \gamma ^3-2 \pi  k^3 \tau
   ^3+\pi  \gamma  k^2 \tau ^2+4 (\gamma -\pi ) \gamma ^2 k \tau \right) e^{6 t \left(\frac{k}{\gamma }+\frac{1}{\tau }\right)}\right.\nonumber\\\left.+\left(-3 (\pi -2 \gamma ) \gamma ^3+\pi  k^3 \tau ^3+(3 \pi -2 \gamma ) \gamma  k^2
   \tau ^2+3 \pi  \gamma ^2 k \tau \right) e^{\frac{4 k t}{\gamma }+\frac{6 t}{\tau }}-\gamma  \left(\Gamma _{\tau }\right){}^* \left(4 \gamma ^2-3 \pi  \gamma +\pi  k \tau \right) e^{\frac{2 k t}{\gamma }+\frac{6
   t}{\tau }}\right.\nonumber\\\left.+\Gamma _{\tau }^2 \left(\gamma ^2+\pi  k \tau \right) e^{\frac{8 k t}{\gamma }+\frac{6 t}{\tau }}+4 \gamma ^2 k \tau  (3 \gamma +k \tau -2 \pi ) e^{\frac{3 k t}{\gamma }+\frac{5 t}{\tau }}-4 \gamma  k
   \tau  \left(-\gamma ^2+\gamma  k \tau -2 \pi  k \tau \right) e^{\frac{7 k t}{\gamma }+\frac{5 t}{\tau }}\right.\nonumber\\\left.+4 (\pi -\gamma ) \gamma  k \tau  \left(\Gamma _{\tau }\right){}^* e^{\frac{k t}{\gamma }+\frac{5 t}{\tau
   }}-\pi  k \tau  \left(\left(\Gamma _{\tau }\right){}^*\right){}^2 e^{4 t \left(\frac{2 k}{\gamma }+\frac{1}{\tau }\right)}+4 \gamma  k \tau  (\pi  (\gamma -3 k \tau )+\gamma  (k \tau -3 \gamma )) e^{5 t
   \left(\frac{k}{\gamma }+\frac{1}{\tau }\right)}\right.\nonumber\\\left.+(\gamma -\pi ) \gamma  \left(\left(\Gamma _{\tau }\right){}^*\right){}^2 e^{\frac{6 t}{\tau }}\right)^2,
\end{eqnarray}

\begin{eqnarray}
    b = \beta ^2 k^2 \Gamma _{\tau }^4 \left(\left(\Gamma _{\tau }\right){}^*\right){}^4 \left(e^{\frac{2 k t}{\gamma }}-1\right)^4 e^{\frac{8 k t}{\gamma }+\frac{12 t}{\tau }},
\end{eqnarray}

\begin{eqnarray}
    c = -\frac{2 F_0^3 e^{-2 t \left(\frac{k}{\gamma }+\frac{1}{\tau }\right)-\frac{4 k t}{\gamma }-\frac{6 t}{\tau }}}{\beta  k^2 \Gamma _{\tau }^4 \left(\left(\Gamma _{\tau }\right){}^*\right){}^3 \left(e^{\frac{2 k t}{\gamma }}-1\right)^2} \left(4 \gamma  \Gamma _{\tau } \left(\Gamma _{\tau }\right){}^* e^{\frac{k t}{\gamma }+\frac{2 t}{\tau
   }}-\Gamma _{\tau }^2 (\gamma +2 k \tau ) e^{2 t \left(\frac{k}{\gamma }+\frac{1}{\tau }\right)}+4 k \tau  \left(\Gamma _{\tau }\right){}^* (k \tau -\gamma ) e^{t \left(\frac{2 k}{\gamma }+\frac{1}{\tau
   }\right)}\right.\nonumber\\\left.-k \tau  \left(\Gamma _{\tau }\right){}^* (\gamma +2 k \tau ) e^{\frac{2 k t}{\gamma }}+4 \gamma  k \tau  (2 \gamma +k \tau ) e^{t \left(\frac{k}{\gamma }+\frac{1}{\tau }\right)}-3 \gamma ^2 \left(\Gamma
   _{\tau }\right){}^* e^{\frac{2 t}{\tau }}\right),
\end{eqnarray}

\begin{eqnarray}
    d = -\frac{F_0^2 e^{-2 t \left(\frac{k}{\gamma }+\frac{1}{\tau }\right)}}{2 k \Gamma _{\tau }^2
   \left(\Gamma _{\tau }\right){}^*} \left(-2 k^2 \tau  \left(\Gamma _{\tau }\right){}^* e^{\frac{2 k t}{\gamma }}+4 k^2 \tau  \left(\Gamma _{\tau }\right){}^* e^{t \left(\frac{2
   k}{\gamma }+\frac{1}{\tau }\right)}+4 \gamma  k^2 \tau  e^{t \left(\frac{k}{\gamma }+\frac{1}{\tau }\right)}+4 k^2 \tau  (k \tau -\gamma ) e^{t \left(\frac{2 k}{\gamma }+\frac{1}{\tau }\right)}\right.\nonumber\\\left.-k^2 \tau  (\gamma
   +2 k \tau ) e^{\frac{2 k t}{\gamma }}-3 \gamma ^2 k e^{\frac{2 t}{\tau }}+\frac{2 t \Gamma _{\tau }^2 (\gamma +2 k \tau ) e^{2 t \left(\frac{k}{\gamma }+\frac{1}{\tau }\right)}}{\tau ^2}-\frac{8 \gamma  t \Gamma
   _{\tau } \left(\Gamma _{\tau }\right){}^* e^{\frac{k t}{\gamma }+\frac{2 t}{\tau }}}{\tau ^2}-2 k \Gamma _{\tau }^2 e^{2 t \left(\frac{k}{\gamma }+\frac{1}{\tau }\right)}\right.\nonumber\\\left.+4 \gamma  k \Gamma _{\tau } e^{\frac{k
   t}{\gamma }+\frac{2 t}{\tau }}+2 k \Gamma _{\tau } (\gamma +2 k \tau ) e^{2 t \left(\frac{k}{\gamma }+\frac{1}{\tau }\right)}-4 \gamma  k \left(\Gamma _{\tau }\right){}^* e^{\frac{k t}{\gamma }+\frac{2 t}{\tau
   }}+4 k \left(\Gamma _{\tau }\right){}^* (k \tau -\gamma ) e^{t \left(\frac{2 k}{\gamma }+\frac{1}{\tau }\right)}\right.\nonumber\\\left.-\frac{4 k t \left(\Gamma _{\tau }\right){}^* (k \tau -\gamma ) e^{t \left(\frac{2 k}{\gamma
   }+\frac{1}{\tau }\right)}}{\tau }-k \left(\Gamma _{\tau }\right){}^* (\gamma +2 k \tau ) e^{\frac{2 k t}{\gamma }}+4 \gamma  k (2 \gamma +k \tau ) e^{t \left(\frac{k}{\gamma }+\frac{1}{\tau }\right)}-\frac{4
   \gamma  k t (2 \gamma +k \tau ) e^{t \left(\frac{k}{\gamma }+\frac{1}{\tau }\right)}}{\tau }\right.\nonumber\\\left.+\frac{6 \gamma ^2 t \left(\Gamma _{\tau }\right){}^* e^{\frac{2 t}{\tau }}}{\tau ^2}\right).
\end{eqnarray}

\begin{eqnarray}
    e = \frac{F_0^2 e^{-2 t \left(\frac{k}{\gamma }+\frac{1}{\tau }\right)} }{2 \Gamma _{\tau }^2 \left(\left(\Gamma _{\tau }\right){}^*\right){}^2}\left(4 \gamma  \Gamma _{\tau } \left(\Gamma _{\tau }\right){}^* e^{\frac{k t}{\gamma }+\frac{2 t}{\tau }}-\Gamma _{\tau }^2 (\gamma +2 k \tau )
   e^{2 t \left(\frac{k}{\gamma }+\frac{1}{\tau }\right)}+4 k \tau  \left(\Gamma _{\tau }\right){}^* (k \tau -\gamma ) e^{t \left(\frac{2 k}{\gamma }+\frac{1}{\tau }\right)}\right.\nonumber\\\left.-k \tau  \left(\Gamma _{\tau }\right){}^*
   (\gamma +2 k \tau ) e^{\frac{2 k t}{\gamma }}+4 \gamma  k \tau  (2 \gamma +k \tau ) e^{t \left(\frac{k}{\gamma }+\frac{1}{\tau }\right)}-3 \gamma ^2 \left(\Gamma _{\tau }\right){}^* e^{\frac{2 t}{\tau
   }}\right)
\end{eqnarray}

\begin{eqnarray}
    f = -\frac{F_0^2 e^{-2 t \left(\frac{k}{\gamma }+\frac{1}{\tau }\right)}}{\Gamma _{\tau }^3 \left(\Gamma _{\tau }\right){}^*} \left(4 \gamma  \Gamma _{\tau } \left(\Gamma _{\tau }\right){}^* e^{\frac{k t}{\gamma }+\frac{2 t}{\tau }}-\Gamma _{\tau }^2 (\gamma +2 k \tau )
   e^{2 t \left(\frac{k}{\gamma }+\frac{1}{\tau }\right)}+4 k \tau  \left(\Gamma _{\tau }\right){}^* (k \tau -\gamma ) e^{t \left(\frac{2 k}{\gamma }+\frac{1}{\tau }\right)}\right.\nonumber\\\left.-k \tau  \left(\Gamma _{\tau }\right){}^*
   (\gamma +2 k \tau ) e^{\frac{2 k t}{\gamma }}+4 \gamma  k \tau  (2 \gamma +k \tau ) e^{t \left(\frac{k}{\gamma }+\frac{1}{\tau }\right)}-3 \gamma ^2 \left(\Gamma _{\tau }\right){}^* e^{\frac{2 t}{\tau
   }}\right)
\end{eqnarray}

\begin{eqnarray}
    g = -\frac{F_0^2 t e^{-2 t \left(\frac{k}{\gamma }+\frac{1}{\tau }\right)} }{k \tau ^2 \Gamma _{\tau }^2 \left(\Gamma _{\tau }\right){}^*}\left(4 \gamma  \Gamma _{\tau } \left(\Gamma _{\tau }\right){}^* e^{\frac{k t}{\gamma }+\frac{2 t}{\tau }}-\Gamma _{\tau }^2 (\gamma +2 k \tau )
   e^{2 t \left(\frac{k}{\gamma }+\frac{1}{\tau }\right)}+4 k \tau  \left(\Gamma _{\tau }\right){}^* (k \tau -\gamma ) e^{t \left(\frac{2 k}{\gamma }+\frac{1}{\tau }\right)}\right.\nonumber\\\left.-k \tau  \left(\Gamma _{\tau }\right){}^*
   (\gamma +2 k \tau ) e^{\frac{2 k t}{\gamma }}+4 \gamma  k \tau  (2 \gamma +k \tau ) e^{t \left(\frac{k}{\gamma }+\frac{1}{\tau }\right)}-3 \gamma ^2 \left(\Gamma _{\tau }\right){}^* e^{\frac{2 t}{\tau
   }}\right),
\end{eqnarray}

\begin{eqnarray}
    h = 2 k \tau  \left(\pi  \left(\gamma ^2+k^2 \tau ^2\right)+2 \gamma ^2 k \tau \right) e^{\frac{6 k t}{\gamma }+\frac{4 t}{\tau }}-k \tau  \left(\pi  \left(\gamma ^2+k^2 \tau ^2-6 \gamma  k \tau \right)+8 \gamma ^2 k
   \tau \right) e^{4 t \left(\frac{k}{\gamma }+\frac{1}{\tau }\right)}\nonumber\\-4 (\pi -\gamma ) \gamma  k^2 \tau ^2 e^{\frac{2 k t}{\gamma }+\frac{4 t}{\tau }}+\left((\pi -4 \gamma ) \gamma ^3-2 \pi  k^3 \tau ^3+\pi  \gamma
    k^2 \tau ^2+4 (\gamma -\pi ) \gamma ^2 k \tau \right) e^{6 t \left(\frac{k}{\gamma }+\frac{1}{\tau }\right)}\nonumber\\+\left(-3 (\pi -2 \gamma ) \gamma ^3+\pi  k^3 \tau ^3+(3 \pi -2 \gamma ) \gamma  k^2 \tau ^2+3 \pi 
   \gamma ^2 k \tau \right) e^{\frac{4 k t}{\gamma }+\frac{6 t}{\tau }}-\gamma  \left(\Gamma _{\tau }\right){}^* \left(4 \gamma ^2-3 \pi  \gamma +\pi  k \tau \right) e^{\frac{2 k t}{\gamma }+\frac{6 t}{\tau
   }}\nonumber\\+\Gamma _{\tau }^2 \left(\gamma ^2+\pi  k \tau \right) e^{\frac{8 k t}{\gamma }+\frac{6 t}{\tau }}+4 \gamma ^2 k \tau  (3 \gamma +k \tau -2 \pi ) e^{\frac{3 k t}{\gamma }+\frac{5 t}{\tau }}-4 \gamma  k \tau 
   \left(-\gamma ^2+\gamma  k \tau -2 \pi  k \tau \right) e^{\frac{7 k t}{\gamma }+\frac{5 t}{\tau }}\nonumber\\+4 (\pi -\gamma ) \gamma  k \tau  \left(\Gamma _{\tau }\right){}^* e^{\frac{k t}{\gamma }+\frac{5 t}{\tau }}-\pi 
   k \tau  \left(\left(\Gamma _{\tau }\right){}^*\right){}^2 e^{4 t \left(\frac{2 k}{\gamma }+\frac{1}{\tau }\right)}+4 \gamma  k \tau  (\pi  (\gamma -3 k \tau )+\gamma  (k \tau -3 \gamma )) e^{5 t
   \left(\frac{k}{\gamma }+\frac{1}{\tau }\right)}\nonumber\\+(\gamma -\pi ) \gamma  \left(\left(\Gamma _{\tau }\right){}^*\right){}^2 e^{\frac{6 t}{\tau }},
\end{eqnarray}

\begin{eqnarray}
    i = \frac{2 F_0 e^{-\frac{4 k t}{\gamma }-\frac{6 t}{\tau }} }{\beta
    k \Gamma _{\tau }^2 \left(\left(\Gamma _{\tau }\right){}^*\right){}^2 \left(e^{\frac{2 k t}{\gamma }}-1\right)^2}\left(2 k \tau  \left(\pi  \left(\gamma ^2+k^2 \tau ^2\right)+2 \gamma ^2 k \tau \right) e^{\frac{6 k t}{\gamma }+\frac{4 t}{\tau }}-k \tau  \left(\pi 
   \left(\gamma ^2+k^2 \tau ^2-6 \gamma  k \tau \right)+8 \gamma ^2 k \tau \right) e^{4 t \left(\frac{k}{\gamma }+\frac{1}{\tau }\right)}\right.\nonumber\\\left.-4 (\pi -\gamma ) \gamma  k^2 \tau ^2 e^{\frac{2 k t}{\gamma }+\frac{4 t}{\tau
   }}+\left((\pi -4 \gamma ) \gamma ^3-2 \pi  k^3 \tau ^3+\pi  \gamma  k^2 \tau ^2+4 (\gamma -\pi ) \gamma ^2 k \tau \right) e^{6 t \left(\frac{k}{\gamma }+\frac{1}{\tau }\right)}\right.\nonumber\\\left.+\left(-3 (\pi -2 \gamma ) \gamma
   ^3+\pi  k^3 \tau ^3+(3 \pi -2 \gamma ) \gamma  k^2 \tau ^2+3 \pi  \gamma ^2 k \tau \right) e^{\frac{4 k t}{\gamma }+\frac{6 t}{\tau }}-\gamma  \left(\Gamma _{\tau }\right){}^* \left(4 \gamma ^2-3 \pi  \gamma +\pi
    k \tau \right) e^{\frac{2 k t}{\gamma }+\frac{6 t}{\tau }}\right.\nonumber\\\left.+\Gamma _{\tau }^2 \left(\gamma ^2+\pi  k \tau \right) e^{\frac{8 k t}{\gamma }+\frac{6 t}{\tau }}+4 \gamma ^2 k \tau  (3 \gamma +k \tau -2 \pi )
   e^{\frac{3 k t}{\gamma }+\frac{5 t}{\tau }}-4 \gamma  k \tau  \left(-\gamma ^2+\gamma  k \tau -2 \pi  k \tau \right) e^{\frac{7 k t}{\gamma }+\frac{5 t}{\tau }}\right.\nonumber\\\left.+4 (\pi -\gamma ) \gamma  k \tau  \left(\Gamma
   _{\tau }\right){}^* e^{\frac{k t}{\gamma }+\frac{5 t}{\tau }}-\pi  k \tau  \left(\left(\Gamma _{\tau }\right){}^*\right){}^2 e^{4 t \left(\frac{2 k}{\gamma }+\frac{1}{\tau }\right)}+4 \gamma  k \tau  (\pi 
   (\gamma -3 k \tau )+\gamma  (k \tau -3 \gamma )) e^{5 t \left(\frac{k}{\gamma }+\frac{1}{\tau }\right)}\right.\nonumber\\\left.+(\gamma -\pi ) \gamma  \left(\left(\Gamma _{\tau }\right){}^*\right){}^2 e^{\frac{6 t}{\tau }}\right).~~~~~~~~
\end{eqnarray}

\begin{eqnarray}
    j = -\frac{e^{-\frac{4 k t}{\gamma }-\frac{6 t}{\tau }} F_0^2}{\left(-1+e^{\frac{2 k t}{\gamma }}\right)^2 k \beta  \tau ^2 \Gamma _{\tau }^2 \left(\left(\Gamma _{\tau }\right){}^*\right){}^2} \left(-4 e^{\frac{3 k t}{\gamma }+\frac{5 t}{\tau }} k \tau  (2 \pi  (5 t-\tau )-5 t (3 \gamma +k \tau )+\tau  (3 \gamma +2 k \tau )) \gamma ^2\right.\nonumber\\\left.-8 e^{\frac{2
   k t}{\gamma }+\frac{4 t}{\tau }} k^2 (\pi -\gamma ) (2 t-\tau ) \tau ^2 \gamma +4 e^{\frac{7 k t}{\gamma }+\frac{5 t}{\tau }} k \tau  \left(5 t \left(\gamma ^2-k \tau  \gamma +2 k \pi  \tau \right)+\tau 
   \left(-\gamma ^2+2 k \tau  \gamma -4 k \pi  \tau \right)\right) \gamma \right.\nonumber\\\left.-4 e^{5 t \left(\frac{k}{\gamma }+\frac{1}{\tau }\right)} k \tau  \left(\pi  \tau  (\gamma -6 k \tau )-5 \pi  t (\gamma -3 k \tau )+\gamma 
   \left(2 k \tau ^2-5 k t \tau -3 \gamma  \tau +15 t \gamma \right)\right) \gamma\right.\nonumber\\\left. -4 e^{\frac{k t}{\gamma }+\frac{5 t}{\tau }} k (\pi -\gamma ) \tau  \left(\tau  (\gamma +2 k \tau )-5 t \left(\Gamma _{\tau
   }\right){}^*\right) \gamma -2 e^{\frac{6 t}{\tau }} (\pi -\gamma ) \left(\Gamma _{\tau }\right){}^* \left(3 t \left(\Gamma _{\tau }\right){}^*-k \tau ^2\right) \gamma \right.\nonumber\\\left.+2 e^{\frac{2 k t}{\gamma }+\frac{6 t}{\tau
   }} \left(2 \left(k \tau ^2-6 t \left(\Gamma _{\tau }\right){}^*\right) \gamma ^2+\pi  \left(k (k \tau -\gamma ) \tau ^2+3 t (3 \gamma -k \tau ) \left(\Gamma _{\tau }\right){}^*\right)\right) \gamma \right.\nonumber\\\left.+2 e^{6 t
   \left(\frac{k}{\gamma }+\frac{1}{\tau }\right)} \left(3 k^3 \pi  \tau ^4-k^2 \pi  (6 k t+\gamma ) \tau ^3+k \gamma  (3 k \pi  t+2 (\pi -\gamma ) \gamma ) \tau ^2+12 k t \gamma ^2 (\gamma -\pi ) \tau +3 t (\pi -4
   \gamma ) \gamma ^3\right)\right.\nonumber\\\left.-2 e^{\frac{6 k t}{\gamma }+\frac{4 t}{\tau }} k \tau  \left(4 k \tau  (\tau -2 t) \gamma ^2-4 \pi  t \left(\gamma ^2+k^2 \tau ^2\right)+\pi  \tau  \left(\gamma ^2+3 k^2 \tau
   ^2\right)\right)\right.\nonumber\\\left.+e^{4 t \left(\frac{k}{\gamma }+\frac{1}{\tau }\right)} k \tau  \left(16 k \tau  (\tau -2 t) \gamma ^2+\pi  \left(3 k^2 \tau ^3-4 k (k t+3 \gamma ) \tau ^2+\gamma  (24 k t+\gamma ) \tau -4 t
   \gamma ^2\right)\right)\right.\nonumber\\\left.+e^{\frac{8 k t}{\gamma }+\frac{6 t}{\tau }} \Gamma _{\tau } \left(k \left(2 \gamma ^2-\pi  \gamma +3 k \pi  \tau \right) \tau ^2+6 t \left(\gamma ^2+k \pi  \tau \right) \Gamma _{\tau
   }\right)+e^{4 t \left(\frac{2 k}{\gamma }+\frac{1}{\tau }\right)} k \pi  \tau  \left(\Gamma _{\tau }\right){}^* \left(\tau  (\gamma +3 k \tau )-4 t \left(\Gamma _{\tau }\right){}^*\right)\right.\nonumber\\\left.+e^{\frac{4 k t}{\gamma
   }+\frac{6 t}{\tau }} \left(4 \left(k^2 \tau ^3-3 k^2 t \tau ^2+9 t \gamma ^2\right) \gamma ^2-3 k \pi  \tau ^2 \left(\left(\Gamma _{\tau }\right){}^*\right){}^2+6 \pi  t \left(-3 \gamma ^3+3 k \tau  \gamma ^2+3
   k^2 \tau ^2 \gamma +k^3 \tau ^3\right)\right)\right),
\end{eqnarray}
\begin{eqnarray}
   k = -\frac{2 F_0^2 e^{-\frac{4 k t}{\gamma }-\frac{6 t}{\tau }}}{\beta
    \Gamma _{\tau }^2 \left(\left(\Gamma _{\tau }\right){}^*\right){}^3 \left(e^{\frac{2 k t}{\gamma }}-1\right)^2} \left(2 k \tau  \left(\pi  \left(\gamma ^2+k^2 \tau ^2\right)+2 \gamma ^2 k \tau \right) e^{\frac{6 k t}{\gamma }+\frac{4 t}{\tau }}-k \tau  \left(\pi 
   \left(\gamma ^2+k^2 \tau ^2-6 \gamma  k \tau \right)+8 \gamma ^2 k \tau \right) e^{4 t \left(\frac{k}{\gamma }+\frac{1}{\tau }\right)}\nonumber\right.\\\left.-4 (\pi -\gamma ) \gamma  k^2 \tau ^2 e^{\frac{2 k t}{\gamma }+\frac{4 t}{\tau
   }}+\left((\pi -4 \gamma ) \gamma ^3-2 \pi  k^3 \tau ^3+\pi  \gamma  k^2 \tau ^2+4 (\gamma -\pi ) \gamma ^2 k \tau \right) e^{6 t \left(\frac{k}{\gamma }+\frac{1}{\tau }\right)}\right.\nonumber\\\left.+\left(-3 (\pi -2 \gamma ) \gamma
   ^3+\pi  k^3 \tau ^3+(3 \pi -2 \gamma ) \gamma  k^2 \tau ^2+3 \pi  \gamma ^2 k \tau \right) e^{\frac{4 k t}{\gamma }+\frac{6 t}{\tau }}-\gamma  \left(\Gamma _{\tau }\right){}^* \left(4 \gamma ^2-3 \pi  \gamma +\pi
    k \tau \right) e^{\frac{2 k t}{\gamma }+\frac{6 t}{\tau }}\right.\nonumber\\\left.+\Gamma _{\tau }^2 \left(\gamma ^2+\pi  k \tau \right) e^{\frac{8 k t}{\gamma }+\frac{6 t}{\tau }}+4 \gamma ^2 k \tau  (3 \gamma +k \tau -2 \pi )
   e^{\frac{3 k t}{\gamma }+\frac{5 t}{\tau }}-4 \gamma  k \tau  \left(-\gamma ^2+\gamma  k \tau -2 \pi  k \tau \right) e^{\frac{7 k t}{\gamma }+\frac{5 t}{\tau }}\right.\nonumber\\\left.+4 (\pi -\gamma ) \gamma  k \tau  \left(\Gamma
   _{\tau }\right){}^* e^{\frac{k t}{\gamma }+\frac{5 t}{\tau }}-\pi  k \tau  \left(\left(\Gamma _{\tau }\right){}^*\right){}^2 e^{4 t \left(\frac{2 k}{\gamma }+\frac{1}{\tau }\right)}+4 \gamma  k \tau  (\pi 
   (\gamma -3 k \tau )+\gamma  (k \tau -3 \gamma )) e^{5 t \left(\frac{k}{\gamma }+\frac{1}{\tau }\right)}\right.\nonumber\\\left.+(\gamma -\pi ) \gamma  \left(\left(\Gamma _{\tau }\right){}^*\right){}^2 e^{\frac{6 t}{\tau }}\right).~~~~~~~~
\end{eqnarray}
\begin{eqnarray}
    l = \frac{2 F_0^2 e^{-\frac{4 k t}{\gamma }-\frac{6 t}{\tau }}}{\beta
    \Gamma _{\tau }^3 \left(\left(\Gamma _{\tau }\right){}^*\right){}^2 \left(e^{\frac{2 k t}{\gamma }}-1\right)^2} \left(2 k \tau  \left(\pi  \left(\gamma ^2+k^2 \tau ^2\right)+2 \gamma ^2 k \tau \right) e^{\frac{6 k t}{\gamma }+\frac{4 t}{\tau }}-k \tau  \left(\pi 
   \left(\gamma ^2+k^2 \tau ^2-6 \gamma  k \tau \right)+8 \gamma ^2 k \tau \right) e^{4 t \left(\frac{k}{\gamma }+\frac{1}{\tau }\right)}\right.\nonumber\\\left.-4 (\pi -\gamma ) \gamma  k^2 \tau ^2 e^{\frac{2 k t}{\gamma }+\frac{4 t}{\tau
   }}+\left((\pi -4 \gamma ) \gamma ^3-2 \pi  k^3 \tau ^3+\pi  \gamma  k^2 \tau ^2+4 (\gamma -\pi ) \gamma ^2 k \tau \right) e^{6 t \left(\frac{k}{\gamma }+\frac{1}{\tau }\right)}\right.\nonumber\\\left.+\left(-3 (\pi -2 \gamma ) \gamma
   ^3+\pi  k^3 \tau ^3+(3 \pi -2 \gamma ) \gamma  k^2 \tau ^2+3 \pi  \gamma ^2 k \tau \right) e^{\frac{4 k t}{\gamma }+\frac{6 t}{\tau }}-\gamma  \left(\Gamma _{\tau }\right){}^* \left(4 \gamma ^2-3 \pi  \gamma +\pi
    k \tau \right) e^{\frac{2 k t}{\gamma }+\frac{6 t}{\tau }}\right.\nonumber\\\left.+\Gamma _{\tau }^2 \left(\gamma ^2+\pi  k \tau \right) e^{\frac{8 k t}{\gamma }+\frac{6 t}{\tau }}+4 \gamma ^2 k \tau  (3 \gamma +k \tau -2 \pi )
   e^{\frac{3 k t}{\gamma }+\frac{5 t}{\tau }}-4 \gamma  k \tau  \left(-\gamma ^2+\gamma  k \tau -2 \pi  k \tau \right) e^{\frac{7 k t}{\gamma }+\frac{5 t}{\tau }}\right.\nonumber\\\left.+4 (\pi -\gamma ) \gamma  k \tau  \left(\Gamma
   _{\tau }\right){}^* e^{\frac{k t}{\gamma }+\frac{5 t}{\tau }}-\pi  k \tau  \left(\left(\Gamma _{\tau }\right){}^*\right){}^2 e^{4 t \left(\frac{2 k}{\gamma }+\frac{1}{\tau }\right)}+4 \gamma  k \tau  (\pi 
   (\gamma -3 k \tau )+\gamma  (k \tau -3 \gamma )) e^{5 t \left(\frac{k}{\gamma }+\frac{1}{\tau }\right)}\right.\nonumber\\\left.+(\gamma -\pi ) \gamma  \left(\left(\Gamma _{\tau }\right){}^*\right){}^2 e^{\frac{6 t}{\tau }}\right).~~~~~~~
\end{eqnarray}
\begin{eqnarray}
    m = \frac{6 F_0^2 t e^{-\frac{4 k t}{\gamma }-\frac{6 t}{\tau }} }{\beta
    k \tau ^2 \Gamma _{\tau }^2 \left(\left(\Gamma _{\tau }\right){}^*\right){}^2 \left(e^{\frac{2 k t}{\gamma }}-1\right)^2}\left(2 k \tau  \left(\pi  \left(\gamma ^2+k^2 \tau ^2\right)+2 \gamma ^2 k \tau \right) e^{\frac{6 k t}{\gamma }+\frac{4 t}{\tau }}-k \tau  \left(\pi 
   \left(\gamma ^2+k^2 \tau ^2-6 \gamma  k \tau \right)+8 \gamma ^2 k \tau \right) e^{4 t \left(\frac{k}{\gamma }+\frac{1}{\tau }\right)}\right.\nonumber\\\left.-4 (\pi -\gamma ) \gamma  k^2 \tau ^2 e^{\frac{2 k t}{\gamma }+\frac{4 t}{\tau
   }}+\left((\pi -4 \gamma ) \gamma ^3-2 \pi  k^3 \tau ^3+\pi  \gamma  k^2 \tau ^2+4 (\gamma -\pi ) \gamma ^2 k \tau \right) e^{6 t \left(\frac{k}{\gamma }+\frac{1}{\tau }\right)}\right.\nonumber\\\left.+\left(-3 (\pi -2 \gamma ) \gamma
   ^3+\pi  k^3 \tau ^3+(3 \pi -2 \gamma ) \gamma  k^2 \tau ^2+3 \pi  \gamma ^2 k \tau \right) e^{\frac{4 k t}{\gamma }+\frac{6 t}{\tau }}-\gamma  \left(\Gamma _{\tau }\right){}^* \left(4 \gamma ^2-3 \pi  \gamma +\pi
    k \tau \right) e^{\frac{2 k t}{\gamma }+\frac{6 t}{\tau }}\right.\nonumber\\\left.+\Gamma _{\tau }^2 \left(\gamma ^2+\pi  k \tau \right) e^{\frac{8 k t}{\gamma }+\frac{6 t}{\tau }}+4 \gamma ^2 k \tau  (3 \gamma +k \tau -2 \pi )
   e^{\frac{3 k t}{\gamma }+\frac{5 t}{\tau }}-4 \gamma  k \tau  \left(-\gamma ^2+\gamma  k \tau -2 \pi  k \tau \right) e^{\frac{7 k t}{\gamma }+\frac{5 t}{\tau }}\right.\nonumber\\\left.+4 (\pi -\gamma ) \gamma  k \tau  \left(\Gamma
   _{\tau }\right){}^* e^{\frac{k t}{\gamma }+\frac{5 t}{\tau }}-\pi  k \tau  \left(\left(\Gamma _{\tau }\right){}^*\right){}^2 e^{4 t \left(\frac{2 k}{\gamma }+\frac{1}{\tau }\right)}+4 \gamma  k \tau  (\pi 
   (\gamma -3 k \tau )+\gamma  (k \tau -3 \gamma )) e^{5 t \left(\frac{k}{\gamma }+\frac{1}{\tau }\right)}\right.\nonumber\\\left.+(\gamma -\pi ) \gamma  \left(\left(\Gamma _{\tau }\right){}^*\right){}^2 e^{\frac{6 t}{\tau }}\right).~~~~~~~~
\end{eqnarray}

\subsection{$P(\Delta_i s)$ and $\mathcal{I}^{\Delta_i s~\delta}_{\{F_0,\tau\}}$ for an Dirac delta initial condition}

For a Dirac delta initial condition, the characteristic function is given by
\begin{eqnarray}
    \log(Z(\lambda)) = a + b + c + d
\end{eqnarray}
with
\begin{equation}
    a = -\frac{e^{\frac{2 t}{\tau }-2 t \left(\frac{k}{\gamma }+\frac{1}{\tau }\right)} \pi  \left(\frac{2 e^{\frac{k t}{\gamma }} \beta  F_0 \Gamma _{\tau } \Theta _-(t) \gamma
   ^2}{\pi }-\frac{2 e^{2 t \left(\frac{k}{\gamma }+\frac{1}{\tau }\right)-\frac{2 t}{\tau }} \beta  F_0 \Gamma _{\tau }^2 \Theta _-(t) \gamma }{\pi }-\frac{2 e^{t
   \left(\frac{2 k}{\gamma }+\frac{1}{\tau }\right)-\frac{2 t}{\tau }} k \beta  \tau  F_0 \Gamma _{\tau } \Theta _-(t) \gamma }{\pi }\right){}^2}{16 k \beta  \gamma  \Gamma
   _{\tau }^4 \Theta _-(t)}
\end{equation}
\begin{eqnarray}
    b = \frac{e^{\frac{2 t}{\tau }-2 t \left(\frac{k}{\gamma }+\frac{1}{\tau }\right)} \pi}{16 k \beta  \gamma  \Gamma
   _{\tau }^4 \Theta _-(t)}  \left(-\frac{2 e^{2 t \left(\frac{k}{\gamma }+\frac{1}{\tau
   }\right)-\frac{2 t}{\tau }} \beta  \gamma  F_0 \Theta _-(t) \Gamma _{\tau }^2}{\pi }+\frac{2 i e^{-\frac{t}{\tau }} \gamma  \lambda  F_0 \Theta _-(t) \Gamma _{\tau
   }^2}{\left(\Gamma _{\tau }\right){}^*}\right.\nonumber\\\left.+4 i \left(-e^{-\frac{k t}{\gamma }}+e^{-\frac{t}{\tau }}\right) \gamma  \lambda  F_0 \Gamma _{\tau }-2 i e^{t \left(\frac{2
   k}{\gamma }+\frac{1}{\tau }\right)-\frac{2 t}{\tau }} \gamma  \lambda  F_0 \Theta _-(t) \Gamma _{\tau }-\frac{2 e^{t \left(\frac{2 k}{\gamma }+\frac{1}{\tau
   }\right)-\frac{2 t}{\tau }} k \beta  \gamma  \tau  F_0 \Theta _-(t) \Gamma _{\tau }}{\pi }\right.\nonumber\\\left.+\frac{2 e^{\frac{k t}{\gamma }} \gamma  \left(2 i k \pi  \lambda  \tau 
   F_0+\beta  \gamma  \left(\Gamma _{\tau }\right){}^* F_0\right) \Theta _-(t) \Gamma _{\tau }}{\pi  \left(\Gamma _{\tau }\right){}^*}\right)^2,
\end{eqnarray}
\begin{eqnarray}
    c = -\frac{1}{4 \Gamma _{\tau }^2}\left(-\frac{\beta  F_0^2 \Theta _-(t) \gamma ^3}{k \pi }+\frac{2 e^{t \left(\frac{k}{\gamma }+\frac{1}{\tau }\right)-\frac{2 t}{\tau }} \beta 
   \tau  F_0^2 \Theta _-(t) \gamma ^2}{\pi }+\frac{2 e^{\frac{k t}{\gamma }} \beta  F_0^2 \Gamma _{\tau } \Theta _-(t) \gamma ^2}{k \pi }\right.\nonumber\\\left.-\frac{e^{\frac{2 k t}{\gamma
   }-\frac{2 t}{\tau }} k \beta  \tau ^2 F_0^2 \Theta _-(t) \gamma }{\pi }-\frac{e^{2 t \left(\frac{k}{\gamma }+\frac{1}{\tau }\right)-\frac{2 t}{\tau }} \beta  F_0^2 \Gamma
   _{\tau }^2 \Theta _-(t) \gamma }{k \pi }-\frac{2 e^{t \left(\frac{2 k}{\gamma }+\frac{1}{\tau }\right)-\frac{2 t}{\tau }} \beta  \tau  F_0^2 \Gamma _{\tau } \Theta _-(t)
   \gamma }{\pi }\right),
\end{eqnarray}
\begin{eqnarray}
    d = \frac{1}{4 \Gamma _{\tau }^2}\left(-\frac{2 i e^{-2 t \left(\frac{k}{\gamma }+\frac{1}{\tau }\right)} \lambda  \left(-e^{t/\tau } \gamma +e^{\frac{k t}{\gamma }}
   k \tau +e^{t \left(\frac{k}{\gamma }+\frac{1}{\tau }\right)} \Gamma _{\tau }\right){}^2 F_0^2}{k}-\frac{e^{\frac{2 k t}{\gamma }-\frac{2 t}{\tau }} (\beta  \gamma +i \pi 
   \lambda ) \tau  (i \pi  \lambda +k \beta  \tau ) \Theta _-(t) F_0^2}{\pi  \beta }\right.\nonumber\\\left.-\frac{\left(\beta ^2 \gamma ^3-k \pi ^2 \lambda ^2 \tau -i k \pi  \beta  \lambda  \tau 
   (k \tau -3 \gamma )\right) \Theta _-(t) F_0^2}{k \pi  \beta }-\frac{2 e^{t \left(\frac{2 k}{\gamma }+\frac{1}{\tau }\right)-\frac{2 t}{\tau }} \beta  \gamma  \tau  \Gamma
   _{\tau } \Theta _-(t) F_0^2}{\pi }\right.\nonumber\\\left.-2 i e^{t \left(\frac{2 k}{\gamma }+\frac{1}{\tau }\right)-\frac{2 t}{\tau }} \lambda  \tau  \Gamma _{\tau } \Theta _-(t)
   F_0^2-\frac{e^{-\frac{2 t}{\tau }} \lambda  \tau  \Gamma _{\tau }^2 \left(\pi  \lambda -i \beta  \left(\Gamma _{\tau }\right){}^*\right) \Theta _-(t) F_0^2}{\beta 
   \left(\left(\Gamma _{\tau }\right){}^*\right){}^2}\right.\nonumber\\\left.-\frac{e^{2 t \left(\frac{k}{\gamma }+\frac{1}{\tau }\right)-\frac{2 t}{\tau }} \Gamma _{\tau }^2 \left(k \pi ^2 \tau 
   \lambda ^2+i k \pi  \beta  \tau  \left(\Gamma _{\tau }\right){}^* \lambda +\beta ^2 \gamma  \left(\left(\Gamma _{\tau }\right){}^*\right){}^2\right) \Theta _-(t) F_0^2}{k
   \pi  \beta  \left(\left(\Gamma _{\tau }\right){}^*\right){}^2}-\frac{2 i e^{-\frac{t}{\tau }} \lambda  \tau  \Gamma _{\tau }^2 \Theta _-(t) F_0^2}{\left(\Gamma _{\tau
   }\right){}^*}\right.\nonumber\\\left.+\frac{2 e^{t \left(\frac{k}{\gamma }+\frac{1}{\tau }\right)-\frac{2 t}{\tau }} \gamma  \tau  \left(2 i \pi  \lambda +\beta  \left(\Gamma _{\tau
   }\right){}^*\right) \left(2 i k \pi  \lambda  \tau  F_0+\beta  \gamma  \left(\Gamma _{\tau }\right){}^* F_0\right) \Theta _-(t) F_0}{\pi  \beta  \left(\left(\Gamma _{\tau
   }\right){}^*\right){}^2} \right.\nonumber\\\left.+\frac{2 e^{\frac{k t}{\gamma }} \gamma  \Gamma _{\tau } \left(2 i k \pi  \lambda  \tau  F_0+\beta  \gamma  \left(\Gamma _{\tau }\right){}^*
   F_0\right) \Theta _-(t) F_0}{k \pi  \left(\Gamma _{\tau }\right)^*}\right)
\end{eqnarray}

The resulting Gaussian probability distribution has mean
\begin{eqnarray}
    \mu_{s}^{\delta} = -\frac{F_0^2 e^{-2 t \left(\frac{k}{\gamma }+\frac{1}{\tau }\right)}}{2 k \Gamma _{\tau }^2 \left(\Gamma
   _{\tau }\right){}^*} \left(4 \gamma  \Gamma _{\tau } \left(\Gamma _{\tau }\right){}^* e^{\frac{k t}{\gamma }+\frac{2 t}{\tau
   }}-\Gamma _{\tau }^2 (\gamma +2 k \tau ) e^{2 t \left(\frac{k}{\gamma }+\frac{1}{\tau }\right)}+4 k \tau  \left(\Gamma _{\tau }\right){}^* (k \tau -\gamma ) e^{t
   \left(\frac{2 k}{\gamma }+\frac{1}{\tau }\right)}\right.\nonumber\\\left.-k \tau  \left(\Gamma _{\tau }\right){}^* (\gamma +2 k \tau ) e^{\frac{2 k t}{\gamma }}+4 \gamma  k \tau  (2 \gamma +k
   \tau ) e^{t \left(\frac{k}{\gamma }+\frac{1}{\tau }\right)}-3 \gamma ^2 \left(\Gamma _{\tau }\right){}^* e^{\frac{2 t}{\tau }}\right),
\end{eqnarray}
and variance
\begin{eqnarray}
    \sigma_{s}^{\delta 2}(t) = \frac{\pi  F_0^2 e^{-2 t \left(\frac{2 k}{\gamma }+\frac{1}{\tau }\right)}}{\beta k \left(\gamma ^2-k^2 \tau ^2\right)^2} \left(4 \gamma  k^2 \tau ^2 e^{\frac{2 k t}{\gamma }}-8 \gamma  k^2 \tau ^2 e^{t \left(\frac{3
   k}{\gamma }+\frac{1}{\tau }\right)}-k \tau  \Gamma _{\tau }^2 e^{2 t \left(\frac{2 k}{\gamma }+\frac{1}{\tau }\right)}+k \tau  \left(\left(\Gamma _{\tau
   }\right){}^*\right){}^2 e^{\frac{4 k t}{\gamma }}\right.\nonumber\\\left.-4 \gamma  k \tau  \left(\Gamma _{\tau }\right){}^* e^{t \left(\frac{k}{\gamma }+\frac{1}{\tau }\right)}-\gamma 
   \left(\Gamma _{\tau }\right){}^* (\gamma -3 k \tau ) e^{2 t \left(\frac{k}{\gamma }+\frac{1}{\tau }\right)}+\gamma  \left(\left(\Gamma _{\tau }\right){}^*\right){}^2
   e^{\frac{2 t}{\tau }}\right).
\end{eqnarray}

In that case, $\alpha$ for $\mathcal{I}_{F_0, F_0}^{\Delta_i s~\delta}$ is given by
\begin{eqnarray}
    \alpha_{\Delta_{i}s~\delta}(t) = -\frac{e^{-\frac{8 k t}{\gamma }-\frac{6 t}{\tau }} a^2}{\pi  \beta  k^3 \Gamma_{\tau}^{6} \Gamma_{\tau}^{* 4}b}
\end{eqnarray}
\begin{eqnarray}
    a = \left(4 \gamma ^2 k \tau -4 k^3 \tau ^3\right) e^{t \left(\frac{2 k}{\gamma }+\frac{1}{\tau }\right)}+3 \gamma ^2
   e^{\frac{2 t}{\tau }} (\gamma +k \tau )-4 \gamma  (\gamma -k \tau ) (\gamma +k \tau ) e^{\frac{k t}{\gamma }+\frac{2 t}{\tau }}\nonumber\\-4 \gamma  k \tau  (2 \gamma +k \tau ) e^{t
   \left(\frac{k}{\gamma }+\frac{1}{\tau }\right)}+(\gamma -k \tau )^2 (\gamma +2 k \tau ) e^{2 t \left(\frac{k}{\gamma }+\frac{1}{\tau }\right)}+k \tau  (\gamma +k \tau )
   (\gamma +2 k \tau ) e^{\frac{2 k t}{\gamma }},\\
   b =  \left(4 \gamma  k^2 \tau ^2 e^{\frac{2 k t}{\gamma }}-8
   \gamma  k^2 \tau ^2 e^{t \left(\frac{3 k}{\gamma }+\frac{1}{\tau }\right)}-k \tau  (\gamma -k \tau )^2 e^{2 t \left(\frac{2 k}{\gamma }+\frac{1}{\tau }\right)}+k \tau 
   (\gamma +k \tau )^2 e^{\frac{4 k t}{\gamma }}\right.\nonumber\\\left.-4 \gamma  k \tau  (\gamma +k \tau ) e^{t \left(\frac{k}{\gamma }+\frac{1}{\tau }\right)}+\gamma  e^{\frac{2 t}{\tau }}
   (\gamma +k \tau )^2-\gamma  (\gamma -3 k \tau ) (\gamma +k \tau ) e^{2 t \left(\frac{k}{\gamma }+\frac{1}{\tau }\right)}\right).
\end{eqnarray}

For $\tau$, the Fisher information is given by
\begin{eqnarray}
    \mathcal{I}_{\tau}^{\Delta_i s~\delta}(t) = a(b + cd(e + f + g + hi)^2)
\end{eqnarray}
\begin{eqnarray}
    b = -\frac{1}{k \beta  \left(\gamma ^2-k^2 \tau ^2\right)^2}2 e^{-2 t \left(\frac{2 k}{\gamma }+\frac{1}{\tau }\right)} \pi  F_0^2 \left(4 e^{\frac{2 k t}{\gamma }} k^2 \gamma  \tau ^2-8 e^{t \left(\frac{3 k}{\gamma
   }+\frac{1}{\tau }\right)} k^2 \gamma  \tau ^2-e^{2 t \left(\frac{2 k}{\gamma }+\frac{1}{\tau }\right)} k \Gamma _{\tau }^2 \tau +e^{\frac{4 k t}{\gamma }} k
   \left(\left(\Gamma _{\tau }\right){}^*\right){}^2 \tau \right.\nonumber\\\left.-4 e^{t \left(\frac{k}{\gamma }+\frac{1}{\tau }\right)} k \gamma  \left(\Gamma _{\tau }\right){}^* \tau +e^{\frac{2
   t}{\tau }} \gamma  \left(\left(\Gamma _{\tau }\right){}^*\right){}^2-e^{2 t \left(\frac{k}{\gamma }+\frac{1}{\tau }\right)} \gamma  (\gamma -3 k \tau ) \left(\Gamma
   _{\tau }\right){}^*\right)\nonumber\\ \left(-\frac{e^{-2 t \left(\frac{k}{\gamma }+\frac{1}{\tau }\right)} t}{k \tau ^2 \Gamma _{\tau }^2 \Gamma_{\tau }^*} \left(-3 e^{\frac{2 t}{\tau }} \left(\Gamma _{\tau }\right){}^* \gamma
   ^2+4 e^{t \left(\frac{k}{\gamma }+\frac{1}{\tau }\right)} k \tau  (2 \gamma +k \tau ) \gamma +4 e^{\frac{k t}{\gamma }+\frac{2 t}{\tau }} \Gamma _{\tau } \left(\Gamma
   _{\tau }\right){}^* \gamma \right.\right.\nonumber\\\left.\left.-e^{2 t \left(\frac{k}{\gamma }+\frac{1}{\tau }\right)} (\gamma +2 k \tau ) \Gamma _{\tau }^2+4 e^{t \left(\frac{2 k}{\gamma }+\frac{1}{\tau
   }\right)} k \tau  (k \tau -\gamma ) \left(\Gamma _{\tau }\right){}^*-e^{\frac{2 k t}{\gamma }} k \tau  (\gamma +2 k \tau ) \left(\Gamma _{\tau }\right){}^*\right)
   F_0^2\nonumber\right.\\\left.-\frac{e^{-2 t \left(\frac{k}{\gamma }+\frac{1}{\tau }\right)}}{\Gamma _{\tau }^3 \left(\Gamma _{\tau }\right){}^*} \left(-3 e^{\frac{2 t}{\tau }}
   \left(\Gamma _{\tau }\right){}^* \gamma ^2+4 e^{t \left(\frac{k}{\gamma }+\frac{1}{\tau }\right)} k \tau  (2 \gamma +k \tau ) \gamma +4 e^{\frac{k t}{\gamma }+\frac{2
   t}{\tau }} \Gamma _{\tau } \left(\Gamma _{\tau }\right){}^* \gamma \right.\right.\nonumber\\\left.\left.-e^{2 t \left(\frac{k}{\gamma }+\frac{1}{\tau }\right)} (\gamma +2 k \tau ) \Gamma _{\tau }^2+4 e^{t
   \left(\frac{2 k}{\gamma }+\frac{1}{\tau }\right)} k \tau  (k \tau -\gamma ) \left(\Gamma _{\tau }\right){}^*-e^{\frac{2 k t}{\gamma }} k \tau  (\gamma +2 k \tau )
   \left(\Gamma _{\tau }\right){}^*\right) F_0^2\right.\nonumber\\\left.+\frac{e^{-2 t \left(\frac{k}{\gamma }+\frac{1}{\tau }\right)}}{2 \Gamma _{\tau }^2 \left(\left(\Gamma _{\tau }\right){}^*\right){}^2} \left(-3
   e^{\frac{2 t}{\tau }} \left(\Gamma _{\tau }\right){}^* \gamma ^2+4 e^{t \left(\frac{k}{\gamma }+\frac{1}{\tau }\right)} k \tau  (2 \gamma +k \tau ) \gamma +4 e^{\frac{k
   t}{\gamma }+\frac{2 t}{\tau }} \Gamma _{\tau } \left(\Gamma _{\tau }\right){}^* \gamma\right.\right.\nonumber\\\left.\left. -e^{2 t \left(\frac{k}{\gamma }+\frac{1}{\tau }\right)} (\gamma +2 k \tau ) \Gamma
   _{\tau }^2+4 e^{t \left(\frac{2 k}{\gamma }+\frac{1}{\tau }\right)} k \tau  (k \tau -\gamma ) \left(\Gamma _{\tau }\right){}^*-e^{\frac{2 k t}{\gamma }} k \tau  (\gamma
   +2 k \tau ) \left(\Gamma _{\tau }\right){}^*\right) F_0^2\nonumber\right.\\\left.-\frac{e^{-2 t \left(\frac{k}{\gamma
   }+\frac{1}{\tau }\right)}}{2 k \Gamma _{\tau }^2 \left(\Gamma _{\tau
   }\right){}^*} \left(4 e^{t \left(\frac{k}{\gamma }+\frac{1}{\tau }\right)} \gamma  \tau  k^2+4 e^{t \left(\frac{2 k}{\gamma }+\frac{1}{\tau }\right)} \tau  (k
   \tau -\gamma ) k^2-e^{\frac{2 k t}{\gamma }} \tau  (\gamma +2 k \tau ) k^2-2 e^{\frac{2 k t}{\gamma }} \tau  \left(\Gamma _{\tau }\right){}^* k^2\right.\right.\nonumber\\\left.\left.+4 e^{t \left(\frac{2
   k}{\gamma }+\frac{1}{\tau }\right)} \tau  \left(\Gamma _{\tau }\right){}^* k^2-3 e^{\frac{2 t}{\tau }} \gamma ^2 k-2 e^{2 t \left(\frac{k}{\gamma }+\frac{1}{\tau
   }\right)} \Gamma _{\tau }^2 k+4 e^{t \left(\frac{k}{\gamma }+\frac{1}{\tau }\right)} \gamma  (2 \gamma +k \tau ) k\right.\right.\nonumber\\\left.\left.-\frac{4 e^{t \left(\frac{k}{\gamma }+\frac{1}{\tau
   }\right)} t \gamma  (2 \gamma +k \tau ) k}{\tau }+4 e^{\frac{k t}{\gamma }+\frac{2 t}{\tau }} \gamma  \Gamma _{\tau } k+2 e^{2 t \left(\frac{k}{\gamma }+\frac{1}{\tau
   }\right)} (\gamma +2 k \tau ) \Gamma _{\tau } k\right.\right.\nonumber\\\left.\left.-4 e^{\frac{k t}{\gamma }+\frac{2 t}{\tau }} \gamma  \left(\Gamma _{\tau }\right){}^* k+4 e^{t \left(\frac{2 k}{\gamma
   }+\frac{1}{\tau }\right)} (k \tau -\gamma ) \left(\Gamma _{\tau }\right){}^* k-\frac{4 e^{t \left(\frac{2 k}{\gamma }+\frac{1}{\tau }\right)} t (k \tau -\gamma )
   \left(\Gamma _{\tau }\right){}^* k}{\tau }-e^{\frac{2 k t}{\gamma }} (\gamma +2 k \tau ) \left(\Gamma _{\tau }\right){}^* k\right.\right.\nonumber\\\left.\left.+\frac{2 e^{2 t \left(\frac{k}{\gamma
   }+\frac{1}{\tau }\right)} t (\gamma +2 k \tau ) \Gamma _{\tau }^2}{\tau ^2}-\frac{8 e^{\frac{k t}{\gamma }+\frac{2 t}{\tau }} t \gamma  \Gamma _{\tau } \left(\Gamma
   _{\tau }\right){}^*}{\tau ^2}+\frac{6 e^{\frac{2 t}{\tau }} t \gamma ^2 \left(\Gamma _{\tau }\right){}^*}{\tau ^2}\right) F_0^2\right){}^2,
   \end{eqnarray}
   \begin{eqnarray}
   c = -\frac{2 \pi  F_0^2 e^{-2 t \left(\frac{2 k}{\gamma }+\frac{1}{\tau }\right)}}{\beta  k \left(\gamma ^2-k^2 \tau ^2\right)^2},
   \end{eqnarray}
   \begin{eqnarray}
    d = 4 \gamma  k^2 \tau ^2 e^{\frac{2 k t}{\gamma }}-8 \gamma  k^2 \tau ^2 e^{t \left(\frac{3 k}{\gamma }+\frac{1}{\tau }\right)}-k \tau  \Gamma _{\tau }^2 e^{2 t \left(\frac{2
   k}{\gamma }+\frac{1}{\tau }\right)}+k \tau  \left(\left(\Gamma _{\tau }\right){}^*\right){}^2 e^{\frac{4 k t}{\gamma }}-4 \gamma  k \tau  \left(\Gamma _{\tau }\right){}^*
   e^{t \left(\frac{k}{\gamma }+\frac{1}{\tau }\right)}\nonumber\\-\gamma  \left(\Gamma _{\tau }\right){}^* (\gamma -3 k \tau ) e^{2 t \left(\frac{k}{\gamma }+\frac{1}{\tau
   }\right)}+\gamma  \left(\left(\Gamma _{\tau }\right){}^*\right){}^2 e^{\frac{2 t}{\tau }},
   \end{eqnarray}
   \begin{eqnarray}
   e = \frac{F_0^2 e^{-2 t \left(\frac{k}{\gamma }+\frac{1}{\tau }\right)}}{2 \Gamma _{\tau }^2
   \left(\left(\Gamma _{\tau }\right){}^*\right){}^2} \left(4 \gamma  \Gamma _{\tau } \left(\Gamma _{\tau }\right){}^* e^{\frac{k t}{\gamma }+\frac{2 t}{\tau
   }}-\Gamma _{\tau }^2 (\gamma +2 k \tau ) e^{2 t \left(\frac{k}{\gamma }+\frac{1}{\tau }\right)}+4 k \tau  \left(\Gamma _{\tau }\right){}^* (k \tau -\gamma ) e^{t
   \left(\frac{2 k}{\gamma }+\frac{1}{\tau }\right)}\right.\nonumber\\\left.-k \tau  \left(\Gamma _{\tau }\right){}^* (\gamma +2 k \tau ) e^{\frac{2 k t}{\gamma }}+4 \gamma  k \tau  (2 \gamma +k
   \tau ) e^{t \left(\frac{k}{\gamma }+\frac{1}{\tau }\right)}-3 \gamma ^2 \left(\Gamma _{\tau }\right){}^* e^{\frac{2 t}{\tau }}\right),\\
   f = -\frac{F_0^2 e^{-2 t \left(\frac{k}{\gamma }+\frac{1}{\tau }\right)}}{\Gamma _{\tau }^3 \left(\Gamma
   _{\tau }\right){}^*} \left(4 \gamma  \Gamma _{\tau } \left(\Gamma _{\tau }\right){}^* e^{\frac{k t}{\gamma }+\frac{2 t}{\tau}}-\Gamma _{\tau }^2 (\gamma +2 k \tau ) e^{2 t \left(\frac{k}{\gamma }+\frac{1}{\tau }\right)}+4 k \tau  \left(\Gamma _{\tau }\right){}^* (k \tau -\gamma ) e^{t\left(\frac{2 k}{\gamma }+\frac{1}{\tau }\right)}\right.\nonumber\\\left.-k \tau  \left(\Gamma _{\tau }\right){}^* (\gamma +2 k \tau ) e^{\frac{2 k t}{\gamma }}+4 \gamma  k \tau  (2 \gamma +k\tau ) e^{t \left(\frac{k}{\gamma }+\frac{1}{\tau }\right)}-3 \gamma ^2 \left(\Gamma _{\tau }\right){}^* e^{\frac{2 t}{\tau }}\right),
   \end{eqnarray}
   \begin{eqnarray}
   g = -\frac{F_0^2 t e^{-2 t \left(\frac{k}{\gamma }+\frac{1}{\tau }\right)} }{k \tau ^2 \Gamma _{\tau }^2
   \left(\Gamma _{\tau }\right){}^*}\left(4 \gamma  \Gamma _{\tau } \left(\Gamma _{\tau }\right){}^* e^{\frac{k t}{\gamma }+\frac{2
   t}{\tau }}-\Gamma _{\tau }^2 (\gamma +2 k \tau ) e^{2 t \left(\frac{k}{\gamma }+\frac{1}{\tau }\right)}+4 k \tau  \left(\Gamma _{\tau }\right){}^* (k \tau -\gamma ) e^{t
   \left(\frac{2 k}{\gamma }+\frac{1}{\tau }\right)}\right.\nonumber\\\left.-k \tau  \left(\Gamma _{\tau }\right){}^* (\gamma +2 k \tau ) e^{\frac{2 k t}{\gamma }}+4 \gamma  k \tau  (2 \gamma +k
   \tau ) e^{t \left(\frac{k}{\gamma }+\frac{1}{\tau }\right)}-3 \gamma ^2 \left(\Gamma _{\tau }\right){}^* e^{\frac{2 t}{\tau }}\right),\\
   h = -\frac{F_0^2 e^{-2 t \left(\frac{k}{\gamma }+\frac{1}{\tau }\right)}}{2 k \Gamma _{\tau }^2 \left(\Gamma _{\tau }\right){}^*},\\
   i = -2 k^2 \tau  \left(\Gamma _{\tau }\right){}^* e^{\frac{2 k t}{\gamma }}+4 k^2 \tau  \left(\Gamma _{\tau }\right){}^* e^{t \left(\frac{2 k}{\gamma }+\frac{1}{\tau }\right)}+4
   \gamma  k^2 \tau  e^{t \left(\frac{k}{\gamma }+\frac{1}{\tau }\right)}+4 k^2 \tau  (k \tau -\gamma ) e^{t \left(\frac{2 k}{\gamma }+\frac{1}{\tau }\right)}\nonumber\\-k^2 \tau 
   (\gamma +2 k \tau ) e^{\frac{2 k t}{\gamma }}-3 \gamma ^2 k e^{\frac{2 t}{\tau }}+\frac{2 t \Gamma _{\tau }^2 (\gamma +2 k \tau ) e^{2 t \left(\frac{k}{\gamma
   }+\frac{1}{\tau }\right)}}{\tau ^2}-\frac{8 \gamma  t \Gamma _{\tau } \left(\Gamma _{\tau }\right){}^* e^{\frac{k t}{\gamma }+\frac{2 t}{\tau }}}{\tau ^2}-2 k \Gamma
   _{\tau }^2 e^{2 t \left(\frac{k}{\gamma }+\frac{1}{\tau }\right)}\nonumber\\+4 \gamma  k \Gamma _{\tau } e^{\frac{k t}{\gamma }+\frac{2 t}{\tau }}+2 k \Gamma _{\tau } (\gamma +2 k\tau ) e^{2 t \left(\frac{k}{\gamma }+\frac{1}{\tau }\right)}-4 \gamma  k \left(\Gamma _{\tau }\right){}^* e^{\frac{k t}{\gamma }+\frac{2 t}{\tau }}+4 k \left(\Gamma{\tau }\right){}^* (k \tau -\gamma ) e^{t \left(\frac{2 k}{\gamma }+\frac{1}{\tau }\right)}\nonumber\\-\frac{4 k t \left(\Gamma _{\tau }\right){}^* (k \tau -\gamma ) e^{t
   \left(\frac{2 k}{\gamma }+\frac{1}{\tau }\right)}}{\tau }-k \left(\Gamma _{\tau }\right){}^* (\gamma +2 k \tau ) e^{\frac{2 k t}{\gamma }}+4 \gamma  k (2 \gamma +k \tau )
   e^{t \left(\frac{k}{\gamma }+\frac{1}{\tau }\right)}\nonumber\\-\frac{4 \gamma  k t (2 \gamma +k \tau ) e^{t \left(\frac{k}{\gamma }+\frac{1}{\tau }\right)}}{\tau }+\frac{6 \gamma
   ^2 t \left(\Gamma _{\tau }\right){}^* e^{\frac{2 t}{\tau }}}{\tau ^2}
\end{eqnarray}

The off-diagonal term is given by
\begin{equation}
    \mathcal{I}_{F_0, \tau}^{\Delta_i s~\delta} = \frac{b(c + d(e + f + g + h))}{a},
\end{equation}
with
\begin{eqnarray}
    a = 2 \pi ^2 F_0^4 \left(4 \gamma  k^2 \tau ^2 e^{\frac{2 k t}{\gamma }}-8 \gamma  k^2 \tau ^2 e^{t \left(\frac{3 k}{\gamma }+\frac{1}{\tau }\right)}-k \tau  \Gamma _{\tau }^2
   e^{2 t \left(\frac{2 k}{\gamma }+\frac{1}{\tau }\right)}+k \tau  \left(\left(\Gamma _{\tau }\right){}^*\right){}^2 e^{\frac{4 k t}{\gamma }}-4 \gamma  k \tau 
   \left(\Gamma _{\tau }\right){}^* e^{t \left(\frac{k}{\gamma }+\frac{1}{\tau }\right)}\nonumber\right.\\\left.-\gamma  \left(\Gamma _{\tau }\right){}^* (\gamma -3 k \tau ) e^{2 t
   \left(\frac{k}{\gamma }+\frac{1}{\tau }\right)}+\gamma  \left(\left(\Gamma _{\tau }\right){}^*\right){}^2 e^{\frac{2 t}{\tau }}\right){}^2,
\end{eqnarray}
\begin{eqnarray}
    b = \beta ^2 k^2 \left(\gamma ^2-k^2 \tau ^2\right)^4 e^{4 t \left(\frac{2 k}{\gamma }+\frac{1}{\tau }\right)},
\end{eqnarray}
\begin{eqnarray}
    c = \frac{2 \pi ^2 F_0^3 e^{-4 t \left(\frac{2 k}{\gamma }+\frac{1}{\tau }\right)}}{\beta ^2 k \tau  \left(k^2 \tau ^2-\gamma ^2\right)^5} \left(-4 \gamma  k^2 \tau ^2 e^{\frac{2 k t}{\gamma }}+8 \gamma  k^2 \tau ^2 e^{t
   \left(\frac{3 k}{\gamma }+\frac{1}{\tau }\right)}+k \tau  \Gamma _{\tau }^2 e^{2 t \left(\frac{2 k}{\gamma }+\frac{1}{\tau }\right)}-k \tau  \left(\left(\Gamma _{\tau
   }\right){}^*\right){}^2 e^{\frac{4 k t}{\gamma }}\right.\nonumber\\\left.+4 \gamma  k \tau  \left(\Gamma _{\tau }\right){}^* e^{t \left(\frac{k}{\gamma }+\frac{1}{\tau }\right)}+\gamma 
   \left(\Gamma _{\tau }\right){}^* (\gamma -3 k \tau ) e^{2 t \left(\frac{k}{\gamma }+\frac{1}{\tau }\right)}-\gamma  \left(\left(\Gamma _{\tau }\right){}^*\right){}^2
   e^{\frac{2 t}{\tau }}\right) \left(2 \gamma  \tau  \left(\Gamma _{\tau }\right){}^* \left(\gamma ^2+3 k^2 \tau ^2\right) e^{2 t \left(\frac{k}{\gamma }+\frac{1}{\tau
   }\right)}\right.\nonumber\\\left.+8 \gamma  k \tau  e^{\frac{2 k t}{\gamma }} \left(\tau  \left(\gamma ^2+k^2 \tau ^2\right)+t \Gamma _{\tau } \left(\Gamma _{\tau }\right){}^*\right)-8 \gamma  k
   \tau  e^{t \left(\frac{3 k}{\gamma }+\frac{1}{\tau }\right)} \left(2 \tau  \left(\gamma ^2+k^2 \tau ^2\right)+t \Gamma _{\tau } \left(\Gamma _{\tau }\right){}^*\right)\right.\nonumber\\\left.-4
   \gamma  \left(\Gamma _{\tau }\right){}^* e^{t \left(\frac{k}{\gamma }+\frac{1}{\tau }\right)} \left(\gamma ^2 \tau +2 k^2 \tau ^3+k \tau ^2 (\gamma -k t)+\gamma ^2
   t\right)+\tau  \Gamma _{\tau }^4 \left(-e^{2 t \left(\frac{2 k}{\gamma }+\frac{1}{\tau }\right)}\right)\right.\nonumber\\\left.+\left(\left(\Gamma _{\tau }\right){}^*\right){}^3 e^{\frac{4 k
   t}{\gamma }} \left(\tau  \left(\Gamma _{\tau }\right){}^*+2 t \Gamma _{\tau }\right)+2 \gamma  \tau  \left(\left(\Gamma _{\tau }\right){}^*\right){}^3 e^{\frac{2 t}{\tau
   }}\right)
\end{eqnarray}

\begin{eqnarray}
    d = -\frac{2 \pi  F_0^3 e^{-\frac{6 k t}{\gamma }-\frac{4 t}{\tau }}}{\beta  k^2 \Gamma _{\tau }^4 \left(\left(\Gamma _{\tau }\right){}^*\right){}^3} \left(\left(4 \gamma ^2 k \tau -4 k^3 \tau ^3\right) e^{t \left(\frac{2 k}{\gamma }+\frac{1}{\tau
   }\right)}-4 \gamma  \Gamma _{\tau } \left(\Gamma _{\tau }\right){}^* e^{\frac{k t}{\gamma }+\frac{2 t}{\tau }}+\Gamma _{\tau }^2 (\gamma +2 k \tau ) e^{2 t
   \left(\frac{k}{\gamma }+\frac{1}{\tau }\right)}\right.\nonumber\\\left.+k \tau  \left(\Gamma _{\tau }\right){}^* (\gamma +2 k \tau ) e^{\frac{2 k t}{\gamma }}-4 \gamma  k \tau  (2 \gamma +k \tau
   ) e^{t \left(\frac{k}{\gamma }+\frac{1}{\tau }\right)}+3 \gamma ^2 \left(\Gamma _{\tau }\right){}^* e^{\frac{2 t}{\tau }}\right) \left(4 \gamma  k^2 \tau ^2 e^{\frac{2 k
   t}{\gamma }}-8 \gamma  k^2 \tau ^2 e^{t \left(\frac{3 k}{\gamma }+\frac{1}{\tau }\right)}\right.\nonumber\\\left.-k \tau  \Gamma _{\tau }^2 e^{2 t \left(\frac{2 k}{\gamma }+\frac{1}{\tau
   }\right)}+k \tau  \left(\left(\Gamma _{\tau }\right){}^*\right){}^2 e^{\frac{4 k t}{\gamma }}-4 \gamma  k \tau  \left(\Gamma _{\tau }\right){}^* e^{t
   \left(\frac{k}{\gamma }+\frac{1}{\tau }\right)}-\gamma  \left(\Gamma _{\tau }\right){}^* (\gamma -3 k \tau ) e^{2 t \left(\frac{k}{\gamma }+\frac{1}{\tau }\right)}+\gamma
    \left(\left(\Gamma _{\tau }\right){}^*\right){}^2 e^{\frac{2 t}{\tau }}\right),
\end{eqnarray}

\begin{eqnarray}
    e = \frac{F_0^2 e^{-2 t \left(\frac{k}{\gamma }+\frac{1}{\tau }\right)} }{2 k \tau ^2 \Gamma _{\tau }^2
   \left(\Gamma _{\tau }\right){}^*}\left(k \tau ^2 \left(\gamma ^2+6 k^2 \tau ^2+6 \gamma  k \tau \right) e^{\frac{2 k t}{\gamma }}+4 k \tau
    e^{t \left(\frac{2 k}{\gamma }+\frac{1}{\tau }\right)} \left(\gamma ^2 \tau -3 k^2 \tau ^3+k^2 t \tau ^2-\gamma ^2 t\right)\right.\nonumber\\\left.+8 \gamma  e^{\frac{k t}{\gamma }+\frac{2
   t}{\tau }} \left(k^2 \tau ^3+t \Gamma _{\tau } \left(\Gamma _{\tau }\right){}^*\right)-2 \Gamma _{\tau } e^{2 t \left(\frac{k}{\gamma }+\frac{1}{\tau }\right)} \left(3
   k^2 \tau ^3+t \Gamma _{\tau } (\gamma +2 k \tau )\right)-3 \gamma ^2 e^{\frac{2 t}{\tau }} \left(2 t \left(\Gamma _{\tau }\right){}^*-k \tau ^2\right)\right.\nonumber\\\left.+4 \gamma  k \tau 
   e^{t \left(\frac{k}{\gamma }+\frac{1}{\tau }\right)} \left(t (2 \gamma +k \tau )-2 \tau  \left(\Gamma _{\tau }\right){}^*\right)\right),
\end{eqnarray}
\begin{eqnarray}
    f = \frac{F_0^2 e^{-2 t \left(\frac{k}{\gamma }+\frac{1}{\tau }\right)}}{2 \left(\gamma ^2-k^2 \tau
   ^2\right)^2} \left(4 \gamma  \Gamma _{\tau } \left(\Gamma _{\tau }\right){}^* e^{\frac{k t}{\gamma }+\frac{2 t}{\tau
   }}-\Gamma _{\tau }^2 (\gamma +2 k \tau ) e^{2 t \left(\frac{k}{\gamma }+\frac{1}{\tau }\right)}+4 k \tau  \left(\Gamma _{\tau }\right){}^* (k \tau -\gamma ) e^{t
   \left(\frac{2 k}{\gamma }+\frac{1}{\tau }\right)}\right.\nonumber\\\left.-k \tau  \left(\Gamma _{\tau }\right){}^* (\gamma +2 k \tau ) e^{\frac{2 k t}{\gamma }}+4 \gamma  k \tau  (2 \gamma +k
   \tau ) e^{t \left(\frac{k}{\gamma }+\frac{1}{\tau }\right)}-3 \gamma ^2 \left(\Gamma _{\tau }\right){}^* e^{\frac{2 t}{\tau }}\right),
\end{eqnarray}
\begin{eqnarray}
    g = \frac{F_0^2 e^{-2 t \left(\frac{k}{\gamma }+\frac{1}{\tau }\right)}}{\Gamma _{\tau }^3 \left(\Gamma _{\tau
   }\right){}^*} \left(\left(4 \gamma ^2 k \tau -4 k^3 \tau ^3\right) e^{t \left(\frac{2 k}{\gamma }+\frac{1}{\tau
   }\right)}-4 \gamma  \Gamma _{\tau } \left(\Gamma _{\tau }\right){}^* e^{\frac{k t}{\gamma }+\frac{2 t}{\tau }}+\Gamma _{\tau }^2 (\gamma +2 k \tau ) e^{2 t
   \left(\frac{k}{\gamma }+\frac{1}{\tau }\right)}\right.\nonumber\\\left.+k \tau  \left(\Gamma _{\tau }\right){}^* (\gamma +2 k \tau ) e^{\frac{2 k t}{\gamma }}-4 \gamma  k \tau  (2 \gamma +k \tau
   ) e^{t \left(\frac{k}{\gamma }+\frac{1}{\tau }\right)}+3 \gamma ^2 \left(\Gamma _{\tau }\right){}^* e^{\frac{2 t}{\tau }}\right),
\end{eqnarray}
\begin{eqnarray}
    h = \frac{F_0^2 t e^{-2 t \left(\frac{k}{\gamma }+\frac{1}{\tau }\right)}}{k \tau ^2 \Gamma _{\tau }^2 \left(\Gamma
   _{\tau }\right){}^*} \left(\left(4 \gamma ^2 k \tau -4 k^3 \tau ^3\right) e^{t \left(\frac{2 k}{\gamma }+\frac{1}{\tau
   }\right)}-4 \gamma  \Gamma _{\tau } \left(\Gamma _{\tau }\right){}^* e^{\frac{k t}{\gamma }+\frac{2 t}{\tau }}+\Gamma _{\tau }^2 (\gamma +2 k \tau ) e^{2 t
   \left(\frac{k}{\gamma }+\frac{1}{\tau }\right)}\right.\nonumber\\\left.+k \tau  \left(\Gamma _{\tau }\right){}^* (\gamma +2 k \tau ) e^{\frac{2 k t}{\gamma }}-4 \gamma  k \tau  (2 \gamma +k \tau
   ) e^{t \left(\frac{k}{\gamma }+\frac{1}{\tau }\right)}+3 \gamma ^2 \left(\Gamma _{\tau }\right){}^* e^{\frac{2 t}{\tau }}\right).
\end{eqnarray}

\section{ Limits of average and variance for the thermodynamical distributions}
\subsection{Irreversible work distribution}
\subsubsection{Gaussian initial condition}
 There are two limits of both $\mu_{w}$~and~$\sigma_{w}$ that are worth taking a look at: the situation where $F(t)$ behaves like a single-shot pulse $\tau \rightarrow 0^+$, and the case wherein the protocol is applied for a very long, time $t \rightarrow \infty $.
For the former we have,
\begin{eqnarray}
    \lim_{\tau\rightarrow 0^{+}}\mu_{w}(t) &=& \frac{F_{0}^2}{2 k}, \\
    && \nonumber \\
   \lim_{\tau\rightarrow 0^{+}} \sigma_{w}^{2} (t) &=& \frac{F_{0}^{2}}{k\beta},
\end{eqnarray}
whereas for the latter,
\begin{eqnarray}
    \lim_{t\rightarrow\infty} \mu_{w} &=& \frac{\gamma  F_{0}^2}{2 k^2 \tau +2 \gamma  k}, \\
    && \nonumber \\
   \lim_{t\rightarrow\infty}\sigma_{w}^{2} (t) &=& \frac{F_{0}^2 \left(\gamma ^2 - \pi  k^2 \tau^2 \right)}{\beta  k (\gamma +k \tau )^2}.
\end{eqnarray}
Last, we can have a look at the other limiting case, the quasistatic protocol where $\tau = \infty$. In that case, $\mu_{w} =0$ precisely because a quasistatic process is reversible and regarding the variance it yields,
\begin{equation}
\lim _{\tau \rightarrow \infty} \sigma ^2 _w (t) = \frac{F_{0}^{2}}{k \, \beta}.
\end{equation}
\subsubsection{Dirac delta initial condition}
Both the average $\mu_{w,\delta}(t)$ and the variance $\sigma_{w,\delta}^{2}(t)$ are time-dependent quantities. Hence, it reads

On the infinite time limit, $\mu_{w,\delta}(t)$ converges to a constant
\begin{equation}
    \lim_{t\rightarrow \infty} \mu_{w,\delta} (t) \rightarrow \frac{\gamma  F_{0}^2}{2 k^2 \tau +2 \gamma  k}.
\end{equation}
On the impulse limit ($\tau \rightarrow 0^{+}$), $\mu_{w,\delta}(t)$ converges to
\begin{equation}
    \lim_{\tau\rightarrow 0^{+}}\mu_{w, \delta}(t) \rightarrow \frac{F_{0}^2}{2 k},
\end{equation}
With both limits identical to the results for the initial equilibrium distribution case.

On infinite time limit, $\sigma_{w,\delta}^{2}(t)$ converges to
\begin{equation}
\lim_{t\rightarrow\infty}\sigma^{2}_{w,\delta}(t) \rightarrow \frac{F_{0}^2\tau}{\beta(\gamma + k\tau)^2}.
\end{equation}
On the impulse limit, $\sigma_{w,\delta}^{2}(t)$ converges to
\begin{equation}
    \lim_{\tau \rightarrow 0^{+}}\sigma_{w,\delta}^{2}(t) \rightarrow 0.
\end{equation}

\subsection{Irreversible entropy distribution}
\subsubsection{Gaussian initial condition}

On the infinite time limit, $\mu_{s}(t)$ and $\sigma_{s}^{2}(t)$ becomes
\begin{eqnarray}
    \lim_{t\rightarrow \infty} \mu_{s}(t) \rightarrow \frac{\gamma  F_{0}^2}{2 k^2 \tau +2 \gamma  k},\\
    \lim_{t\rightarrow \infty}\sigma_{s}(t) \rightarrow \frac{F_{0}^2 \left(\gamma ^2+\pi  k^2 \tau^2 \right)}{\beta  k (\gamma +k \tau )^2}.
\end{eqnarray}

On the kick-force limit, $\mu_{s}(t)$ and $\sigma_{s}^{2}(t)$ converges to
\begin{equation}
    \lim_{\tau \rightarrow 0^{+}} \mu_{s}(t) \rightarrow \frac{F_{0}^2 \left(1-e^{-\frac{2 k t}{\gamma }}\right)^2}{2 k},
\end{equation}
\begin{equation}
    \lim_{\tau \rightarrow 0^{+}} \sigma^{2}_{s}(t) \rightarrow 2 F_{0}^2 \left(\gamma +(\pi -\gamma ) e^{-\frac{2 k t}{\gamma }}\right).
\end{equation}
\subsubsection{Dirac delta initial condition}
which in the infinite time limit is given by
\begin{equation}
    \lim_{t\rightarrow\infty}\mu_{s,\delta}(t) \rightarrow \frac{F_{0}^2\gamma}{2k(\gamma + k\tau)}.
\end{equation}

which in the infinite time converges to
\begin{equation}
\lim_{t\rightarrow\infty}\sigma_{s,\delta}^{2}(t) \rightarrow \frac{F_{0}^{2}\pi\tau}{\beta(\gamma + k\tau)^2}.
\end{equation}
For kick-time limit ($\tau \rightarrow 0$), $\mu_{s,\delta} (t)$ and $\sigma^{2}_{s,\delta} (t)$ converges to
\begin{eqnarray}
    \lim_{\tau \rightarrow 0^{+}}\mu_{s,\delta}(t) \rightarrow \frac{F_{0}^2 \left(1-e^{-\frac{2 k t}{\gamma }}\right)^2}{2 k}, \\ 
    \lim_{\tau \rightarrow 0^{+}}\sigma^{2}_{s,\delta}(t) \rightarrow -\frac{\pi  F_{0}^2 \left(e^{-\frac{2 k t}{\gamma }}-e^{-\frac{4 k t}{\gamma }}\right)}{\beta  \gamma  k}.
\end{eqnarray}